\documentclass[useAMS,usenatbib,twocolumn]{mn2e}
\usepackage{natbib, graphics, epsfig}

\usepackage{times}
\usepackage{amsmath}
\usepackage{amssymb}
\usepackage{textcomp}
\newcommand{\sm}{\small}

\def\JCAP{J. Cosmol.  Astropart. Phys.}
\def\aap{A\&A}
\def\apj{ApJ}

\def\apjl{ApJ}
\def\mnras{MNRAS}
\def\araa{ARA\&A}
\def\aj{AJ}

\def\physrep{Phys. Rep.}
\def\nat{Nature}

\def\apjs{ApJS}
\def\prd{Phys. Rev. D}

\title[Moving mesh cosmology: numerical techniques and global statistics]
      {Moving mesh cosmology: numerical techniques and global statistics}
      \author[M. Vogelsberger et al.] {\parbox{18.5cm}{ 
          Mark Vogelsberger$^{1}$\thanks{mvogelsberger@cfa.harvard.edu},
          Debora Sijacki$^{1}$, 
          Du\v{s}an Kere\v{s}$^{2,3}$, 
          Volker Springel$^{4,5}$,
          Lars Hernquist$^{1}$
        }\vspace{0.3cm}\\ 
        $^1$Harvard-Smithsonian Center for Astrophysics, 60 Garden Street, Cambridge, MA 02138, USA\\
        $^2$Department of Astronomy and Theoretical Astrophysics Center, University of California, Berkeley, CA 94720-3411, USA\\
        $^3$Department of Physics, Center for Astrophysics and Space Sciences, University of California at San Diego, 9500 Gilman Drive, La Jolla, CA 92093, USA\\
        $^4$Heidelberg Institute for Theoretical Studies, Schloss-Wolfsbrunnenweg 35, 69118 Heidelberg, Germany\\ 
        $^5$Zentrum f\"ur Astronomie der Universit\"at Heidelberg, ARI, M\"{o}nchhofstr. 12-14, 69120 Heidelberg, Germany}

\begin{document}

\pubyear{2011}

\pagerange{\pageref{firstpage}--\pageref{lastpage}} 

\maketitle

\label{firstpage} 
\begin{abstract}

  We present the first hydrodynamical simulations of structure
  formation using the new moving mesh code {\sm AREPO} and compare the
  results with {\sm GADGET} simulations based on a traditional
  smoothed particle hydrodynamics (SPH) technique.  The two codes
  share the same Tree-PM gravity solver and include identical
  sub-resolution physics for the treatment of star formation, but
  employ different methods to solve the equations of
  hydrodynamics. This allows us to assess the impact of hydro-solver
  uncertainties on the results of cosmological studies of galaxy
  formation.  In this paper, we focus on predictions for global baryon
  statistics, such as the cosmic star formation rate density, after we
  introduce our simulation suite and numerical methods.  Properties of
  individual galaxies and haloes are examined by \cite{Keres2011},
  while a third paper by \cite{Sijacki2011} uses idealised simulations
  to analyse in more detail the differences between the hydrodynamical
  schemes.  We find that the global baryon statistics differ
  significantly between the two simulation approaches.  {\sm AREPO}
  shows systematically higher star formation rates at late times,
  lower mean temperatures averaged over the simulation volume, and
  different gas mass fractions in characteristic phases of the
  intergalactic medium, in particular a reduced amount of hot gas.
  Although both hydrodynamical codes use the same implementation of
  cooling and yield an identical dark matter halo mass function, more
  gas cools out of haloes in {\sm AREPO} compared with {\sm GADGET}
  towards low redshifts, which results in corresponding differences in
  the late-time star formation rates of galaxies.  We show that this
  difference is caused by a higher heating rate with SPH in the outer
  parts of haloes, owing to viscous dissipation of SPH's inherent
  sonic velocity noise and SPH's efficient damping of subsonic
  turbulence injected in the halo infall region, and because of a
  higher efficiency of gas stripping in {\sm AREPO}. As
  a result of such differences, AREPO leads also to more disk-like morphologies 
  in the moving mesh calculation compared to {\sm GADGET}. Our results hence
  indicate that inaccuracies in hydrodynamic solvers can lead to
  comparatively large systematic differences even at the level of
  predictions for the global state of baryons in the universe.

\end{abstract}


\section{Introduction}
 
Current theories of galaxy formation are based on the view that the
dominant mass contribution to the Universe is in the form of cold dark
matter (DM), which clusters under gravity and builds up the backbone
of all cosmic structure. Together with the discovery of a large dark
energy (DE) component \citep[][]{1999AJ....118.2675R,
1999ApJ...517..565P}, this led to the emergence of the concordance
$\Lambda$ cold dark matter ($\Lambda$CDM) cosmogony.  The third
component to the energy density in this scenario, the baryons,
condense via radiative cooling at the centres of a population of
hierarchically merging DM haloes, forming galaxies
\citep[][]{Silk1977, 1977MNRAS.179..541R, 1978MNRAS.183..341W,
1984Natur.311..517B}. Although the precise physical nature of DM and
DE is not yet known \citep[but see][for a review of possible DM
candidates]{2005PhR...405..279B}, large-scale predictions of this
$\Lambda$CDM theory show good agreement with a wide range of
observations, among them cosmic microwave background (CMB)
fluctuations \citep[][]{2009ApJS..180..306D}, large-scale clustering
of galaxies in the low-redshift universe
\citep[][]{2010MNRAS.401.2148P}, and the redshift $z\sim 2-3$ power
spectrum probed by the Lyman-$\alpha$ forest
\citep[][]{2009MNRAS.399L..39V}. It is however crucial to further
constrain and test the $\Lambda$CDM paradigm, especially on the
smaller scales of galaxies, which remains a theoretical and
observational frontier.

Structure formation is a highly non-linear process and only the early
stages can be described analytically using linear theory
\citep[][]{1970A&A.....5...84Z}.  However, some statistical properties
of the evolved DM field, in particular the halo mass function, can
also be predicted analytically by combining linear theory and
spherical collapse models, through the Press-Schechter formalism
\citep[][]{1974ApJ...187..425P} and subsequent extensions
\citep[e.g.][]{1991ApJ...379..440B}. These predictions ultimately rely
on calibration through more accurate calculations, which typically can
only be done numerically.  Indeed, while Press-Schechter theory has
been quite successful in describing the abundances of haloes, once
computer models reached high precision it became clear that the
original analytic treatment was not sufficiently accurate for the
quantitative work required in today's era of precision cosmology
\citep[e.g.][]{2001MNRAS.321..372J}.

Over the last two decades, numerical simulations have played a key
role in guiding our knowledge of structure and galaxy formation. This
is especially evident for the CDM component, where the relevant
calculations have reached a high level of sophistication
\citep[e.g.][]{2005Natur.435..629S, 2008MNRAS.391.1685S,
  2008Natur.454..735D, 2010MNRAS.406..896B, 2010arXiv1002.3660K}.
Many important insights have been gained from purely collisionless
simulations, for example the nearly universal density profiles of DM
haloes \citep[][]{1996ApJ...462..563N, 1997ApJ...490..493N,
  2010MNRAS.402...21N}, assembly bias \citep[][]{2005MNRAS.363L..66G},
and the internal structure of haloes on small scales
\citep[][]{2009MNRAS.395..797V, 2011MNRAS.tmp..168V}.  An important
reason for today's reliability of CDM predictions is that the initial
conditions are unambiguously specified in the $\Lambda$CDM model, with
parameters that are tightly constrained by CMB observations
\citep[][]{2011ApJS..192...18K}.  Moreover, the computational problem
is well-posed and comparatively simple, with equations of motion that
for DM involve only gravity. Efficient new algorithms and the rapid
growth of computing power over the few last decades have allowed ever
more detailed theoretical predictions for the DM distribution.  It is
worthwhile, however, to recall that initially such collisionless DM
simulations suffered from numerical artifacts, leading to issues like
the infamous ``over-merging problem'', which resulted in featureless
and smooth DM haloes primarily due to insufficient mass-, force- and
time-resolution \citep[][]{1998ApJ...499L...5M,1999ApJ...524L..19M}.

To understand properties of the observable universe one needs to
relate the dark matter distribution to that of the baryonic
material. This can be done in a number of different ways: (i) using
so-called semi-analytical modelling of galaxy formation
\citep[e.g.][]{1991ApJ...379...52W, 1993MNRAS.264..201K,
  2006RPPh...69.3101B, 2006MNRAS.365...11C, 2010arXiv1008.1786B,
  2010PhR...495...33B,2011MNRAS.tmp..164G} or the closely related
approach of halo occupation distributions \citep{Benson2000}, or (ii)
using direct hydrodynamical simulations
\citep[e.g][]{1989ApJS...70..419H, 1992ApJ...399L.109K,
  1998ApJ...495...80B, 1999ApJ...511..521D, 2005MNRAS.361..776S,
  2010Natur.463..203G}. Approach (i) allows a fast exploration of a
large parameter space of the underlying coarse description of galaxy
formation physics, whereas (ii) makes possible the calculation of a
self-consistent model that requires far fewer assumptions about the
gas dynamics. Nevertheless, both methods share the common problem that
complicated physics on very small scales, like star formation and
associated feedback processes, need to be accounted for in a rather
crude way. This propagates into uncertainties in the interpretation of
simulation results for the hydrodynamic sector. One possible strategy
to get better control over this problem is to calibrate sub-resolution
treatments in semi-analytic models and hydrodynamical simulations
observationally \citep[e.g.][]{2010MNRAS.402.1536S,
  2011MNRAS.tmp..164G}. In addition, numerical artifacts in
hydrodynamical simulation techniques need to be better understood,
hopefully allowing one to reduce or completely eliminate them.

State-of-the-art cosmological hydrodynamical simulations include a
prescription for star formation \citep[e.g.][]{2003MNRAS.339..289S},
radiative cooling \citep[][]{1996ApJS..105...19K,2008MNRAS.385.1443S},
chemical enrichment \citep[e.g.][]{2009MNRAS.399..574W} and various
feedback processes relevant for low- and high-mass systems
\citep[e.g.][]{2007MNRAS.380..877S, 2008MNRAS.389.1137S}. Furthermore,
some simulation codes can also include magnetic fields
\citep[][]{2010ApJS..186..308C,2005JCAP...01..009D}, radiative
transfer \citep[][]{2011MNRAS.411.1678C, 2010MNRAS.tmp.1851P}, cosmic
ray physics \citep[][]{2008A&A...481...33J}, thermal conduction
\citep{Jubelgas2004, Dolag2004}, or black hole physics
\citep{DiMatteo2005, 2005MNRAS.361..776S, 2008ApJ...676...33D}.

Hydrodynamical simulations have been successfully used to study the
Lyman-$\alpha$-forest \citep[e.g.][]{1996ApJ...457L..51H,
  1996ApJ...457L..57K, 2005PhRvD..71f3534V} and the detailed physics
of the intracluster medium, where cluster scaling relations were
obtained that agreed with X-ray observations \citep{Puchwein2008}.
There have also been numerous studies that focused on the formation of
a representative galaxy population
\citep[e.g.][]{Pearce1999,Murali2002,Weinberg2004,Nagamine2005,Crain2009}.
While some successes have been achieved here as well, a general
finding of these simulations is that it appears to be difficult to
reproduce the observed shallow slope of the faint-end of the galaxy
luminosity function and the observed morphological mix of galaxies,
with its high abundance of disk-like galaxies. Although there has been
some considerable progress over the last few years on the latter issue
\citep[e.g.][]{2004ApJ...606...32R, 2007MNRAS.374.1479G,
  2008MNRAS.389.1137S, 2011MNRAS.410.1391A, 2010Natur.463..203G,
  2011arXiv1103.6030G}, one of the outstanding problems is to produce a statistical sample
of galaxies that agrees reasonably well with observations and reproduces the Hubble sequence. But modifications
in the detailed feedback and star formation prescriptions proved to be important and successful
in forming more realistic late type spiral galaxies.
We note that cosmological hydrodynamic
simulations have also made it clear that a proper inclusion of the
baryonic physics can lead to interesting back reactions on the DM
distribution \citep[][]{Duffy2010, 2010ApJ...709.1138D}, potentially
eliminating or reducing the central dark matter cusp
\citep{2010Natur.463..203G}, or forming a dark disk
\citep[][]{2008MNRAS.389.1041R}. It is hence clear that it is
ultimately not sufficient to work with dark-matter only simulations.

Most hydrodynamical simulations of galaxy formation carried out thus
far have used the smoothed particle hydrodynamics (SPH) technique
\citep[][]{1977AJ.....82.1013L, 1977MNRAS.181..375G,
  1992ARA&A..30..543M, 2005RPPh...68.1703M}, where the gas is
discretised into a set of particles for which appropriate equations of
motion can be derived.  The SPH method is well-suited for cosmological
applications owing to its pseudo-Lagrangian character, which
automatically brings resolution elements to regions where they are
needed most, i.e.~collapsing objects like clusters and galaxies. Also,
the conservation properties of SPH in terms of simultaneous
conservation of energy, momentum, mass, entropy and angular momentum
are excellent. A further convenient feature is that a particle-based
gravity solver (needed for the DM component anyway) can be readily
applied to SPH.  These properties have made the method also popular
for applications other than cosmic structure growth, for example for
isolated merger simulations \citep[][]{1992ARA&A..30..705B,
  1991ApJ...370L..65B, 1996ApJ...471..115B, 1994ApJ...437..611M,
  1996ApJ...464..641M, 2003ApJ...597..893N, 2006ApJ...650..791C}.
 
However, different methods for solving the equations of hydrodynamics
in a cosmological context have been employed as well, most of which
are descendants of classic Eulerian methods
\citep[][]{1992ApJS...80..753S} on regular meshes.  The fixed
Cartesian grids originally available in the corresponding codes are
insufficient to capture the large dynamic ranges encountered in galaxy
formation, but finite volume schemes incorporated in modern adaptive
mesh refinement (AMR) codes \citep[][]{1989JCoPh..82...64B,
  2002A&A...385..337T, 2004astro.ph..3044O} can alleviate this
problem. Interestingly, these mesh-based methods have led in some
cases to significantly different results compared with SPH. A
prominent example of these discrepancies is highlighted in the Santa
Barbara cluster comparison project \citep[][]{1999ApJ...525..554F},
where it was found that the entropy profiles of clusters are
significantly and systematically different between AMR and SPH codes,
the former yielding a large entropy core which is absent in the SPH
calculations. Only recently has some progress been made in identifying
the cause for this difference \citep{Mitchell2009}, but the issue is
still not fully understood; hence we return to it in one of our companion
papers \citep{Sijacki2011}.

It thus appears that there are at least two important challenges in
the field of cosmological hydrodynamics simulations. One lies in an
adequate treatment of all relevant physics of galaxy formation, in
particular with respect to the reliability of the parameterisations of
sub-resolution physics, and the other having to do with the accuracy
and efficiency of the underlying hydrodynamics solver.  It is of
course crucial to strive for a more faithful treatment of the physics
in the simulations, but it should also be evident that a proper
modelling of the physics relies on a correct solution of the basic
hydrodynamical equations in the first place. It is therefore
problematic to tune the sub-resolution physics without first verifying
in detail the quality of the hydro solver, because this risks
incorrectly absorbing deficiencies of a numerical technique into
distorted physics models.  Here, in this initial study we shall
primarily be concerned with an assessment of the extent to which
uncertainties in numerical methodology are reflected in the predicted
galaxy properties, for a fixed physics model.  In future works, we
will explore the consequences of various treatments of feedback
effects, which are known to be crucial for the galaxy formation
problem \citep{Scannapieco2012}.
 
We remark that it is not obvious whether SPH or AMR is more accurate
in all simulation regimes, as both methods are known to have
shortcomings.  For example, SPH suffers from relatively poor shock
resolution, noise on the scale of the smoothing kernel, and
low-order accuracy for the treatment of contact discontinuities.
Furthermore, some hydrodynamic instabilities like the Kelvin-Helmholtz
instability can be suppressed in SPH \citep[][]{2007MNRAS.380..963A}.
Recently, \cite{Bauer2011} also showed that conventional implementations
of SPH do not properly properly subsonic turbulence, which is
generically present in cosmological haloes.  Astrophysical Eulerian
methods on the other hand (usually realised as AMR finite-volume
schemes) may suffer from over-mixing due to advection errors in the
presence of bulk flows.  One of the principal differences between SPH
and AMR lies in their handling of mixing at the level of individual
fluid elements.  This is absent in SPH by construction (unless added
in crudely by hand somehow), while in AMR it occurs implicitly through the
averaging of the evolved Riemann solutions over the scale of the
grid-cells at the end of each timestep.  It is not always clear which
scheme yields a more accurate result \citep[][]{2007MNRAS.377..997T,
  2009MNRAS.395..180M}.

Another issue with classical AMR codes is that their handling of the
discrete equations of motion can lead to errors with the fluid is
moving rapidly across the mesh \citep[][hereafter
  S10]{2010MNRAS.401..791S}.  Because the truncation error of Eulerian
codes depends on the fluid velocity relative to the grid, the results
can thus be sensitive to the presence of bulk velocities \citep[see
  also][]{2008MNRAS.390.1267T, 2008MNRAS.387..427W}.  Finally, a
further challenge for AMR schemes in cosmic structure formation arises
from the need to accurately follow the gravitational growth of even
very small structures.  The mesh-based Poisson solvers typically
employed by AMR codes have been shown to lack sufficient small-scale
force accuracy and to produce too few low mass haloes
\citep[][]{2005ApJS..160....1O,2008CS&D....1a5003H} potentially
corrupting the solution on even well-resolved scales. While this can
be addressed with more aggressive refinement strategies, the
discontinuous changes in gravity resolution brought about by such
refinements introduce non-Hamiltonian perturbations into the dark
matter dynamics which are in principle undesirable.

Recently, S10 introduced a new moving-mesh approach as embodied in the
{\sm AREPO} code. The principal goal is similar to the earlier
implementations of \cite{1995ApJS...97..231G} and
\cite{1998ApJS..115...19P}, but these moving-mesh algorithms suffered
from grid distortions which limited their applicability. {\sm AREPO}
does not use coordinate transformations like previous moving mesh
codes in cosmology, but instead employs an unstructured Voronoi
tessellation of the computational domain. The mesh-generating points
of this tessellation are allowed to move freely, offering significant
flexibility for representing the geometry of the flow. As discussed in
detail in S10, this technique avoids several of the weaknesses of SPH
and AMR schemes while it retains their most important advantages.  For
example, if the mesh motion is tied to the gas flow, the results are
Galilean-invariant (like in SPH), while at the same time a high
accuracy for shocks and contact discontinuities is achieved (like in
Eulerian schemes).

The aim of this paper is to compare the novel cosmological
hydrodynamics code {\sm AREPO} with the widely used SPH code {\sm
  GADGET} \citep[][]{2005MNRAS.364.1105S}. In this first paper of a
series we give an overview of the numerical techniques used for
our galaxy formation simulations, introduce our simulation set,
present an analysis of the global baryon characteristics, and evaluate
the performance and efficiency of the new simulation scheme.  In one
companion paper \citep[][hereafter Paper II]{Keres2011}, we discuss
the properties and statistics of individual galaxies and haloes in the
simulations. A further companion paper \citep[][hereafter Paper
  III]{Sijacki2011} analyses idealised test problems to highlight
various differences of the two simulation schemes and to elucidate how
they affect galaxy formation.

The outline of this paper is as follows. In Section~2, we provide a
discussion of the initial conditions and numerical details which are
particularly important for the goal of our comparison project.  There
we also discuss some modifications of the {\sm AREPO} code that were
adopted during the course of this work. Section~3 presents first
results and discusses the large-scale structure and global statistics
of the baryon content in the simulations at different resolutions. We
turn to an analysis of the origin of the cooling difference we find
between the codes in Section~4. In Section~5 we discuss some generic
issues and problems of SPH. Finally, we give a summary of our results
and our conclusions in Section~6.  In an Appendix, we provide data on
the mesh geometry, analyse the performance of the codes in different
parts of the calculations and present scaling tests.

\section{Numerical methods and simulation set}

\subsection{Implemented physics}

Our {\sm GADGET} and {\sm AREPO} simulations follow the same physics,
consisting of a collisionless dark matter fluid and an
ideal baryonic gas. Both components act as sources of gravity and
are evolved in a Newtonian approximation on top of an expanding
Friedman-Lemaitre background model. We describe the gas dynamics with
the ordinary inviscid Euler equations, augmented with additional terms
that account for ``extra physics'' related to radiative processes and
star formation. In particular, we account for optically thin
radiative cooling of a primordial mixture of helium and hydrogen, with
a hydrogen mass fraction of $0.76$. We also assume a spatially
uniform, time-dependent ionising UV background with an amplitude and
time evolution according to \cite{2009ApJ...703.1416F}.

As a result of cooling, gas can collapse to high density and turn into
stars.  Both simulation codes describe this process with the simple
star formation and supernova feedback model introduced in
\cite{2003MNRAS.339..289S}, which has been used in many SPH
simulations in recent years.  In this approach, every sufficiently dense
gas particle/cell is treated as a representative region of the
interstellar medium (ISM) and is assumed to exhibit a multiphase
structure consisting of cold and hot gas in pressure
equilibrium. Stars form out of the cold gas and return energy back
into the medium though supernova explosions. This energy feedback
pressurises the multiphase medium, the effect of which is described in
terms of an effective equation of state that is imposed onto the gas
once it has overcome the density threshold for star formation. 
The value of the density threshold we use
in this work is $n_{\rm H}=0.13~{\rm cm}^{-3}$, set such that isolated
disk galaxies at $z=0$ are characterised with a relation between disk
surface gas density and star formation density that agrees with the
observed Kennicutt law \citep[][]{1998ApJ...498..541K}.  To reach
these densities gas needs to fall into dark matter haloes and
dissipate energy via radiative cooling to eventually form galaxies
\citep[][]{1978MNRAS.183..341W,1991ApJ...379...52W}.  Once the gas
density is higher than the threshold value, particles/cells are
eligible to form stars and can be converted into collisionless stellar
particles.  We treat this conversion as a stochastic process that
generates one generation of stellar particles per hydrodynamic
resolution element (SPH particle or Voronoi cell, respectively) at an
average rate equal to the star formation rate predicted by the
subresolution multi-phase model.

We stress again that for the purposes of our study it is crucial to
have an identical implementation of this extra physics in both codes
in order to allow a straightforward comparison of the two
hydrodynamical schemes. Fortunately, the quasi-Lagrangian nature of
{\sm AREPO} typically allows an easy adoption of methods originally
developed for particle-based schemes like SPH. We could hence adopt
most of the ``extra physics'' implemented in {\sm GADGET} in a largely
unmodified form in {\sm AREPO}, helping to ensure that possible
implementation differences for this physics do not affect our
conclusions.  We also emphasise that in our present context it is
preferable to employ a relatively simple subresolution model like that
of \cite{2003MNRAS.339..289S} in order to facilitate a clear
assessment of the differences between hydro solvers. In particular,
for now we have not included the effects of strong feedback capable of
inducing galactic winds and outflows.  We are thus not attempting to
solve the galaxy formation problem at this stage, but are instead
concerned with identifying and understanding issues related to the
treatment of the hydrodynamics.

\subsection{Simulation Codes}

In the following two subsections we highlight the most relevant
features and parameter settings of the codes we used, as well as any
particular changes we made for this project. We note that full details
are presented in substantial depth in the corresponding code papers,
so in the interest of brevity we here restrict ourselves only to those
aspects directly pertinent to our study. We are also aided by the fact
that the {\sm AREPO} and {\sm GADGET} codes share the same gravity
solver, something that was not the case in previous comparison studies
of SPH and AMR \citep[e.g.][]{Regan2007}.  Consequently, the evolution
of the collisionless DM component is identical in both codes in the
limit of vanishing hydrodynamical forces. This similar handling of
self-gravity is important for isolating differences caused primarily
by the hydro solvers.

\subsubsection{\sm GADGET}

{\sm GADGET} is a widely used and well-tested SPH-code which employs a
formulation of SPH that simultaneously conserves energy and entropy
despite the use of fully adaptive smoothing lengths
\citep[][]{2002MNRAS.333..649S}. The gravity calculation is split into
short- and long-range components, where short-range forces are
calculated with an hierarchical oct-tree algorithm
\citep{1986Natur.324..446B, 1987ApJS...64..715H} and the long-range
forces are evaluated with a particle mesh (PM) method
\citep[e.g.][]{1981csup.book.....H}.  Our {\sm GADGET} runs are based
on a default setting of the numerical SPH parameters, using 32
neighbours in smoothed estimates and an artificial viscosity parameter
of $\alpha = 1.0$, combined with Balsara's switch
\citep[][]{1995JCoPh.121..357B} to reduce the viscosity in the
presence of strong shear.

\subsubsection{\sm AREPO}

{\sm AREPO} is a second-order accurate finite volume method that
solves the Euler equations using piece-wise linear reconstruction and
a calculation of hydrodynamical fluxes at each cell interface with an
exact Riemann solver. The basic solution strategy of the code is that
of the well-known MUSCL-Hancock scheme. What makes {\sm AREPO} unusual
is that it employs an unstructured mesh based on a Voronoi
tessellation using a set of mesh-generating points. Furthermore, the
mesh is allowed to move freely as a function of time.  In our runs, we
tie the motion of the mesh-generating points to the hydrodynamical
flow, which gives the scheme a quasi-Lagrangian character and makes
the mesh automatically adaptive. As S10 demonstrated, such an approach
offers a number of advantages compared with ordinary Eulerian or
Lagrangian codes, including a reduction of numerical diffusivity and
advection errors and an improved handling of contact
discontinuities.  In the following, we discuss some details and
modifications of {\sm AREPO} relevant for our study.

{\bf(i) Mesh regularisation:} It is desirable to have a Voronoi mesh
geometry where the geometric centre of each Voronoi cell lies
reasonably close to the mesh-generating point of that cell. This
reduces the size of errors in the linear reconstruction step of the
MUSCL-Hancock scheme and also limits the rate at which mesh faces turn
their orientation during the mesh motion. {\sm AREPO} therefore
incorporates a method to regularise the mesh, as described in S10. This
scheme uses a modified Lloyd algorithm to drift the mesh a bit towards
its centroidal configuration at each time step.  To this end the
motion of the mesh-generating points contains an additional velocity
component for certain cells, designed to move a mesh-generating
point closer towards the geometric centre of its cell. The default
setup of {\sm AREPO} uses the spatial offset between the actual
location of a vertex and its cell's geometric centre in units of the
cell's size to decide whether or not this corrective motion should
be added to the cell's vertex velocity.

However, a number of test simulations revealed that this
regularisation scheme can introduce some unwanted side
effects. Specifically, systematic differences in the average mass of
gaseous cells can develop between the star-forming gas and the lower
density material in the surrounding halo, with the former tending to
become systematically more massive than the latter. That such a
tendency exists in principle is to be expected because the Lloyd
algorithm would ultimately converge to a homogeneous Voronoi mesh with
cells of equal volume if it applied to every cell centre with full
intensity all the time.  The mesh regularisation motions in the
default scheme have hence the tendency to move the mesh away from the
densest regions (but leaving the gas there -- in {\small AREPO}, the
gas motion and the mesh motion are independent), such that the mass
per cell will tend to increase in the densest regions.  This is
somewhat undesirable because ideally we want to have the highest mass
resolution in the central parts of galaxies and not in the surrounding
low density gas.
 
There are at least two solutions to this problem (besides disabling
the regularisation scheme altogether). One is to simply make the
settings of the standard regularisation less aggressive, thereby
reducing the systematic effects described above. Another possibility
is to change the criterion that decides when such a mesh correction
motion should be applied.  We have found that the simple criterion
originally implemented in {\sm AREPO}, based on the displacement of a
mesh-generating point from the geometric centre of its cell, is prone
to invoke mesh regularisation drifts even when they are not
needed. This happens especially in regions with strong density
gradients, where such displacements naturally occur even if the mesh
geometry is acceptable. As an alternative criterion we have therefore
considered the maximum opening angle under which a cell face is seen
from the mesh-generating point. This identifies problematic cell
geometries more directly, which are in fact those where a point lies
close to a wall, or equivalently, where this opening angle is
large. We find that this opening angle criterion is more tolerant to
rapid changes in spatial resolution, and is hence less prone to
triggering unwanted mesh-correction motions in the presence of large
density gradients.  Because of this it virtually eliminates the mass
segregation effect mentioned above.

For definiteness, in the simulations presented here, we invoke
mesh-regularisation motions if the maximum face angle of a Voronoi
cell (defined as the maximum of $\sqrt{A_i / \pi} / h_i$, where $A_i$
is the area of a face and $h_i$ its distance to the vertex) is larger
than $1.68$. In this case we add a velocity component up to half the
sound speed of the corresponding cell to the vertex velocity, which
moves the mesh-generating point closer towards the geometric centre of
the Voronoi cell. Ideally, this scheme should then guarantee
reasonably regular cells where the bulk of all cells will have a
maximum face angle less than $1.68$.  Our tests confirm that this is
the case, while at the same time the mass bias between star forming
high density cells and the lower density cells in the surrounding halo
is significantly reduced. The face angle scheme applies the mesh-correction
motions more rarely than the original approach but still frequently
enough to maintain a regular mesh. In the Appendix, we discuss some
measurements of the mesh statistics, which also show that this new
regularisation scheme still ensures that the offsets between
mesh-generating points and the geometric centres of the corresponding
cells remain small, as is desired to minimise errors in the linear
reconstruction step.

{\bf(ii) Refinement and de-refinement:} {\sm AREPO} is
quasi-Lagrangian in the sense that the vertices of the Voronoi mesh
follow the flow of the fluid such that an approximately constant mass
resolution is automatically achieved.  The code is not purely
Lagrangian, however, because mass can be exchanged between cells in a
manner consistent with the equations of motion, in order to
prevent the mesh from becoming highly distorted.  Moreover,
the code can also make use of re- and derefinement of individual cells
in order to improve the local resolution beyond that offered by
the dynamical motion of the mesh or to improve efficiency when
high resolution is no longer required in certain regions.  The method
employed by {\sm AREPO} is more general than that used by standard AMR
methods, which typically employ subgrids to refine a certain section of
the parent grid, thereby creating a hierarchy of overlapping grid
patches. In {\sm AREPO}, we are not required to use subgrids, but 
can split or
merge individual cells, such that a single global mesh with a smooth
transition from low- to high-resolution parts results. The criteria
for when a change of the local resolution should be introduced are
very flexible, just as in the AMR technique.  Indeed, because the
initial distribution of mesh generating points and the manner in
which they are updated in time are arbitrary, {\sm AREPO} offers
the capability of running in an AMR mode.

In our cosmological galaxy formation simulations it is desirable to
maintain a roughly constant mass resolution. This is because in runs
with star formation whole gas cells can turn into collisionless
stellar particles, and it is best to create them with approximately
the same mass to avoid potential two-body heating effects among these
star particles.  A relatively homogeneous mass resolution in the gas
phase is also warranted to guarantee a more proper comparison to the
SPH runs, where a fixed mass resolution is present by construction.
(We note, however, that this fixed mass resolution in SPH is a
consequence of this method's inability to correctly represent the
flow on small scales, as we discuss in Section~5.3.)

We have therefore introduced a new re-/derefinement scheme in {\sm
  AREPO} that ensures that the mass resolution in the gas and the
spawned star particles never deviates too much from the value of the
SPH particle mass in the corresponding {\sm GADGET} simulation.
Specifically, we ``force'' cells to have a mass in the range $1/2
\lesssim m_{\rm cell}/m_{\rm target} \lesssim 2$, where we set the
target gas mass $m_{\rm target}$ equal to the mean cell mass assuming
a homogeneous gas distribution in the simulation volume with equal
volume cells. This corresponds then exactly to the mass of an
individual SPH particle in the {\sm GADGET} calculation at the same
particle/cell number.  We allow only quite ``roundish'' cells to be
refined, specifically we only refine a cell if the maximum face angle
is smaller than $3.38$.  This protects against introducing large local
errors by splitting very irregular cells (which are usually cells that
have just been split in the previous timestep). As a result of this
constraint our simulations may contain at any given time a few cells
which are more massive than $2 \times m_{\rm target}$. We have also
done runs where only cells with a sufficiently shallow temperature
gradient across them were allowed to be derefined, but this did not
make any significant difference to our results.

We also slightly modified the star formation implementation to take
this re-/derefinement scheme into account.  Specifically, if a star
particle is formed in a cell with mass less than $2 \times m_{\rm
  target}$, it inherits the full mass and the cell is removed.
Otherwise, a star particle is created with a mass equal to $m_{\rm
  target}$, leaving the rest of the mass in the (surviving) cell.
Cells that do not contain at least $1/4 \times m_{\rm target}$ do not
produce stellar particles, which prevents very low mass
collisionless particles from being formed. We note that such low mass
cells are exceedingly rare, because cells with less than $1/2 \times
m_{\rm target}$ are normally all removed during the derefinement
process.  However, since two neighbouring cells will never be dissolved
in the same timestep (see the discussion in S10), a cell can in rare
cases briefly scatter below $1/2 \times m_{\rm target}$ and
potentially also attempt to spawn a stellar particle during this time.

This modification of the star formation implementation guarantees that
star particle masses are bounded by the interval $1/2 \lesssim
m_\star/ m_{\rm target} < 2 $.  Our {\sm GADGET} and {\sm AREPO}
simulations both use only one generation of stellar particles, which
implies that the total number of baryonic resolution elements in the
{\sm GADGET} simulations is strictly conserved, while in {\sm AREPO} it
can fluctuate slightly around the initial value.

{\bf(iii) Dual entropy formalism:} As discussed in S10, finite volume
hydro solvers can suffer from spurious heating in highly supersonic,
cold parts of the flow, where the kinetic energy dominates
over the thermal energy.  Especially at high redshifts in cosmological
simulations, this can lead to an incorrect temperature evolution in
low-density gas as a result of discretisation errors in the advection
of the kinetic energy.  Owing to its quasi-Lagrangian character {\sm
  AREPO} is somewhat less prone to this effect than fixed mesh
techniques, but it can still be present at some level.  To circumvent
this problem, {\sm AREPO} can optionally force the entropy to be
conserved instead of the energy during a timestep. In this case, no
spurious heating will be introduced. However, entropy is conserved only
outside of shocks in adiabatic parts of the flow, so a criterion is
required to decide on a cell-by-cell basis whether energy or entropy
conservation should be given precedence for a cell's evolution over
the current timestep.

As S10 showed, this dual entropy formalism is especially useful for
non-radiative simulations, because here a spurious heating cannot be
compensated by radiative cooling.  Since our simulations include
radiative cooling due to a primordial gas mixture we expect this to be
less problematic for our applications. Indeed, extensive test
simulations showed that our results are not sensitive to whether or
not the dual entropy formalism is used.  We have therefore decided to
exclusively use strict energy conservation in all our production
calculations, without employing the dual entropy formalism.

{\bf(iv) Gravitational softening length:} {\sm GADGET} uses a global
gravitational softening length for collisional and collisionless
particles, i.e.~for gas and stellar/dark matter particles. {\sm AREPO}
also uses a global gravitational softening length for collisionless
particles (which we set equal to that used in our {\small GADGET}
simulations). However, as the code is a finite volume method which
uses gas cells of variable size to represent the fluid, we have used
individual Plummer-equivalent gravitational softening lengths
according to $\epsilon_{\rm cell}=\epsilon_{\rm fac} \times r_{\rm
  cell}$ for each gaseous cell, where $r_{\rm cell}=(3 V_{\rm cell}/ 4
\pi)^{1/3}$ is the cell's fiducial radius assuming the cell volume is
spread over a sphere, and $\epsilon_{\rm fac}=2.5$ is an overall
scaling factor. In addition, we imposed a lower floor on the
gravitational softening lengths equal to that of the SPH particles in
the corresponding {\small GADGET} simulation. This was done in order
to safeguard against a potentially higher gravitational resolution in
{\sm AREPO} in the densest gas, which may have distorted our
comparison. However we note that we do not expect any significant
difference due to the slightly different treatment of the
gravitational softening length of the gas even if such a floor was not
used.  Indeed, as described in Appendix~B, we have repeated lower
resolution {\sm AREPO} simulations with a fixed softening length for
all gas cells and obtained nearly identical results. Also, we have
repeated simulations with adaptive gravitational softening without a
lower floor and again obtained equivalent results. We note that the
presence of an effective equation of state for highly dense gas is
ultimately responsible for this insensitivity.  Without it, we would
get fragmentation of very dense gas where the adaptive softening 
could make a difference \citep{Bate1997}.

\begin{table*}
\begin{tabular}{cccccccc}
\hline Name         & Code                        & Boxsize $[h^{-3}\,{\rm Mpc}^3]$         & hydro elements           & DM particles    & $m_{\rm target/SPH} [h^{-1}\,{\rm M}_\odot]$    & $m_{\rm DM}[h^{-1}\,{\rm M}_\odot]$       & $\epsilon$ [$h^{-1}\,{\rm kpc}$]\\
\hline
\hline
       A\_L20n512  & {\sm AREPO}                 & $20^3$                             & $512^3$                  & $512^3$         & $7.444\times 10^5$                        & $3.722\times 10^6$                  & $1$ \\
       G\_L20n512   & {\sm GADGET}                & $20^3$                             & $512^3$                  & $512^3$         & $7.444\times 10^5$                        & $3.722\times 10^6$                  & $1$ \\
\hline
       A\_L20n256  & {\sm AREPO}                 & $20^3$                             & $256^3$                  & $256^3$         & $5.955\times 10^6$                        & $2.977\times 10^7$                  & $2$ \\
       G\_L20n256   & {\sm GADGET}                & $20^3$                             & $256^3$                  & $256^3$         & $5.955\times 10^6$                        & $2.977\times 10^7$                  & $2$ \\
\hline
       A\_L20n128  & {\sm AREPO}                 & $20^3$                             & $128^3$                  & $128^3$         & $4.764\times 10^7$                        & $2.382\times 10^8$                  & $4$ \\
       G\_L20n128   & {\sm GADGET}                & $20^3$                             & $128^3$                  & $128^3$         & $4.764\times 10^7$                        & $2.382\times 10^8$                  & $4$ \\
\hline
\end{tabular}
\caption{Names and basic setup of our different simulations. All
  calculations were performed in a periodic box with sidelength
  $20\,h^{-1}\,{\rm Mpc}$. The number of baryonic resolution elements
  (SPH particles/Voronoi cells, and star particles) is fixed in the
  {\sm GADGET} runs, but can vary due to re- and derefinement
  operations in the {\sm AREPO} simulations.  We enforce a refinement
  scheme that keeps cell masses close to a target gas mass $m_{\rm
    target}$, which we set to the SPH particle mass of the {\sm
    GADGET} run at the same particle/cell number. The comoving Plummer
  equivalent gravitational softening length $\epsilon$ is constant in
  {\sm GADGET}, but adaptive (with a floor) for gas cells in {\sm AREPO}.
  Additional
  {\sm AREPO} simulations with a fixed gravitational softening length
  yielded essentially equivalent results, as described in Appendix~B.}
\label{table:simulations}
\end{table*}

\subsection{Code comparison strategy}

In any comparison between codes as different as {\sm GADGET} and {\sm
  AREPO}, a number of different benchmarks can be used to gauge their
relative performance, and various criteria can be defined to align
runs that are deemed comparable with each other.  In our study, we
have chosen to compare {\sm GADGET} and {\sm AREPO} at matching mass
resolution, using the same initial number of baryonic and dark matter
resolution elements.  This provides a well-defined comparison strategy
and is advantageous for a number of reasons:

(1) First, we can use the same initial conditions for the simulations
with both codes.  This would not be possible if we performed the runs
with different initial mass resolutions. (2) Second, our choice makes
it straightforward to relate the same objects between the different
runs (and they are resolved in dark matter in the same way).  (3)
Third, as we discuss below, for the same initial mass resolution, the CPU time
requirements for {\sm GADGET} and {\sm AREPO} are similar, at least
for cosmological simulations, with the moving mesh runs being only
some $\sim 30\%$ more costly.  In many cases, the limiting factor
for the simulation size that can be achieved is the amount of computing
time required, so our choice effectively enables a comparison for
fixed computational resources.

Of course, in principle one may also attempt a comparison at
equivalent spatial resolution in the gas between {\sm GADGET} and {\sm
  AREPO}, or at the same accuracy in the solution obtained.  We have
not attempted this for the following reasons.  First, while the spatial
resolution that can be achieved for a given number of resolution
elements is expected to be worse in SPH compared to {\sm AREPO}, the
degree to which this matters is likely very problem-dependent. For
example, for our treatment of star formation, the mass resolution is
much more important than a precise treatment of
fluid discontinuities. Aligning the spatial resolutions such that 
discontinuities are broadened over the same scale in both codes would
lead to drastically different mass resolutions in the star-forming
phases and hence likely to large systematic resolution effects in the
smallest galaxies.
 
Second, and even more important in our view, is that the convergence
properties of SPH are not well understood, making a comparison at
fixed accuracy a highly problematic undertaking from the start.  As we
discuss at length in Section~5, formal convergence in SPH ultimately
requires that the number of neighbouring particles used in smoothed
estimates around a particular location be increased as the total
number of particles is made larger.  This has typically not been done
in cosmological applications, and is also in practice not readily
possible due to clumping instabilities
\citep[e.g.][]{Schuessler1981}.

Moreover, while SPH is
often described as being ``Lagrangian,'' this is not formally correct
because individual fluid elements, as represented by SPH particles,
are not allowed to deform arbitrarily
in response to e.g.~shearing motions, as we
illustrate in Section~5.  SPH should, therefore, properly be referred
to as ``pseudo-Lagrangian.''  The errors that this feature of SPH
entails depend on the detailed properties of the flow, in addition to
the local spatial resolution, and consequently cannot be assessed in
general.  In its most basic formulation, SPH may simply lack a formal
convergence condition, making it unclear how a comparison at fixed
accuracy should be defined.

Finally, in our study, we have chosen a particular implementation of
``standard SPH'', as implemented by the {\sm GADGET} code.  
Moreover, by choosing {\sm GADGET} for our SPH simulations,
we are able to use the same gravity solver and the same
physics in our comparisons between SPH and {\sm AREPO}, which might
not be possible otherwise.  This enables us to isolate differences in
runs that owe primarily to the differences in the hydro solvers
between {\sm GADGET} and {\sm AREPO}. As we discuss in Section~5,
various modifications have been proposed to the basic structure of SPH
in order to improve its reliability under some circumstances.  It is,
of course, possible that we would have obtained somewhat different
results had we employed such formulations of SPH.  However, we believe
that many limitations of SPH are generic and are not specific to {\sm
  GADGET}.  Indeed, as we discuss in Section~5, it is difficult to see
how SPH could be modified to entirely eliminate, for example, the
sources of error that owe to its pseudo-Lagrangian nature without
radical modifications to the underlying algorithm.

We also stress again that the primary goal of our comparison is not to
arrive at the best fit to observational data, or to achieve the
highest possible resolution for single galaxy models.  While we
acknowledge that important strides have been made recently in studying
disk formation numerically using zoom-in procedures
in cosmological SPH simulations
\citep[e.g.][]{2011MNRAS.410.1391A, 2011arXiv1103.6030G},
we do not think these successes provide much evidence for the
numerical reliability of SPH, although this misconception appears
widespread.  Our ultimate aim is not merely to model single objects,
but entire populations of galaxies so that we can statistically
constrain uncertain processes associated with star formation and
feedback by using, e.g., observations of the distribution of galaxies
with respect to their Hubble type. This requires the identification of
accurate and computationally efficient techniques that are best
suitable for the task. In particular, our strategy is designed to
detect systematic inaccuracies in current simulation techniques that
may easily be concealed by the adjustment of heuristic feedback
parameters to match a certain observation, as it is commonly done,
while at the same time these effects may have serious consequences in
other places.

\subsection{Simulation Suite}

For our simulation set we adopt the cosmological parameters
$\Omega_{m0}=0.27$, $\Omega_{\Lambda0}=0.73$, $\Omega_{b0}=0.045$,
$\sigma_8=0.8$, $n_s=0.95$ and $H_0=100\,h\,{\rm km}\,{\rm s}^{\rm
  -1}\,{\rm Mpc}^{\rm -1}= 70\,{\rm km}\,{\rm s}^{\rm -1}\,{\rm
  Mpc}^{\rm -1}$ ($h=0.7$). These parameters are consistent with the
most recent WMAP-7 measurements \citep[][]{2011ApJS..192...18K} as
well as with a host of other cosmological constraints.  We create a
realisation of this cosmology in a periodic box with a sidelength
$20\,h^{-1}\,{\rm Mpc}$, at three different mass resolutions corresponding
to $2 \times 128^3$, $2 \times 256^3$ and $2 \times 512^3$
particles/cells. A comoving gravitational softening length of $4\,h^{-1}\,{\rm
  kpc}$, $2\,h^{-1}\,{\rm kpc}$ or $1\,h^{-1}\,{\rm kpc}$, respectively, is used
for the dark matter, and for the gas particles in the {\sm GADGET}
runs. For {\sm AREPO}, we start initially with the same number of
cells as there are particles in the corresponding SPH run, but we use
an adaptive gravitational softening with a floor
as described earlier. Note that
due to re- and derefinement the number of baryonic resolution elements
(cells plus star particles) is not strictly conserved during the
course of a simulation but can fluctuate slightly around the initial
number.  Initial conditions were generated at $z=99$ based on the
power spectrum fit of \cite{1999ApJ...511....5E}, with gas
particles/cells added to the initial conditions by splitting each
original particle into a dark matter and gas particle/cell pair,
displacing them with respect to each other such that two interleaved
grids are formed, keeping the centre-of-mass of each pair fixed.

\begin{figure*}
\centering
  \includegraphics[width=0.495\textwidth]{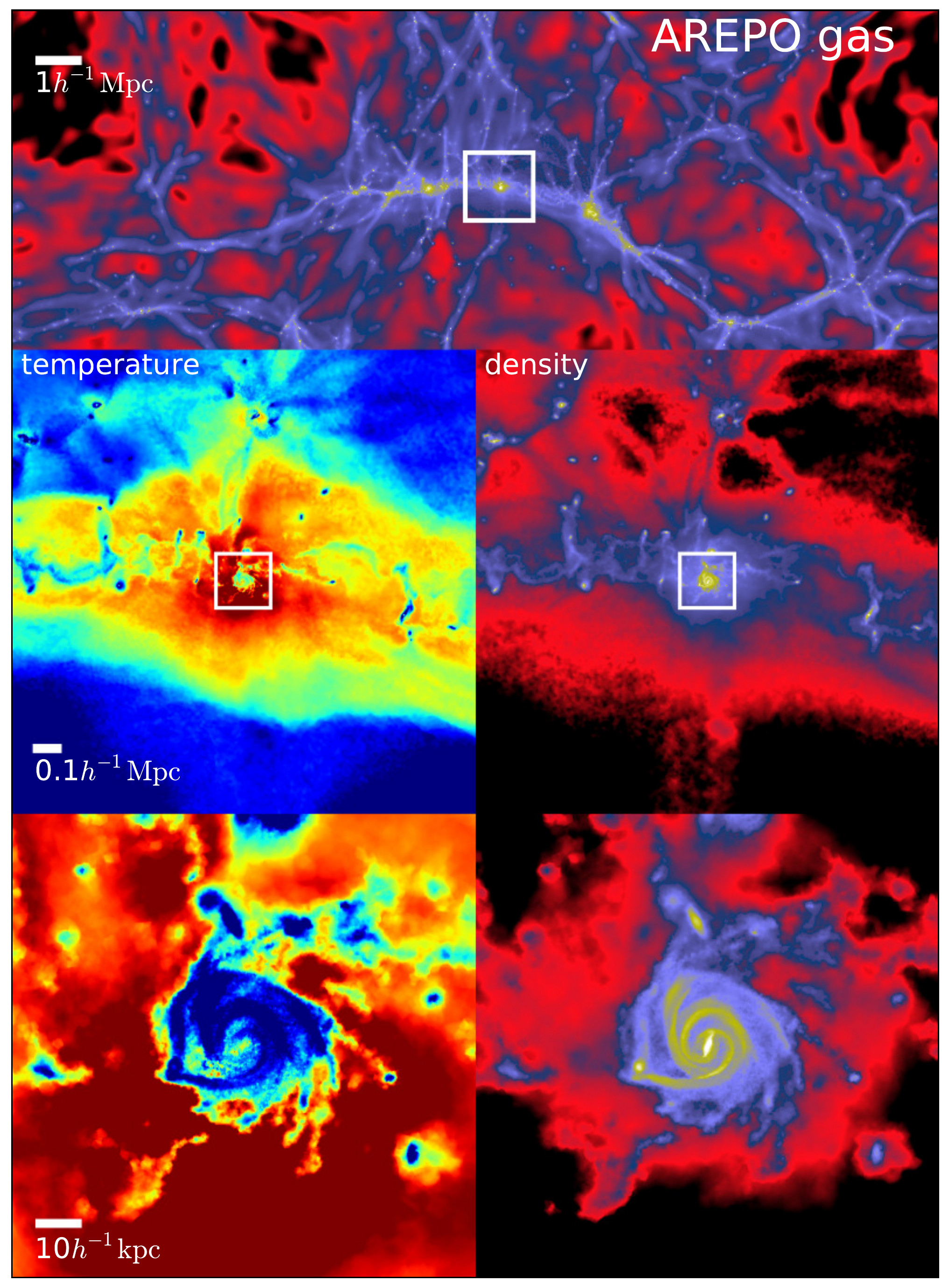}
  \includegraphics[width=0.495\textwidth]{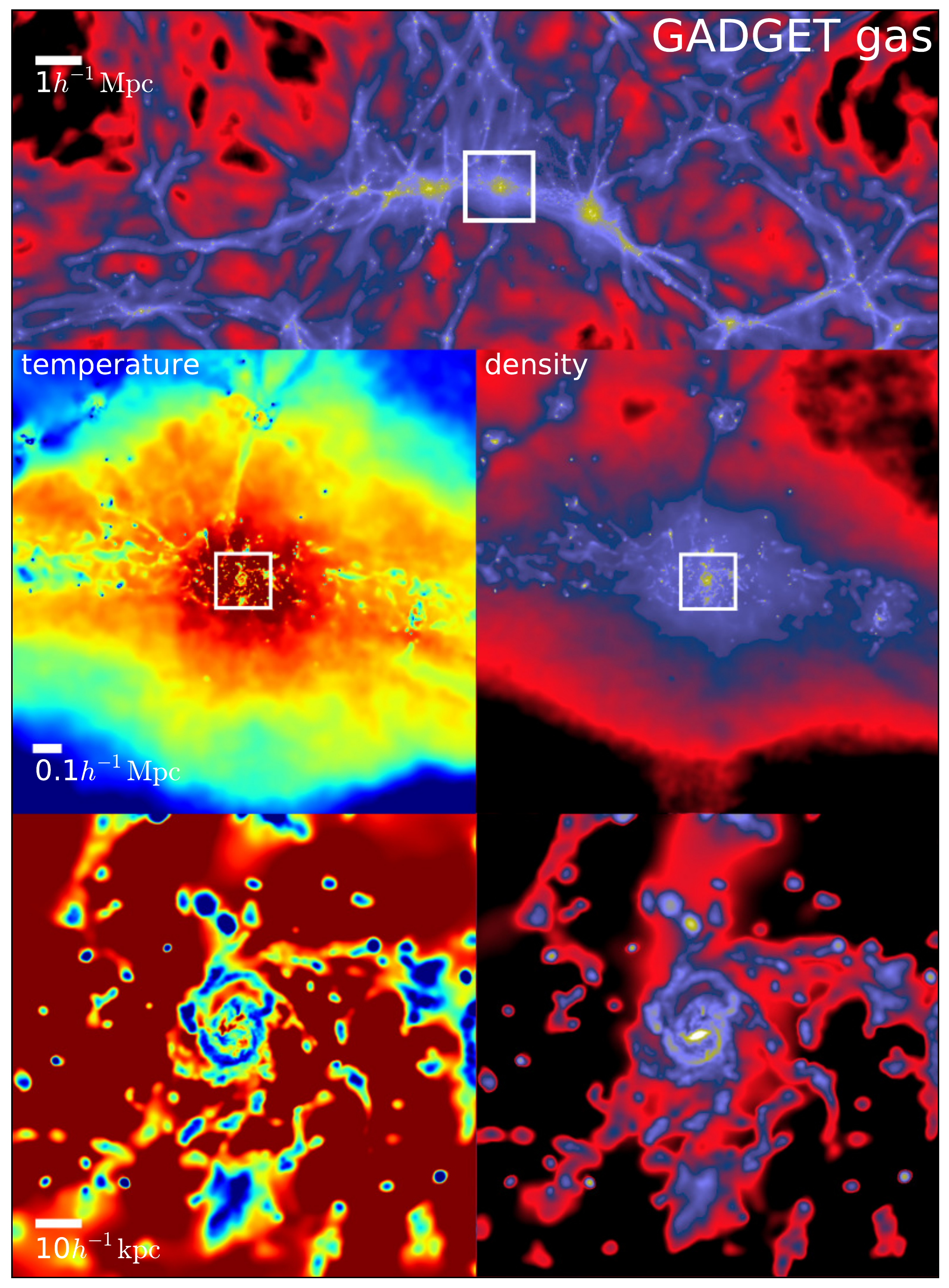}
\caption{Projected gas density and temperature maps for the highest
  resolution {\sm AREPO} (left) and {\sm GADGET} (right) simulations
  ($2 \times 512^3$ particles/cells) at redshift $z=2$. All slices
  have a thickness of $2\,h^{-1}\,{\rm Mpc}$ (comoving). The upper two
  panels show a large part of the full simulation volume. This region
  extends $20\,h^{-1}\,{\rm Mpc}$ in the $x$-direction and contains one of
  the largest filaments in the box. The other eight panels form a
  zoom-in onto one object. Each zoomed region is indicated by a white
  square in the previous step, and the corresponding comoving length
  scales are indicated by the small white bars in the lower left of
  each image. On the left side of the zoom-images we show the
  temperature, and on the right the density fields.  The colour maps
  at the different zoom levels change, but they are the same for both
  codes. At the highest zoom-in factor in the lower panels, a disk
  galaxy is visible. The disk is very extended in {\sm AREPO}, and
  also features a bar. The SPH calculation produces a significantly
  smaller gas disk in the same halo, and the stellar bar is smaller
  compared to the moving-mesh simulation.}
\label{fig:projections}
\end{figure*}

All our simulations have been evolved until redshift $z=0$ even though
the fundamental mode of our small box becomes mildly non-linear at low
redshift. However, our primary goal here is not to obtain
quantitatively accurate large-scale statistics, but rather to compare
the properties of galaxies formed by two different hydrodynamical
schemes. As we start all simulations with the same phases, this
relative comparison is unaffected by box-size effects down to $z=0$.
To easily refer to the different runs in our simulation suite we use
the following naming convention.  A\_L20nX denotes {\sm AREPO} runs
(indicated by the leading `A'), where  $X=128$, $256$, or $512$
encodes the numerical resolution and the tag `L20' stands for the box
size of $L = 20\,h^{-1}\,{\rm Mpc}$. Likewise, G\_L20nX refers to our
{\sm GADGET} runs (indicated by the `G'). In
Table~\ref{table:simulations}, we provide an overview of our
simulation suite and list its most important parameters.

\begin{figure*}
\centering
  \includegraphics[width=0.94\textwidth]{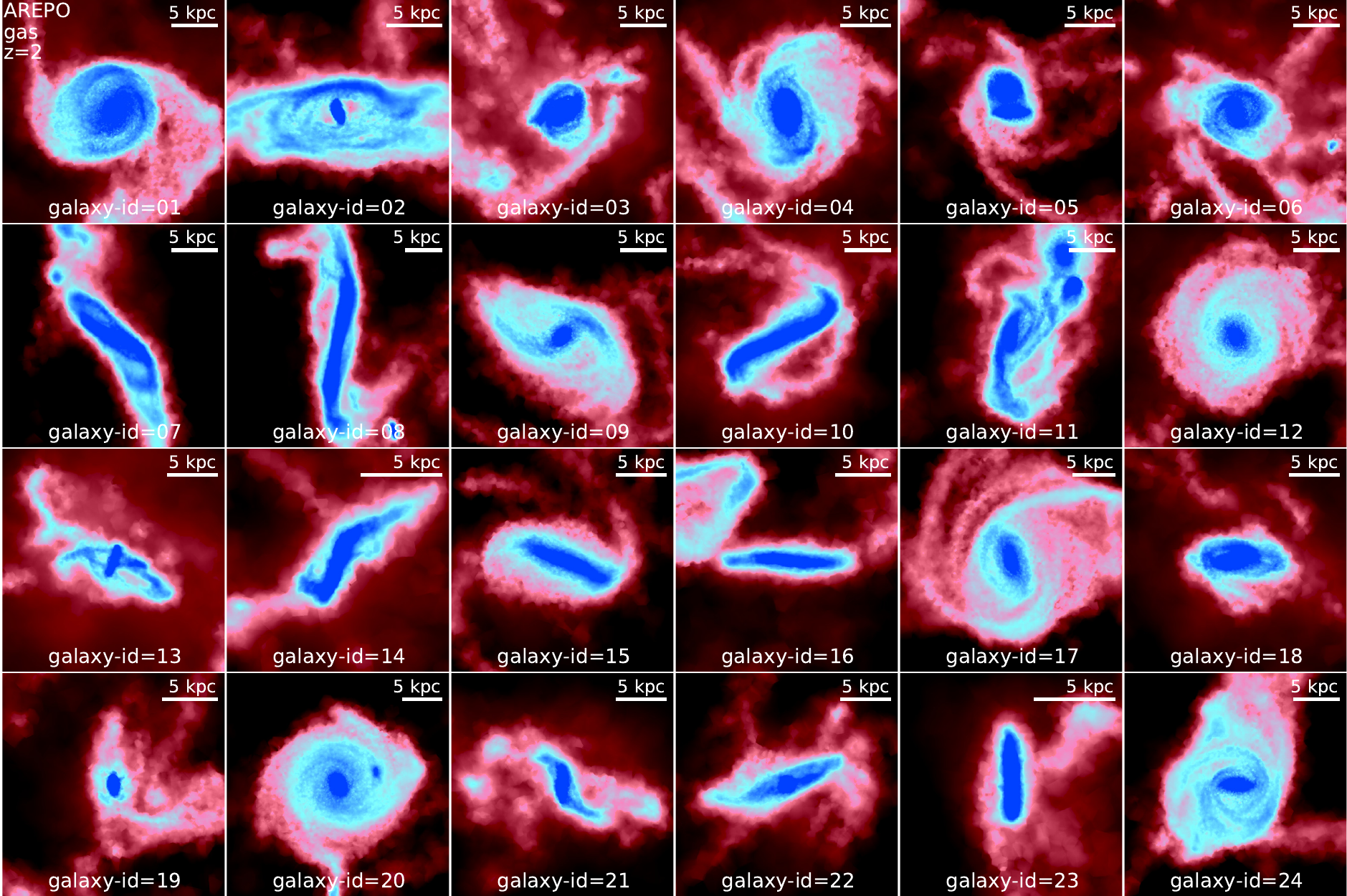}
  \includegraphics[width=0.94\textwidth]{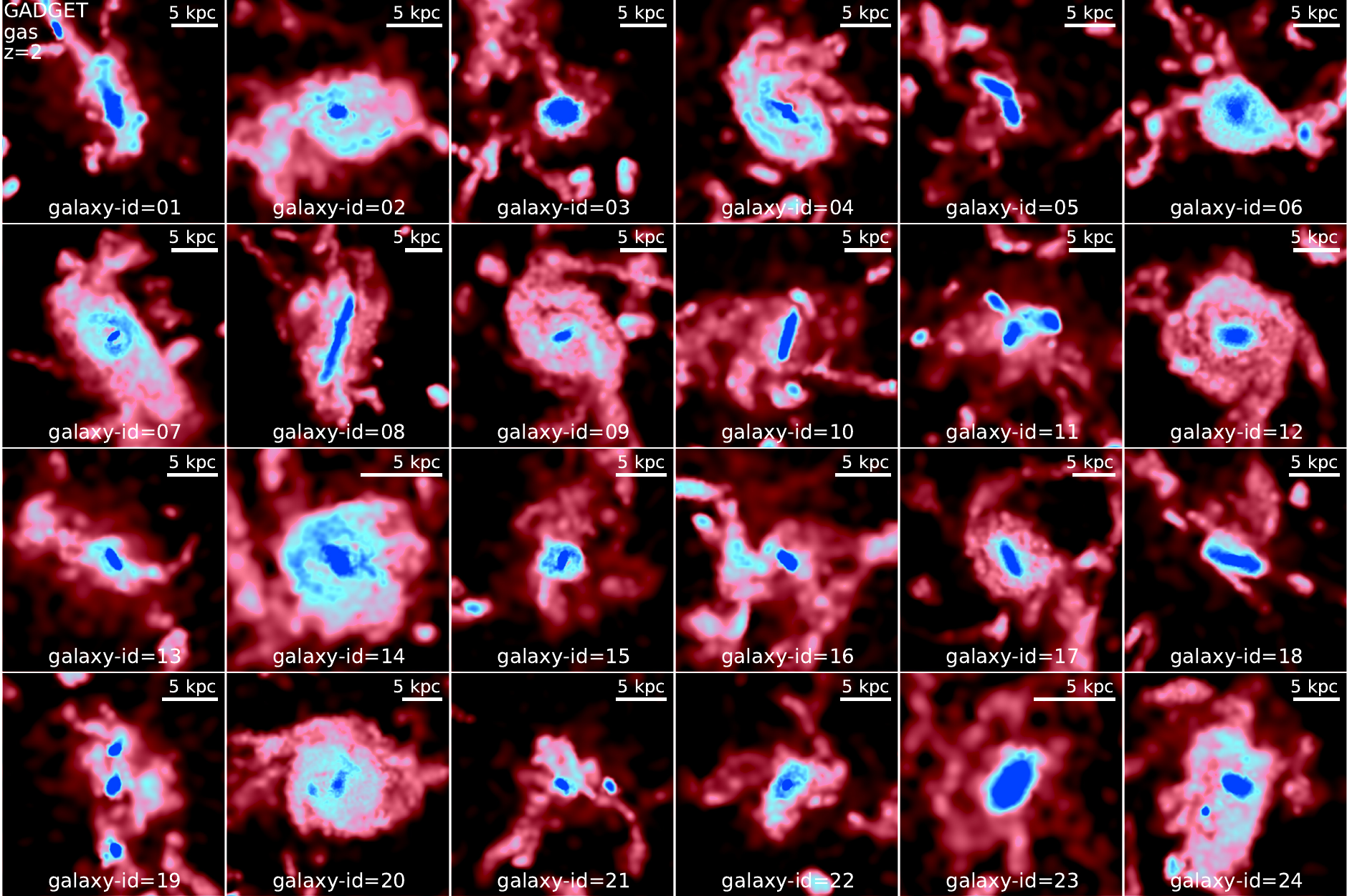}
\caption{Gas density projections of the central galaxies in the 24
  most massive haloes in A\_L20n512 at $z=2$, together with the corresponding
  galaxies in the G\_L20n512 simulation (top four rows: {\sm AREPO}, bottom four rows: {\sm GADGET}). The
  small white bars indicate $5\,{\rm kpc}$ (physical). We
  project along random directions to show various viewing angles,
  similar to a real galaxy field of view. The galaxy with number $8$
  is the object shown in the zoomed-in region of
  Fig.~\ref{fig:projections}, but here now shown more edge-on.
  Essentially in all of the 24 cases the {\sm AREPO} gas distributions
  are more extended than those of the equivalent {\sm GADGET}
  galaxies.  This demonstrates that the galaxy shown in
  Fig.~\ref{fig:projections} is not an exception, but rather a typical
  case in our simulations.}
\label{fig:collage_gas_z2}
\end{figure*}

\begin{figure*}
\centering
  \includegraphics[width=0.94\textwidth]{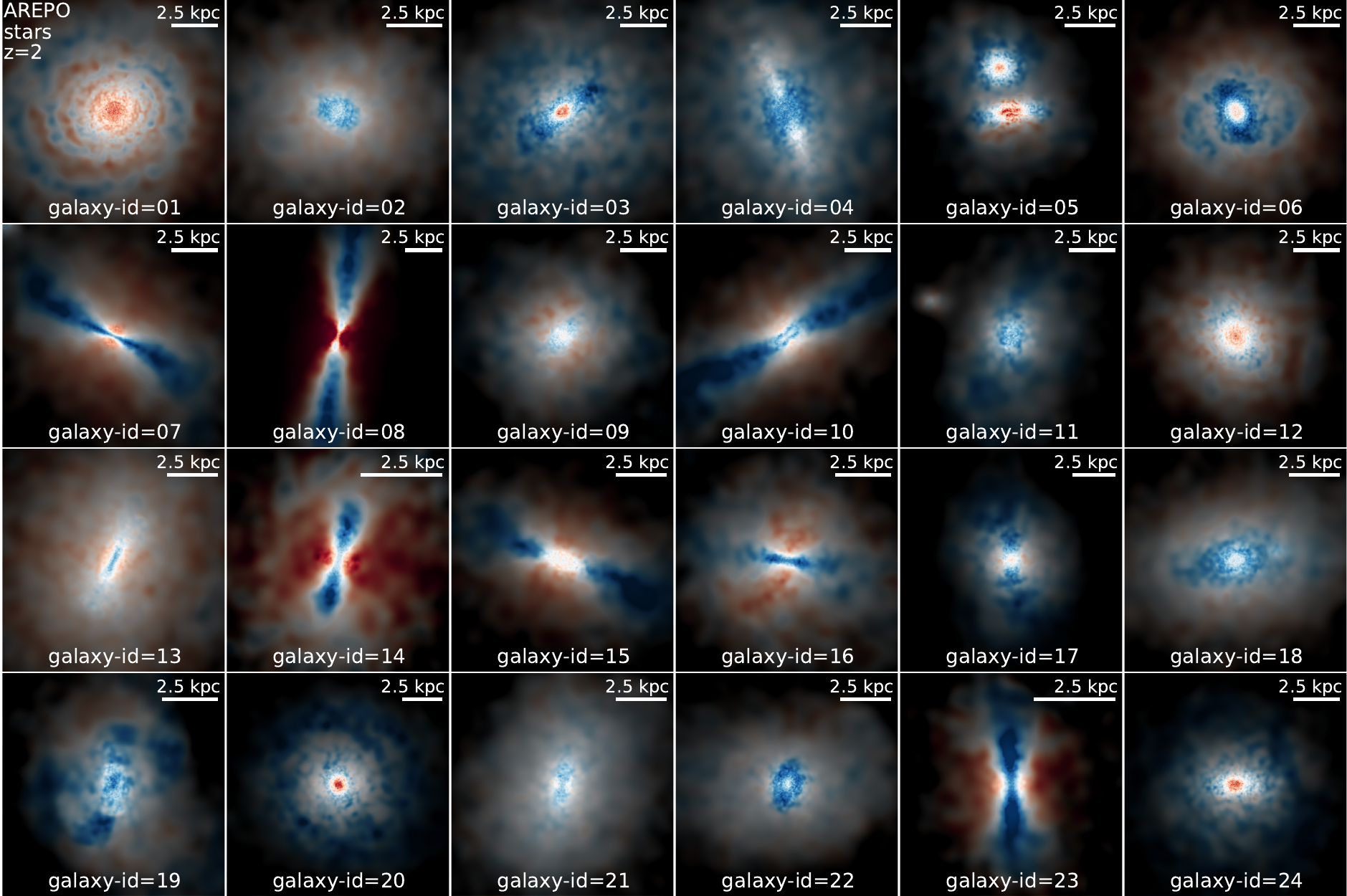}
  \includegraphics[width=0.94\textwidth]{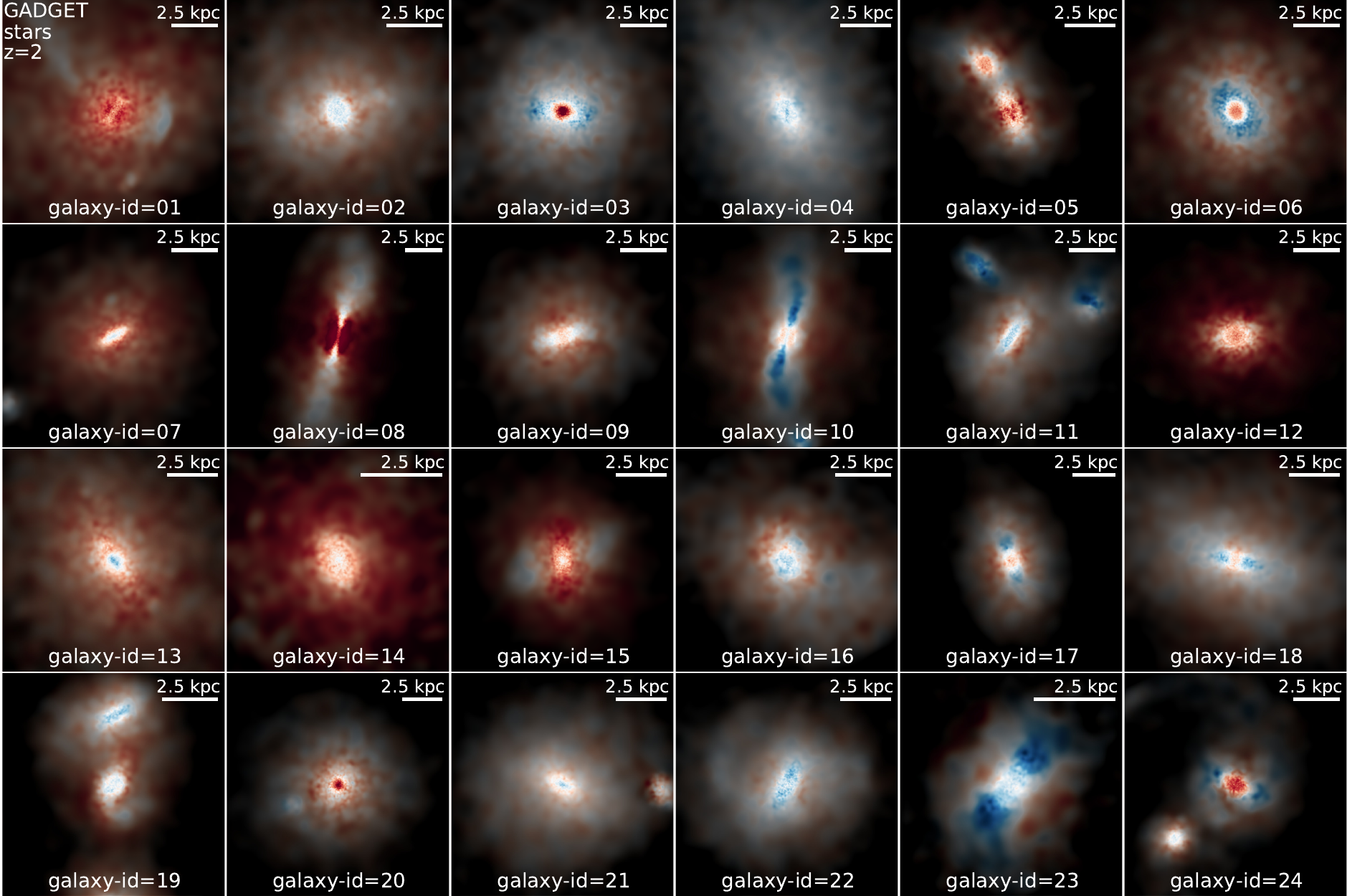}
\caption{Stellar density projections of the central galaxies in the 24
  most massive haloes in A\_L20n512 at $z=2$, together with the corresponding
  galaxies in the G\_L20n512 simulation (top four rows: {\sm AREPO}, bottom four rows: {\sm GADGET}). The
  small white bars indicate $2.5\,{\rm kpc}$ (physical). Density is encoded as intensity, whereas the colour scale
  (blue-red) indicates the stellar age (young-old).
  Galaxies are shown from the same viewing angle as in
  Fig.~\ref{fig:collage_gas_z2}, but  with reduced sidelength.
  The differences in stellar radii are not as large
  as those found for the gas distribution. In most cases the stellar distribution is
  more disky in {\rm AREPO} compared to the galaxies in the SPH
  runs, and the stellar population in the {\sm AREPO} simulations
  is younger making the galaxies on average bluer.}
\label{fig:collage_stars_z2}
\end{figure*}

\begin{figure*}
\centering
  \includegraphics[width=1.0\textwidth]{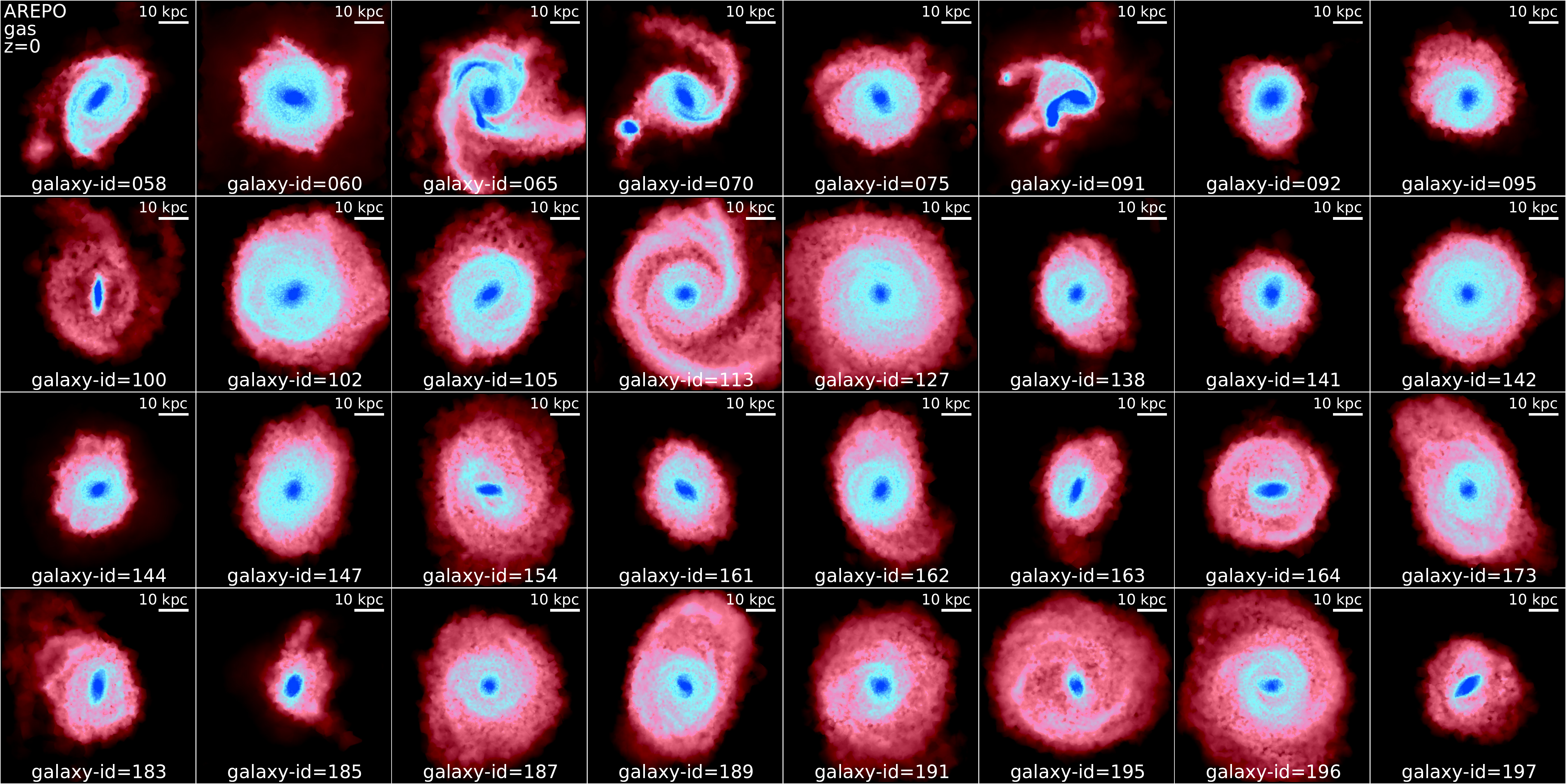}
  \includegraphics[width=1.0\textwidth]{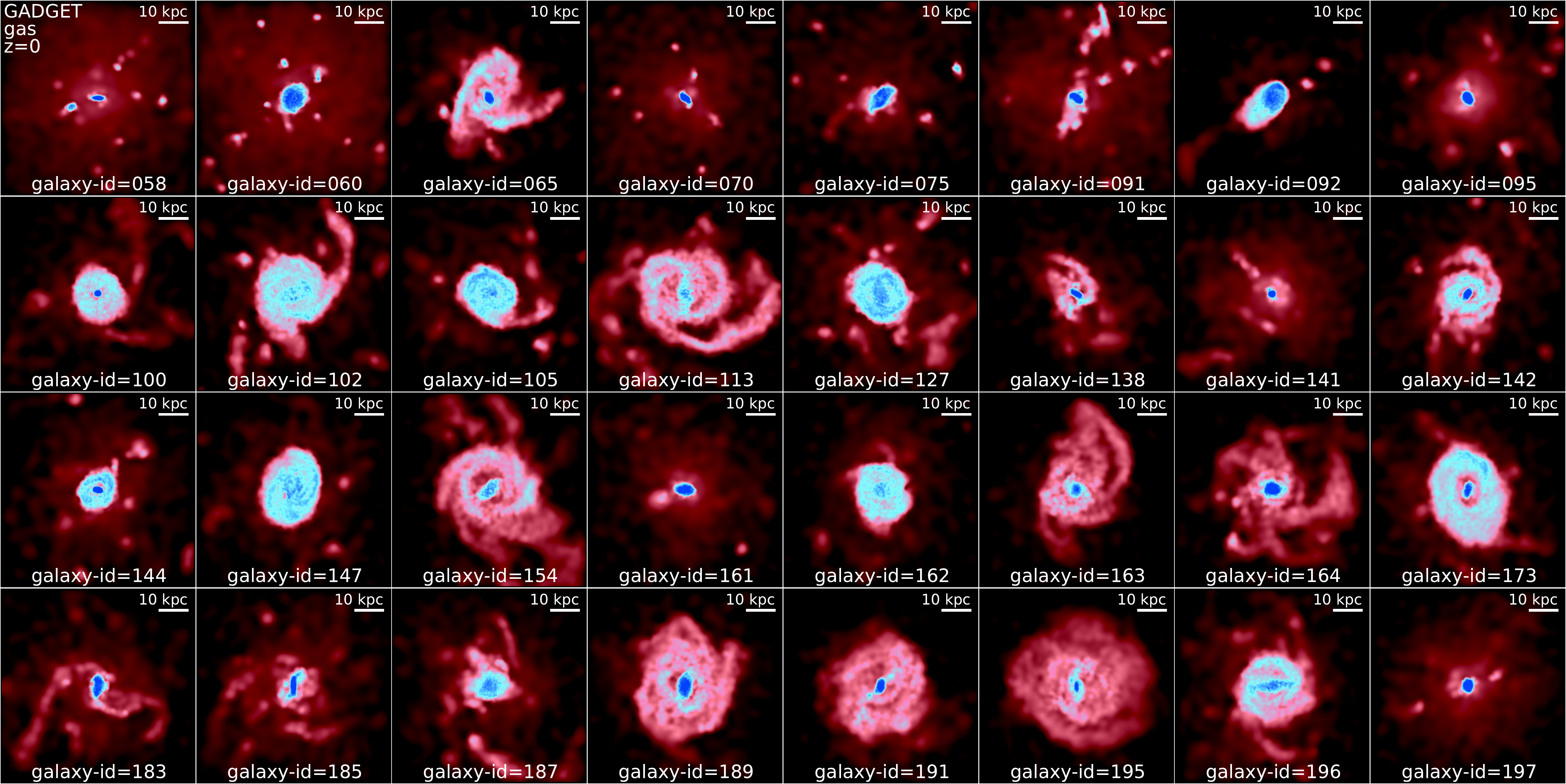}
\caption{Gas density projections of $32$ galaxies at $z=0$ randomly
selected within the halo mass range from $\sim 1\times 10^{11}\,h^{-1}\,\mathrm{M}_\odot$
to $\sim 6\times 10^{11}\,h^{-1}\,\mathrm{M}_\odot$. The top four rows show the {\sm AREPO}
galaxies and the bottom four rows the matched {\sm GADGET} galaxies. The galaxies
are shown face-on for better visual inspection of the disk structure.
The small white bars indicate $10\,{\rm kpc}$ (physical). The
difference in the disk morphology is very striking and follows the same
trend found at $z=2$ in Fig.~\ref{fig:collage_gas_z2}: {\sm AREPO} gas
disks are significantly more extended than those in the SPH simulations.
  }
\label{fig:collage_gas_z0}
\end{figure*}

\begin{figure*}
\centering
  \includegraphics[width=1.0\textwidth]{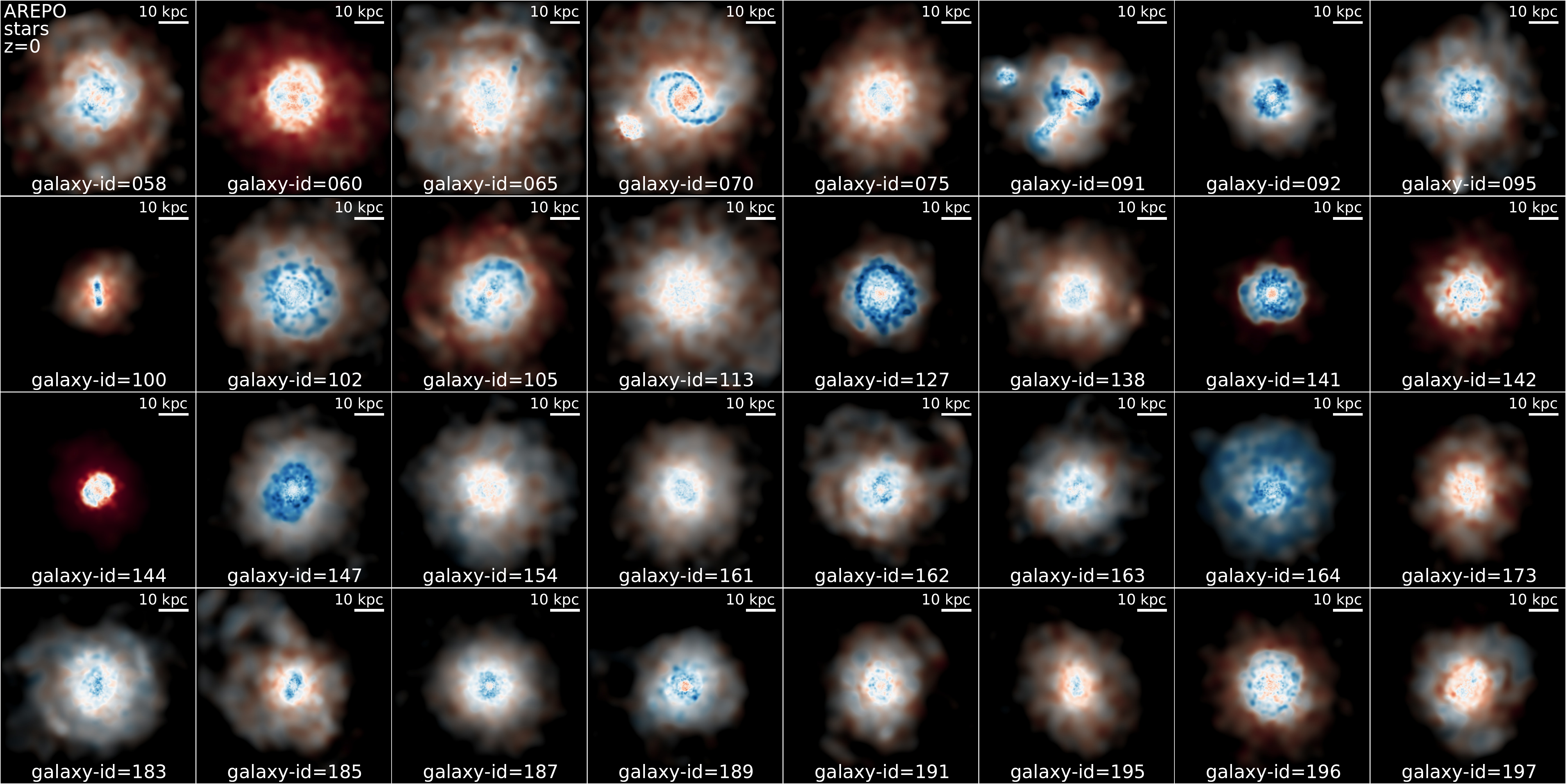}
  \includegraphics[width=1.0\textwidth]{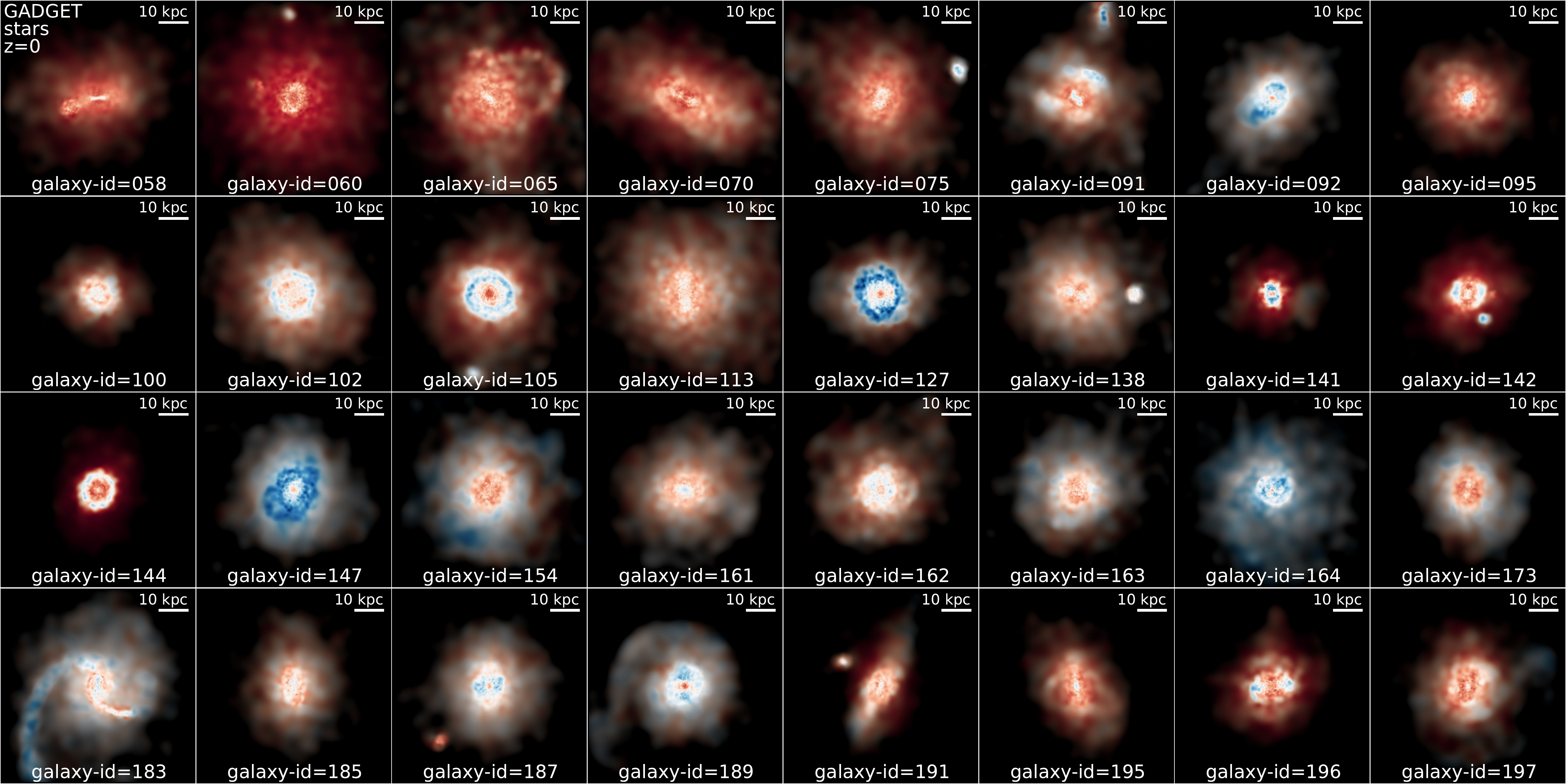}
\caption{Stellar density projections of the $32$ galaxies shown in
 Fig.~\ref{fig:collage_gas_z0}. The small white bars indicate $10\,{\rm kpc}$ (physical).
 As in Fig.~\ref{fig:collage_stars_z2}
 the density is encoded as intensity, whereas the colour scale
 (blue-red) encodes the stellar age (young-old).
 The viewing angle is the same as in Fig.~\ref{fig:collage_gas_z0}.
 Stellar radii differ not as much as the gas disk scale length found
 in Fig.~\ref{fig:collage_gas_z0}. As in Fig.~\ref{fig:collage_stars_z2}
 the {\sm GADGET} galaxies are slightly redder than the galaxies found
 in the {\sm AREPO} simulations.
  }
\label{fig:collage_stars_z0}
\end{figure*}

\begin{figure*}
\centering
  \includegraphics[width=0.33\textwidth]{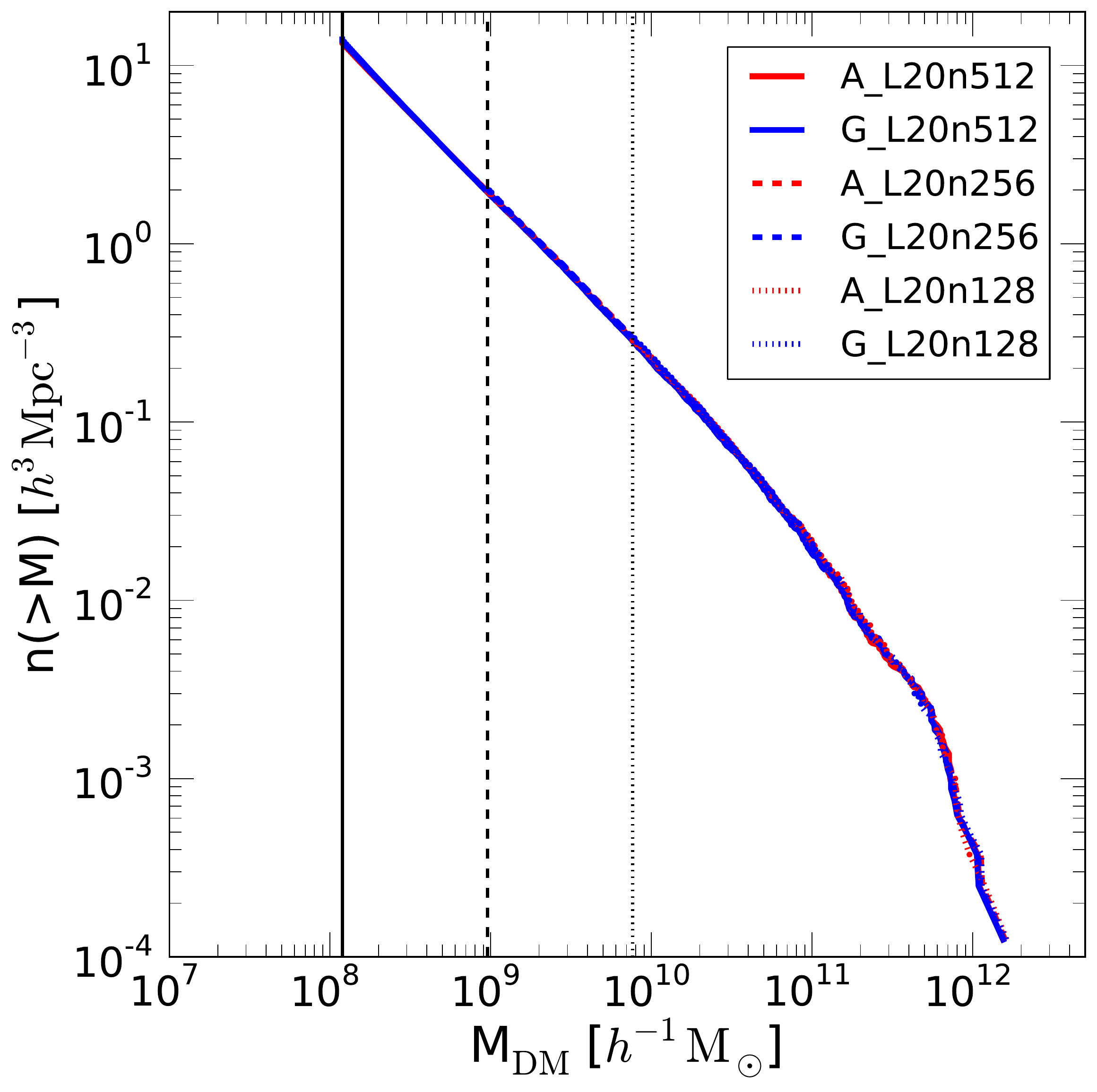}
  \includegraphics[width=0.33\textwidth]{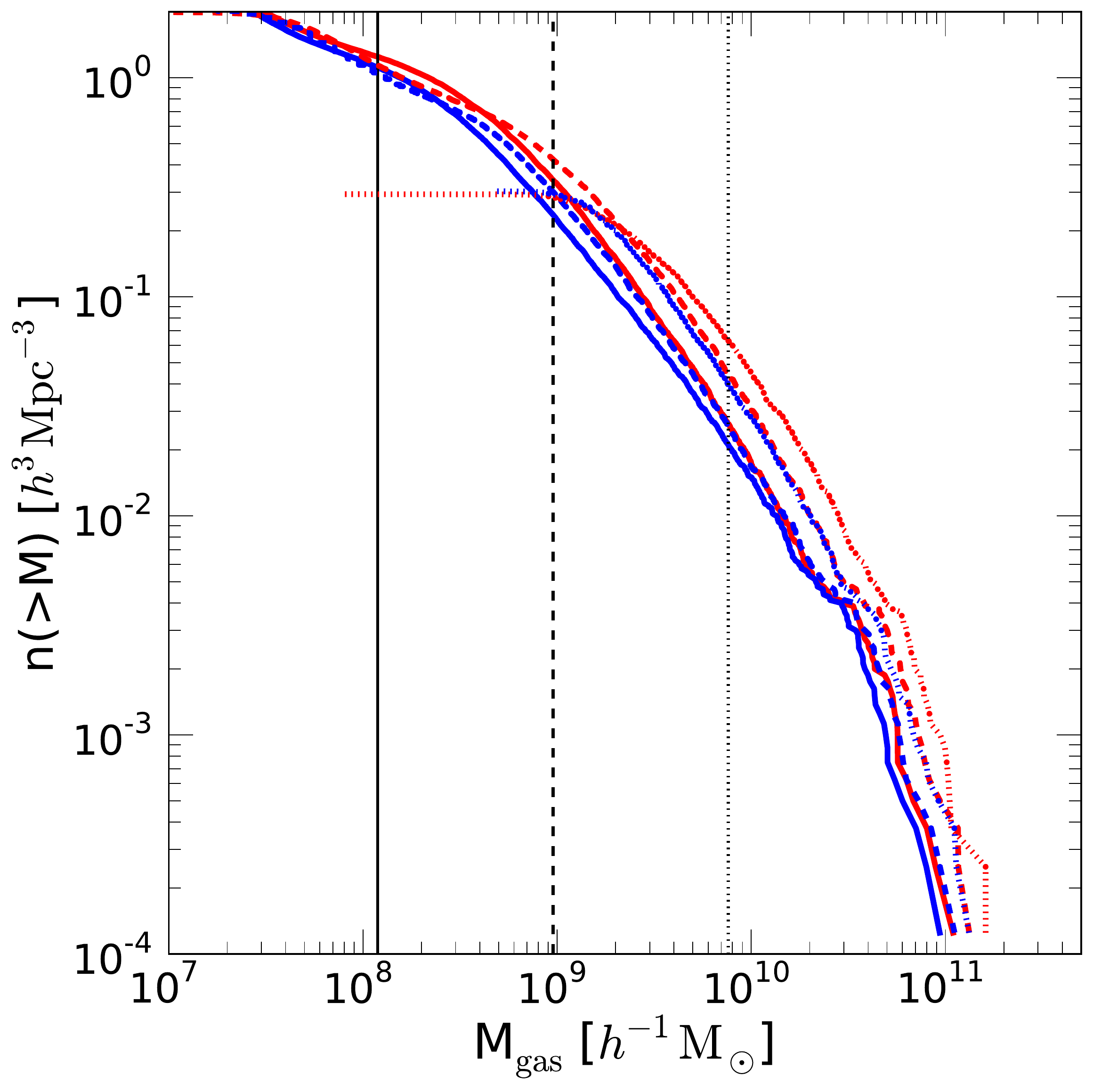}
  \includegraphics[width=0.33\textwidth]{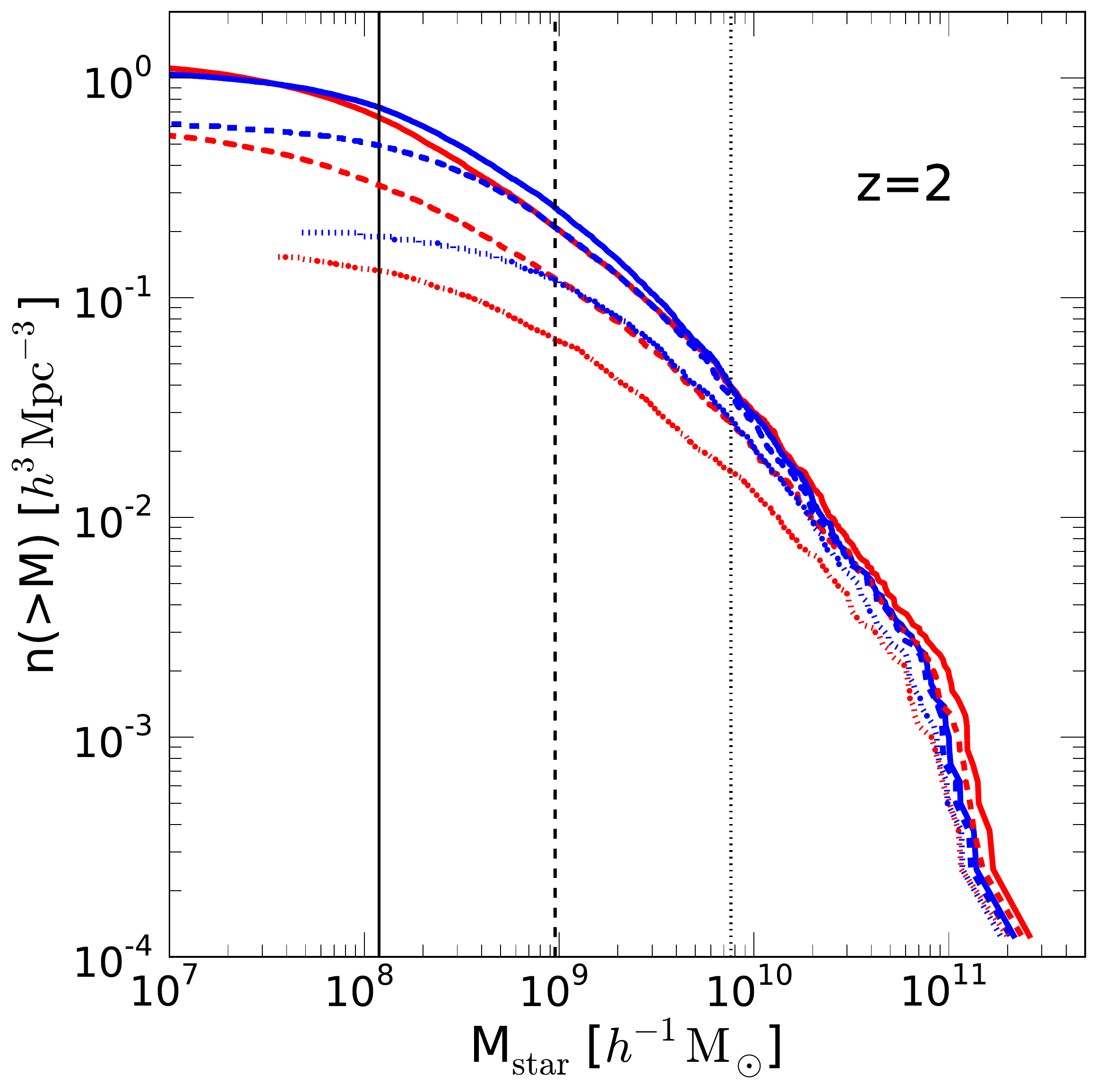}
  \includegraphics[width=0.33\textwidth]{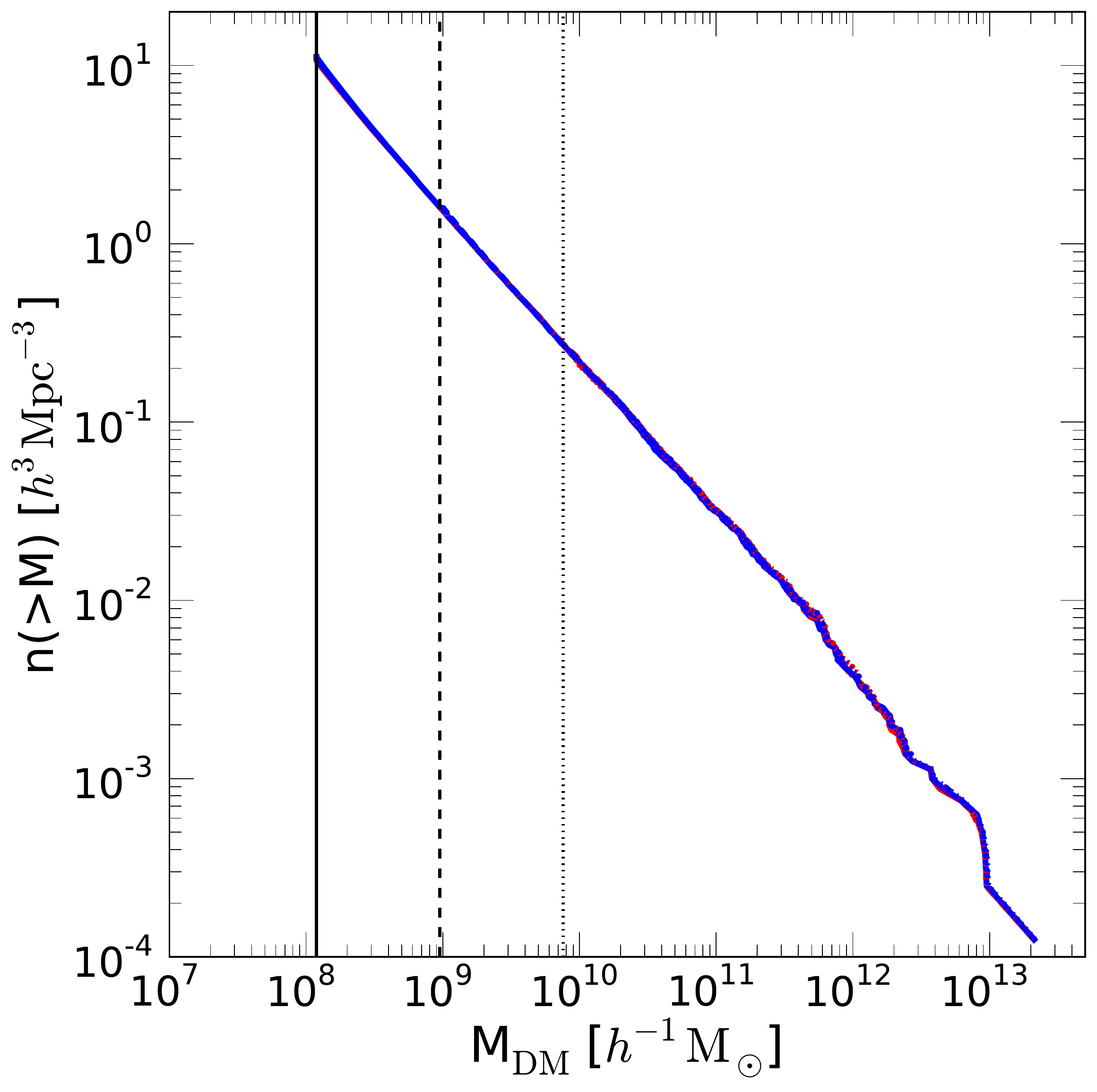}
  \includegraphics[width=0.33\textwidth]{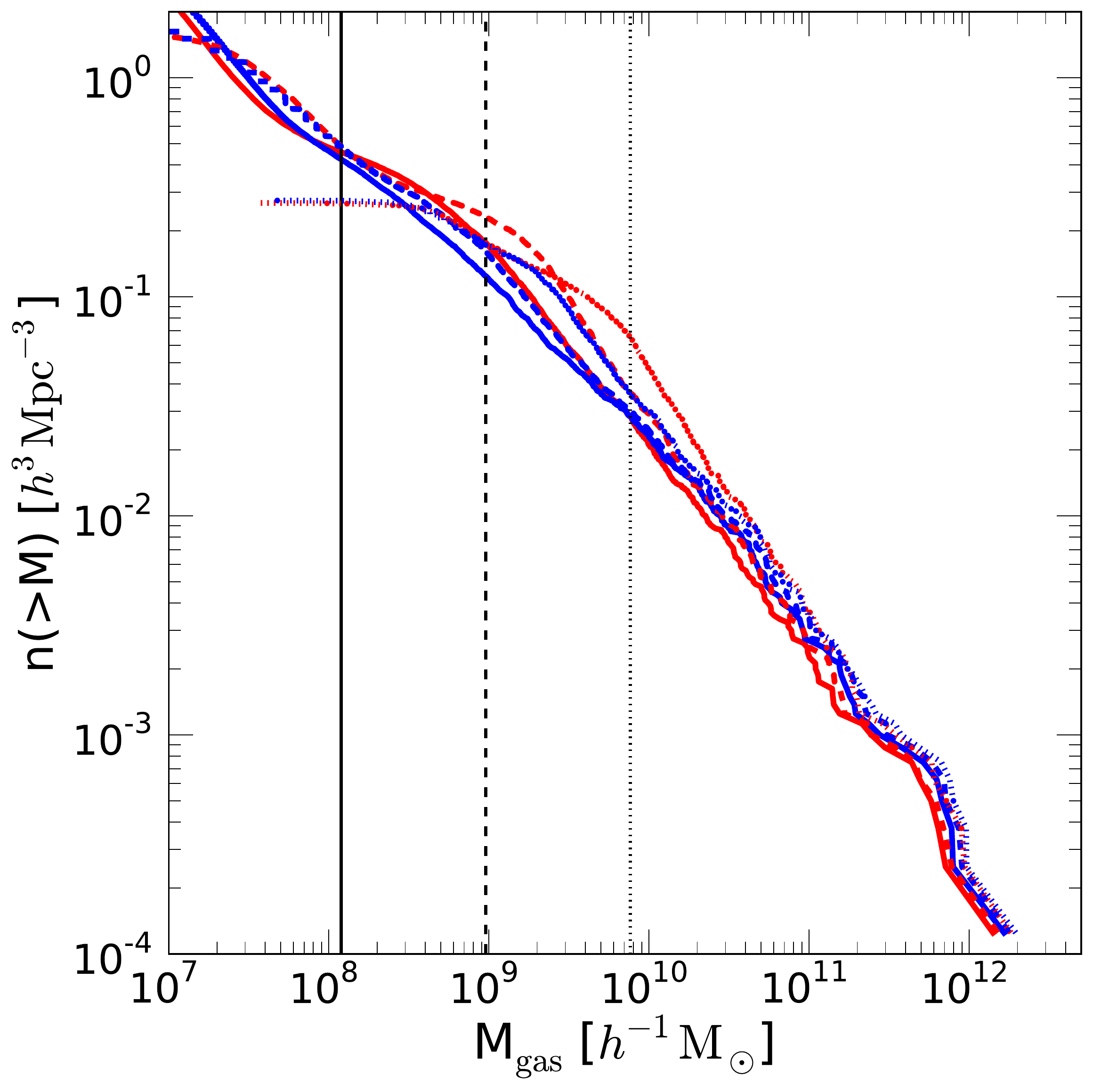}
  \includegraphics[width=0.33\textwidth]{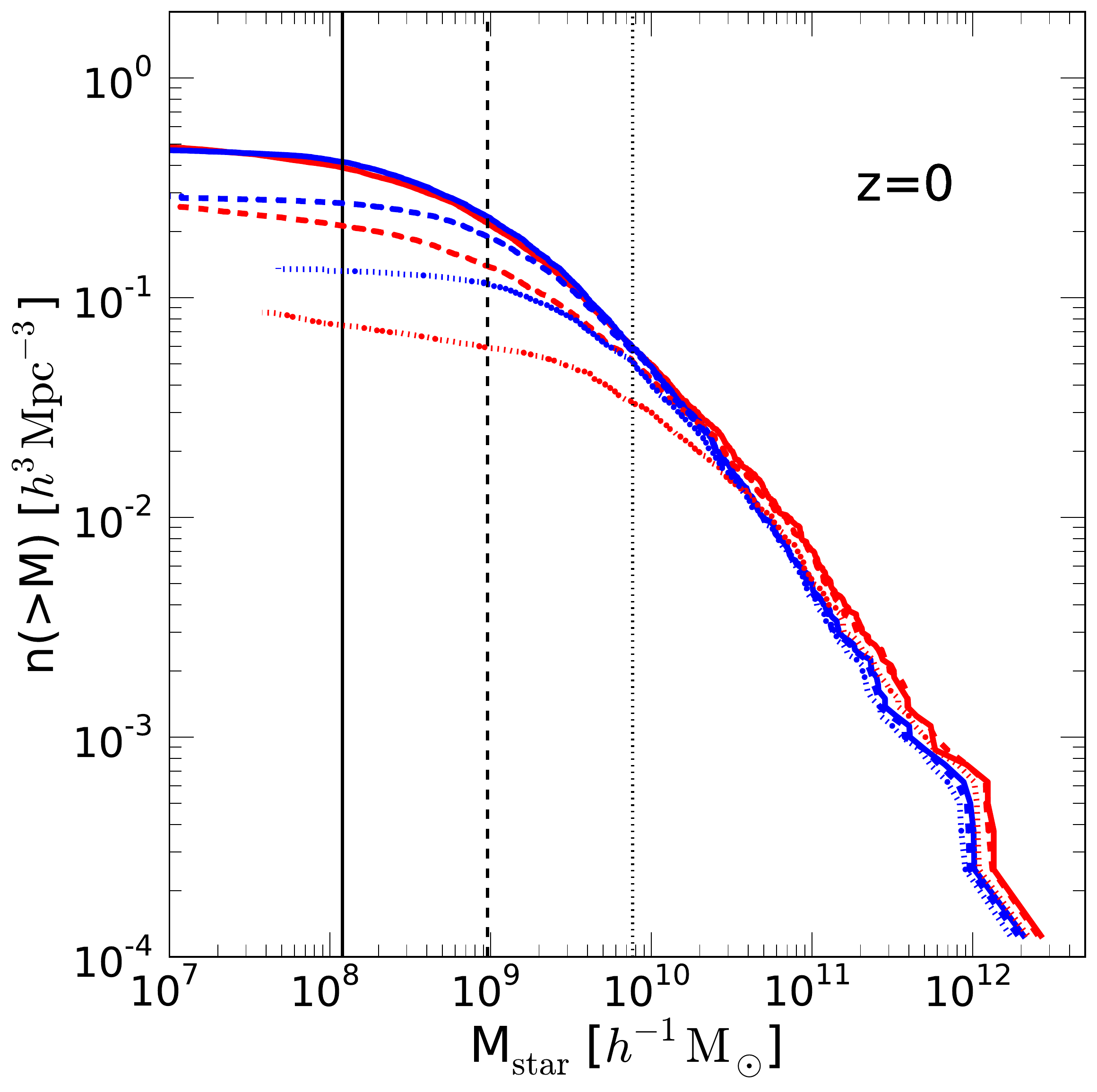}
\caption{Mass functions of FoF groups at $z=2$ (upper panels) and $z=0$
  (lower panels). Left column: dark matter, middle column: gas, right
  column: stars. Vertical lines indicate the 32 dark matter particle
  mass limit. The dark matter mass functions agree and converge very
  well, demonstrating that the gravitational evolution is indeed treated
  on an equal footing, whereas gas and stellar mass functions deviate  due
  to the different hydro-solvers.  Towards lower redshifts there are
  typically fewer massive gas systems and more massive stellar systems
  in {\sm AREPO} compared to {\sm GADGET}. }
\label{fig:fof_MF}
\end{figure*}

\begin{figure*}
\centering
  \includegraphics[width=0.9\textwidth]{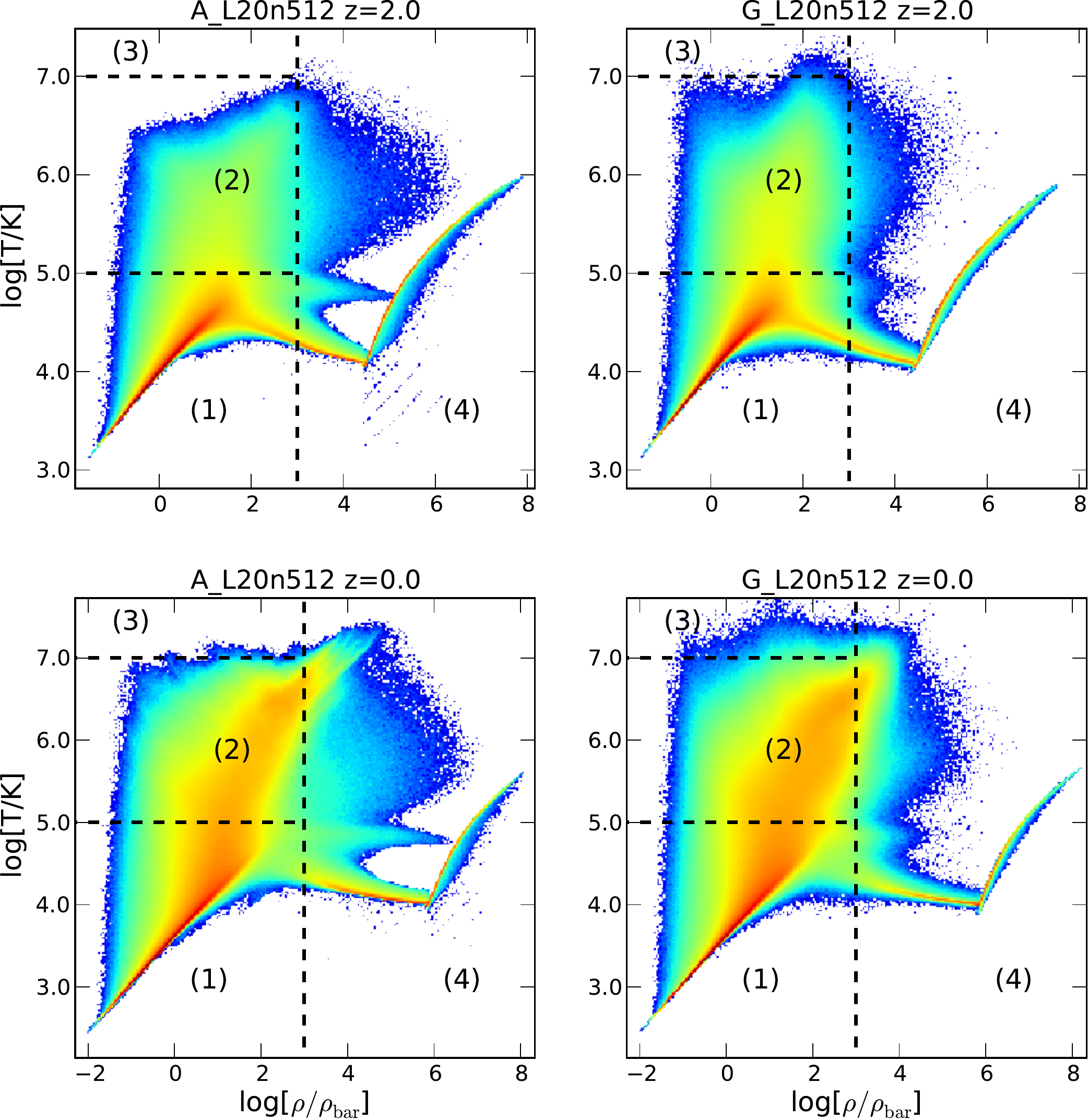}
\caption{Gas density--temperature phase diagrams of {\sm AREPO} and
  {\sm GADGET} at the highest resolution (top: $z=2$; bottom: $z=0$;
  left: {\sm AREPO}; right: {\sm GADGET}).  Each bin of the maps
  contains the total gas mass at the corresponding density and
  temperature.  The different characteristic phases are populated
  similarly in the different runs, showing that the overall global gas
  properties agree. But there are also some visible differences
  between the SPH and moving mesh calculations.  {\sm AREPO} typically
  produces more high temperature gas at large densities, but the
  overall amount of hot gas is reduced, which can be seen by the more
  extended yellow region in {\sm GADGET} at high temperatures.
  The dashed lines and numbers indicate the gas cuts used for quantifying
  the mass and volume fractions in Fig.~\ref{fig:gas_phases_mass} and
  Fig.~\ref{fig:gas_phases_volume}, respectively.}
\label{fig:rho_T}
\end{figure*}

\section{Results for global baryon statistics}

\subsection{Density and temperature maps}

To obtain a first impression of the simulations, it is instructive to
examine maps of the density and temperature fields. In
Fig.~\ref{fig:projections}, we show projections of the gas density
and temperature distributions for our highest resolution
simulations A\_L20n512 and G\_L20n512 at $z=2$.  The left panel gives
the {\sm AREPO} results and the right panel those obtained with {\sm
  GADGET}. In both cases, the image on top depicts a large part of the
full box, while the zoom images below show two successive zooms onto a
disk galaxy (as indicated by the white squares). In these enlarged
images, the left half depicts the temperature field while the right
half gives the density field. All slices have a thickness of
$2\,h^{-1}\,\mathrm{Mpc}$ and their extent in the image plane is
specified by the scale-bars included in the images. Note that we did
not centre the two image sequences individually onto 
the galaxies, so the zoom sequence also illustrates the typical
magnitude of coordinate offsets that can develop between the different
simulations.

Inspection of the maps in Fig.~\ref{fig:projections} demonstrates that
both codes produce essentially identical results on large scales. This
is reasonable since we do not expect large-scale changes induced by a
different hydrodynamical scheme. However, already at the intermediate
zoom level one can identify some significant differences. The
distribution of hot gas is clearly quite different,
with {\sm GADGET} showing a more extended hot atmosphere compared to
{\sm AREPO}. On still smaller scales, the differences between the two
hydrodynamical codes become even more pronounced: whereas {\sm AREPO}
produces an extended disk, {\sm GADGET} forms a significantly smaller
and clumpier disk. The difference can be best appreciated in the
temperature map, where the cold gas disk stands out clearly as a large
blue region in the moving mesh calculation. {\sm AREPO} forms a spiral
disk with a central bar, which is visible in the density map.  For a
more detailed analysis of the structural properties of this galaxy we
refer the reader to Paper II of this series.  We note that although
some spiral features are visible in the disk, these are partly seeded by
Poisson noise in the potential due to the limited number of DM
particles in the halo \citep[see][for details of this
  process]{Donghia2011}, and have hence only limited physical 
significance.
 
\begin{figure*}
\centering
  \includegraphics[width=0.9\textwidth]{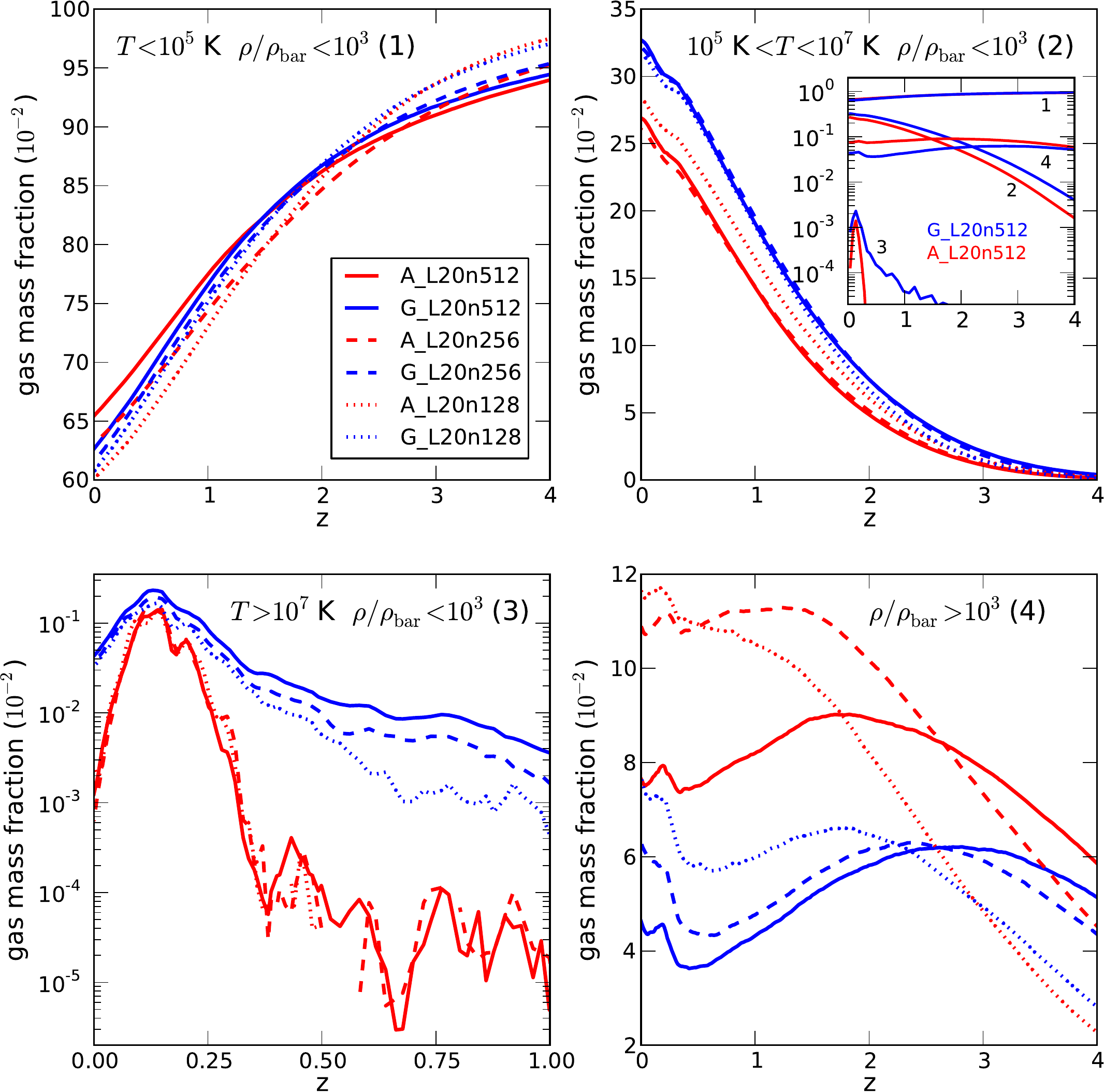}
\caption{Fraction of gas mass in various phases; i.e. the
  different panels correspond to the mass fraction in the parts
  of the phase diagrams in Fig.~\ref{fig:rho_T}. We do not include
  stellar mass in this plot. The four different gas selections add up
  to the total gas mass in the simulation volume. Neither code
  shows very good convergence in all of these fractions. This is true
  both for the SPH and the moving-mesh calculations.  Nevertheless,
  the different mass fractions do not vary dramatically between the
  various resolutions, and there are some characteristic and
  systematic differences between {\sm GADGET} and {\sm AREPO}.  {\sm
    GADGET} produces more low density hot gas and {\sm AREPO} shows a
  larger fraction of high density gas ($\rho/\rho_{\rm bar}>10^5\,{\rm
    K}$). This can already be seen from Fig.~\ref{fig:rho_T} by
  comparing the plume of hot particles/cells.  Also, the fraction of
  mass in the warm to hot intergalactic medium (top right) is lower in
  {\sm AREPO} compared to {\sm GADGET}.  We note that the redshift
  range in the lower left panel is reduced compared to the other
  panels. The inset in the upper right panel shows the fraction of all
  four main panels for the highest resolution runs together. The
  numbers indicate the cut they belong to. These cuts and numbers are
  also shown in Fig.~\ref{fig:rho_T}.}
\label{fig:gas_phases_mass}
\end{figure*}

In order to demonstrate that differences like those seen in
Fig.~\ref{fig:projections} do not appear only in unusual cases, we
show in Fig.~\ref{fig:collage_gas_z2} gas density projections of the
central galaxies of the 24 most massive haloes at $z=2$ of A\_L20n512 
(top panels), and G\_L20n512 (bottom panels), along
random projection directions. We did not impose any constraint on the
galaxy selection other than halo mass; as a result, we can also have
galaxies in our sample that are disturbed or undergo a merger, such as
the system labelled `galaxy-id~5'.  The object `galaxy-id~8' of
Fig.~\ref{fig:collage_gas_z2} corresponds to the galaxy shown in
Fig.~\ref{fig:projections}, but happens to be oriented edge-on
here. The random orientations of the galaxy set give an impression of
how a population of galaxies would look like in these simulations.
Clearly, the gas distributions of most of the galaxies in
Fig.~\ref{fig:collage_gas_z2} are more extended and more reminiscent of
spiral galaxies in {\sm AREPO} than in {\sm GADGET}.

Inspection of animated sequences from the {\sm AREPO} simulations
reveals a population of interacting and merging systems. Many of them show
thin bridges and tails, as expected if the galaxies are rotationally supported disks
\citep[e.g.][]{1972ApJ...178..623T,2010ApJ...725..353D}\footnote{Volume-rendering movies of {\sm AREPO} galaxies (at redshift $z=1$ and $z=2$) and high-resolution images
are available for download at the website {http://www.cfa.harvard.edu/itc/research/movingmeshcosmology.}}.

The differences in the gas properties are also reflected in the
corresponding stellar distributions of these galaxies, which we show in
Fig.~\ref{fig:collage_stars_z2}. Although the stellar distributions are
more similar between the two numerical schemes, in most cases one
still notices that {\sm AREPO}'s disks are slightly larger and appear
typically more disky, an impression that is also supported by the
quantitative analysis of galaxy properties that we present in Paper II.
Furthermore, the stellar populations of the {\sm AREPO} galaxies are bluer,
i.e. younger, than those found in the {\sm GADGET} run.

In Fig.~\ref{fig:collage_gas_z0} and Fig.~\ref{fig:collage_stars_z0}
we show face-on projections of 32 other galaxies at $z=0$ with Milky Way-like
DM halo masses in the mass range from $\sim 1\times 10^{12}\,h^{-1}\,\mathrm{M}_\odot$
to $\sim 6\times 10^{12}\,h^{-1}\,\mathrm{M}_\odot$. The gas disk structure shown
in Fig.~\ref{fig:collage_gas_z0} follows the trend already seen at $z=2$,
and demonstrates that {\sm AREPO} produces significantly larger disks at
all times and at all halo masses compared to the {\sm GADGET} simulations.
Fig.~\ref{fig:collage_stars_z0} demonstrates that the stellar disks are also
slightly larger, but not as significant as the gas disks. But similar to the $z=2$
result, the {\sm AREPO} stellar disk populations are typically younger than
those found in {\sm GADGET}.

A self-evident conclusion from these visual comparisons of
Figs.~\ref{fig:collage_gas_z2}, \ref{fig:collage_stars_z2} 
and Figs.~\ref{fig:collage_gas_z0}, \ref{fig:collage_stars_z0}
is that the morphology of forming galaxies in cosmological hydrodynamics
simulations can be strongly affected by the underlying hydrodynamics
solver.  SPH has significantly more problems producing disk-like
galaxies than the moving mesh approach for a similar computational
cost.
Note however that we do not
expect our simulations to produce a completely realistic $z=2$ galaxy
population, because we here refrained from including strong feedback
capable of producing galactic winds and outflows. However, since both
simulations followed exactly the same physics, we expect this
qualitative difference to be present in simulations with stronger
feedback as well.

\begin{figure*}
\centering
  \includegraphics[width=0.9\textwidth]{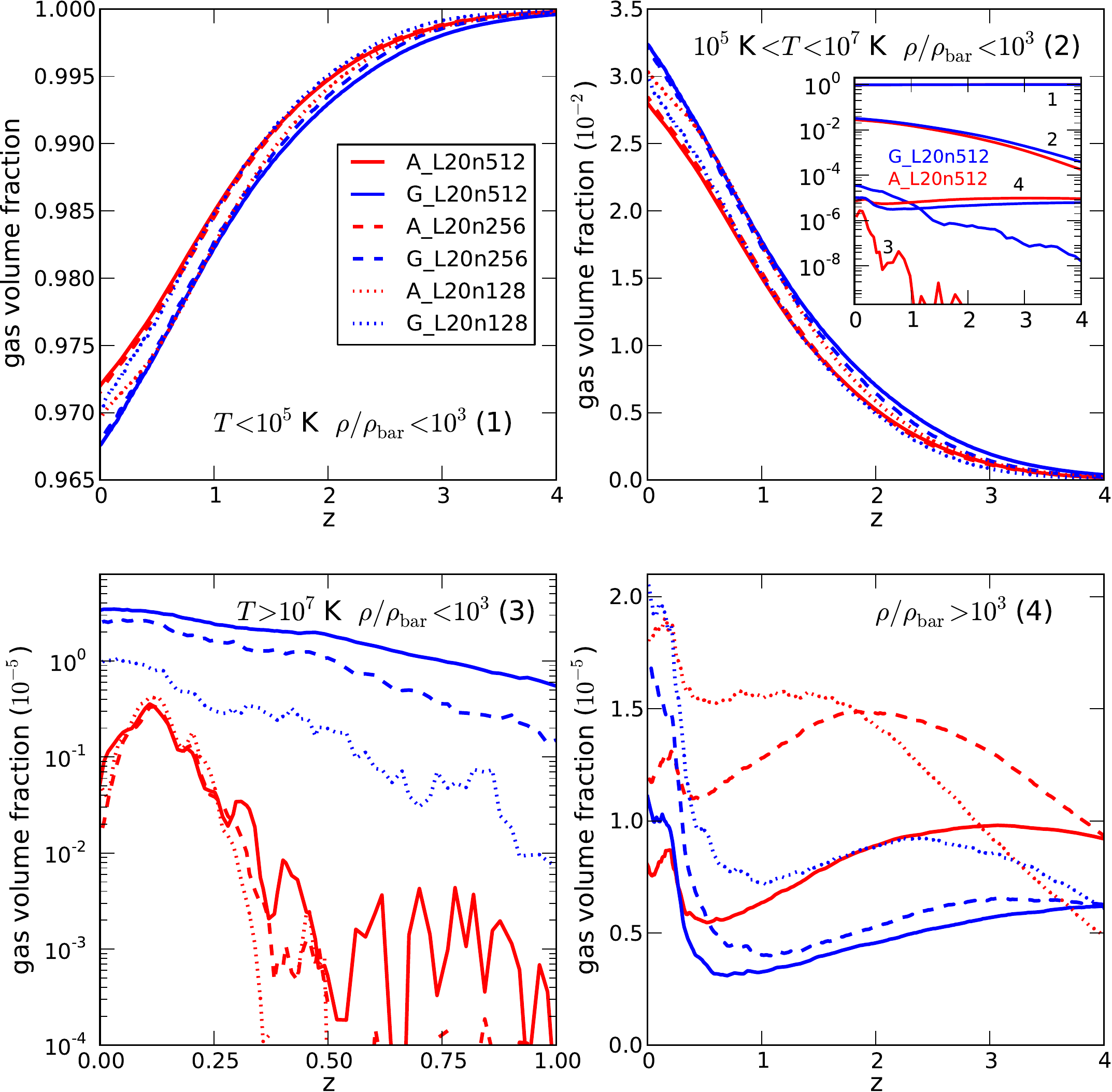}
\caption{Fraction of gas volume in the same gas phases as in
  Fig.~\ref{fig:gas_phases_mass}. For {\sm AREPO} we naturally use the
  volumes of Voronoi cells to calculate these fractions, and for the
  {\sm GADGET} (SPH) simulations we assigned the specific volume
  $m_{\rm SPH}/\rho_{\rm SPH}$ to each SPH particle. We note that gas volume fractions are
  not very well defined in SPH, because the scheme is not volume
  conserving, i.e. the sum over $m_{\rm SPH}/\rho_{\rm SPH}$ does not yield in the
  correct total gas volume. We therefore take as the total volume
  always the sum of all specific volumes. The differences between the
  two hydro-schemes are qualitatively similar to those found in
  Fig.~\ref{fig:gas_phases_mass}, especially the hot gas phase shows a
  very different volume fraction. We note that the redshift range in the
  lower left panel is reduced compared to the other panels. The inset in the upper right panel
  shows the fraction of all four main panels for the highest resolution
  runs together. The numbers indicate the cut they belong to. These cuts
  and numbers are also shown in Fig.~\ref{fig:rho_T}.}
\label{fig:gas_phases_volume}
\end{figure*}

\begin{figure*}
\centering
  \includegraphics[width=0.45\textwidth]{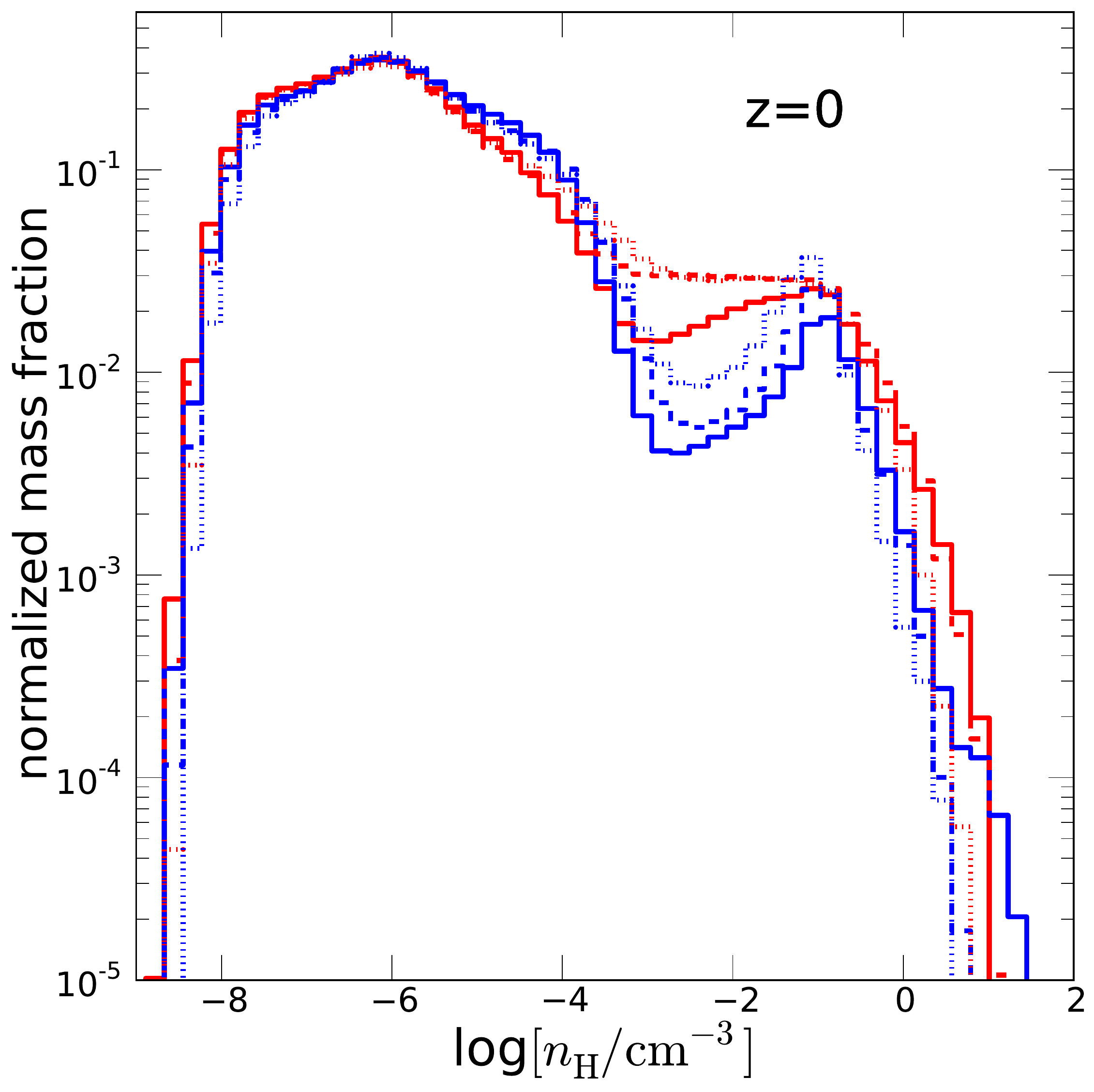}
  \includegraphics[width=0.447\textwidth]{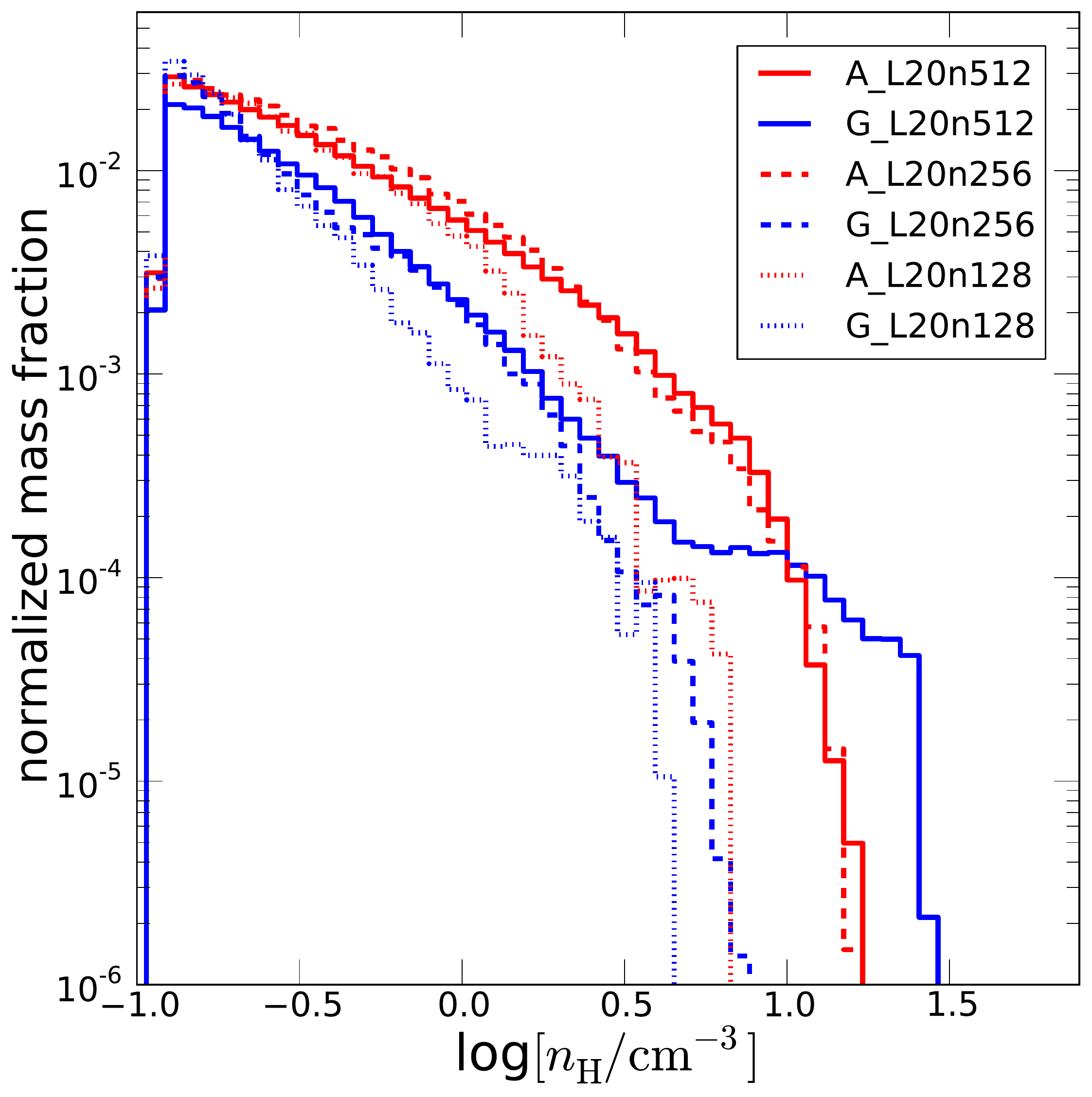}
\caption{Density probability distribution function of the gas in the
  simulation volume in units of hydrogen number density, assuming a
  hydrogen mass fraction of 0.76.  The left panel shows all gas
  particles/cells, whereas the right panel shows only those with
  active star formation. In our star formation model the density
  threshold for the onset of star formation is set to $n_{\rm
    H}=0.13~{\rm cm}^{-3}$, and therefore the right hand panel
  occupies only this high density region. Interestingly, {\sm GADGET}
  shows at all resolutions a significantly reduced gas population
  slightly below the star formation density threshold. This gas is
  about to overcome the density threshold for star formation if it
  contracts a bit more via cooling. The rest of the hydrogen density
  distributions look rather similar for the two simulation
  schemes. The reason for the difference slightly below the threshold
  lies in the more extended disk in {\sm AREPO}. This gas fills the
  ``gap'' present in the SPH calculations.}
\label{fig:density_PDF} 
\end{figure*}

\subsection{Mass functions}

As we emphasised earlier, {\sm AREPO} and {\sm GADGET} use the same
gravity solver based on a high-resolution Tree-PM scheme. Because the
gravitational field is largely dominated by the collisionless DM
component, we expect that the DM distribution in both simulations
should hence be similar, at least on scales where baryonic effects do
not change the structure of DM haloes. We explicitly verify this
expectation in the left column of Fig.~\ref{fig:fof_MF}, where we
compare the DM halo mass function of friends-of-friends (FoF) groups
with a linking length of $0.2$ of the mean particle separation down to a particle number limit of 32 particles at $z=0$ and $z=2$.
The codes show excellent agreement in the DM halo mass functions at both
redshifts. Also, the convergence for the different resolutions is in
both cases very good. We note that previous comparisons of
hydrodynamical mesh codes and high resolution tree codes
\citep{2005ApJS..160....1O,2008CS&D....1a5003H} have shown a deficit
of low-mass haloes in the mesh-based hydrodynamical codes, an effect that
has its origin in their AMR-based gravity solver, which typically does
not refine small DM haloes sufficiently early. As evidenced here,
{\sm AREPO} does not have this problem.

The other panels of Fig.~\ref{fig:fof_MF} count haloes in terms of
their baryonic content, separately for gas and stars. In order to
relate baryons to dark matter haloes, we determine for each
stellar and gaseous cell/particle the closest dark matter particle,
and then associate the baryonic mass element with the FoF group this
DM particle resides in \citep[see][for more
  details]{2009MNRAS.399..497D}. The middle column of
Fig.~\ref{fig:fof_MF} shows the cumulative halo gas mass functions at
redshifts $z=2$ and $z=0$. While both codes converge equally well for
the gas mass functions as a function of resolution, they do not agree
on the result.  At $z=2$, {\sm AREPO} has more objects of a given gas
mass than {\sm GADGET} over nearly the full range of halo masses. This
behaviour changes towards lower redshifts, where fewer very gas-rich
objects are found in {\sm AREPO}. We will show below that this has to
do with more efficient cooling and a higher star formation rate in the
mesh code in massive haloes at late times.

The right column in Fig.~\ref{fig:fof_MF} gives the mass functions for
the stellar component of haloes, and here the situation is slightly
different. At $z=2$, we find fewer objects with low stellar mass in {\sm
  AREPO} compared to {\sm GADGET}, but the situation reverses at the
high mass end where we find {\sm AREPO} has more haloes at a given
stellar mass. This behaviour  changes somewhat towards $z=0$,
where the low mass end agrees now very well between the two
codes. However, there are still more stellar systems of high mass in {\sm
  AREPO}, and this difference becomes more pronounced and extends
over a larger range of masses. Again, this can be attributed to the
more efficient star formation at late times in massive systems in {\sm
  AREPO} that we analyse in detail below. Since the baryonic mass
contributes only a small amount to the total mass of the entire box,
these differences in the baryonic mass functions are neither imprinted
in the DM halo mass functions nor visible in the large-scale structure
shown in Fig.~\ref{fig:projections}.

\subsection{Gas phases}

Because we use two completely different hydro solvers, we may expect
some deviations in the breakdown of various gas phases.  A first
simple way to obtain an overview of the global state of the gas in the
simulation volume is the construction of density--temperature
phase-space diagrams. We show in Fig.~\ref{fig:rho_T} mass-weighted
$\rho$-$T$ phase-space diagrams of the simulations at the highest
resolution for $z=2$ (top row) and $z=0$ (bottom row). The left and
right columns show results for {\sm AREPO} and {\sm GADGET},
respectively.  We can readily identify a number of well-known features
in the gas phase-space \citep{Dave2001}, which show up in both
runs. The narrow ridge of gas with an upward slope and $\rho/\rho_{\rm
  bar}<10$ and $T<10^5\,\mathrm{K}$ consists of low density, highly
photo-ionised gas in the intergalactic medium (IGM). The tight
relation between temperature and density in this regime is maintained
by the competition between adiabatic expansion cooling and
photo-ionisation heating. The plume of gas with $\rho/\rho_{\rm
  bar}\sim10-10^4$ and $T\sim 10^5-10^7\mathrm{K}$ is comprised on the
other hand of shock-heated gas in virialised haloes and, at the lower
density end, in and around filaments. The narrow, downward sloping
locus of gas at $\rho/\rho_{\rm bar}>10^3$ and $T \cong 10^4 \,{\rm
  K}$ represents radiatively cooled, dense gas in galaxies. The
cooling times at these densities are short, so that gas remains close
to its equilibrium temperature where photo-ionisation heating balances
radiative cooling.  This temperature is a slowly decreasing function
of density and lies close to $10^4 \,{\rm K}$. Finally, once the gas
becomes very dense and reaches the star formation threshold, we
describe the gas by an effective equation of state representing the
mean thermal energy density of a two-phase medium of hot and cold
gas. This effective equation of state results in the upward sloping
high density line of gas, which represents the ISM in galaxies. The
pressurisation by supernova feedback in our subresolution model
prevents this gas from fragmenting under self-gravity.

\begin{figure*}
\centering
  \includegraphics[width=0.9\textwidth]{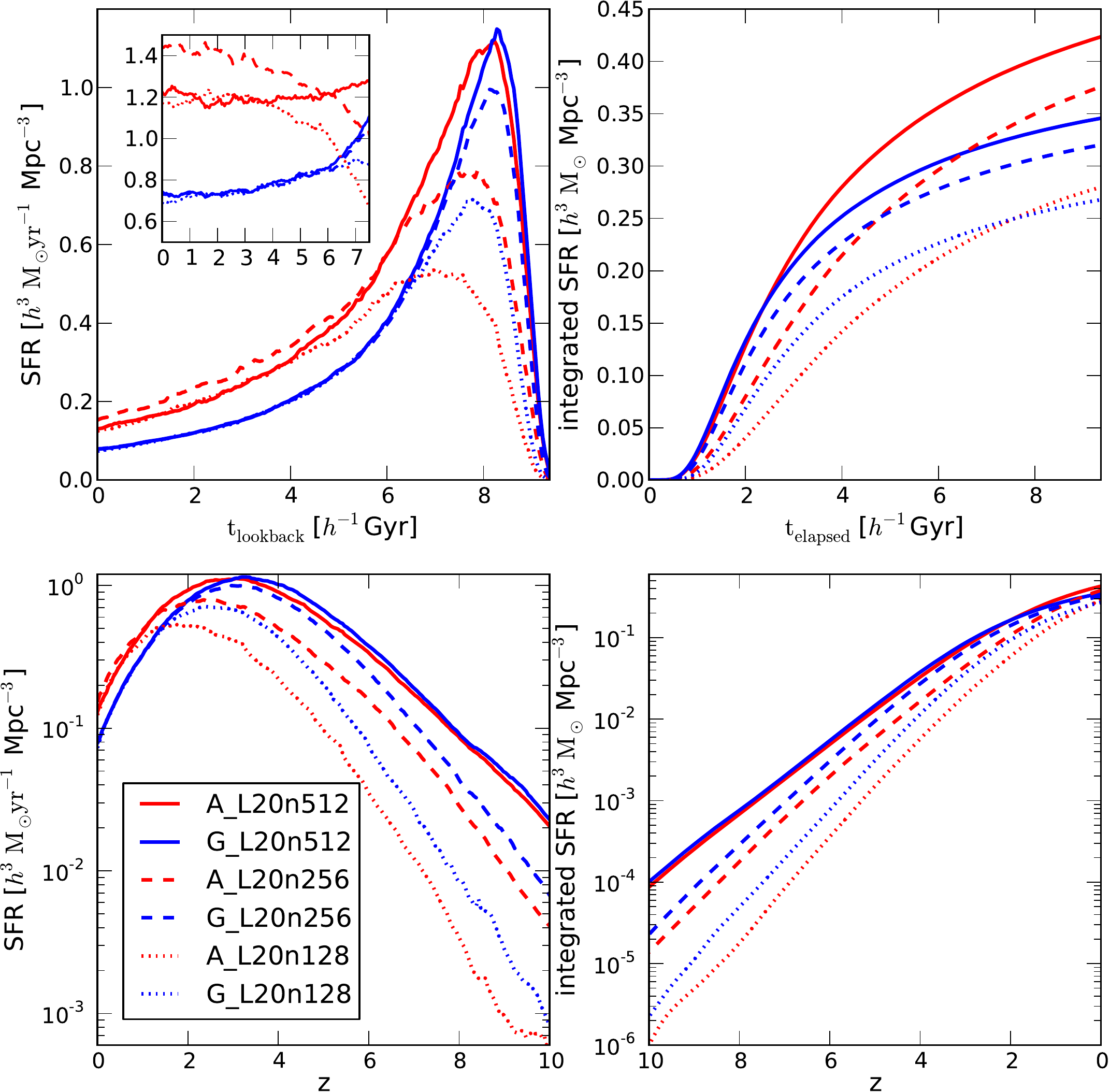}
\caption{Star formation rates per unit volume as a function of
  lookback time (top left) and redshift (bottom left). The panels on
  the right show the corresponding integrated star formation rates as
  a function of lookback time (top right) and redshift (bottom
  right). The inset in the upper left panel shows the relative
  deviations from the inferred mean SFR. Although the {\sm AREPO} star
  formation rates do not converge as well as the SPH results obtained
  with {\sm GADGET}, {\sm AREPO} shows at all resolution levels a
  significantly higher star formation rate at late times. The high
  redshift star formation peak agrees in amplitude quite well
  between the different codes if we focus only on the highest
  resolution (L20n512). }
\label{fig:sfr_hist} 
\end{figure*}

As Fig.~\ref{fig:rho_T} demonstrates, {\sm AREPO} and {\sm
  GADGET} lead to a qualitatively very similar $\rho-T$ phase-space
diagram. This demonstrates that the properties of the gas
distributions are overall quite comparable in both codes. However,
although there is general broad agreement between the simulations,
there are also some striking differences. First of all, the
distribution of hot gas extends to higher densities in the {\sm AREPO}
runs compared to the {\sm GADGET} simulations; i.e.~there is more hot
gas at higher densities in the {\sm AREPO} simulations. We note however
that in terms of total mass this is not very significant.  More
important, {\sm GADGET} appears to have more hot gas in general, which
can be inferred from the slightly more extended yellow region in the
phase diagrams. This is consistent with the temperature structure in
the surroundings of the galaxy shown in Fig.~\ref{fig:projections}.
The {\sm AREPO} runs also exhibit a more pronounced cooling feature
around $T\sim 10^{4.7}\,\mathrm{K}$, which corresponds to a local
minimum of the cooling curve between the two line peaks of hydrogen
and helium.  Although this feature is readily visible in the
phase-space diagrams of the {\sm AREPO} runs, the actual gas mass
populating that feature is very small. The faint stripe-like features 
seen in the top-left and bottom-left panels of Fig. 7 below the effective 
equation of state arise from numerical inaccuracies in the temperature 
evolution of some cells around star-formation sites. Here the temperature 
can sometimes scatter to very low values, which is one incarnation of the 
accuracy problems associated with supersonic cold fluid motions. 
In our multi-phase model, 
the temperature of these cells is then relaxed back to the effective equation of 
state temperature $T_{\rm eff}$ on a timescale $\tau_h$ given by equation 12 of 
of \cite{2003MNRAS.339..289S}. The simulation timestep $\Delta t$ of these cells is 
small against $\tau_h$, such that the temperature of the cell at the end of a 
relaxation step is $T' \simeq T_{\rm eff} \Delta t / \tau_h$. Because $T_{\rm eff} / \tau_h \propto \rho^{0.5}$ 
\citep[see][]{2003MNRAS.339..289S} and due to the power-of-two hierarchy of 
possible values for $\Delta t$, a set of parallel stripes of cells spaced 
by a factor of 2 results. The mass in these cells is however negligibly small, 
so that this effect has no influence on the dynamics.

To quantify the gas mass content in the different regions of the
$\rho$--$T$ phase-space diagrams in Fig.~\ref{fig:rho_T}, we introduce
the following cuts in the $\rho$-$T$ plane: (1) diffuse cool gas with
$\rho/\rho_{\rm bar}<10^3$ and $T<10^5\,{\rm K}$, (2) diffuse warm gas
with $\rho/\rho_{\rm bar}<10^3$ and $10^5\,{\rm K}<T<10^7\,{\rm K}$,
(3) diffuse hot gas with $\rho/\rho_{\rm bar}<10^3$ and $T>10^7\,{\rm
  K}$, and finally, (4) dense gas with $\rho/\rho_{\rm bar}>10^3$.  In
Fig.~\ref{fig:gas_phases_mass}, we first show the different gas mass
fractions of each of these four phases. All runs show the behaviour we
anticipated earlier; i.e.~there is the general trend that the cold gas
fraction decreases and the warm/hot gas fraction increases with
time. The hot gas fraction (lower left panel) increases due to shock
heating in haloes. Interestingly the details vary quite strongly
between the different numerical schemes. The {\sm GADGET} runs
typically show more hot and warm gas, and accordingly produce less
dense gas. Especially the mass fractions in the hot phase are very
different between the two schemes: for most of the time the mass
difference here is larger than one order of magnitude. We will show
below that these two findings are also consistent with the results for
the star formation histories of the different runs. We note that the
same trends are also found for the lower resolution simulations, but
the quantitative numbers differ slightly.  Although not fully
converged, there is more very dense gas (lower right panel) in {\sm
  AREPO} than in {\sm GADGET} at all resolutions, and also the amount
of cold gas (upper left panel) is larger in {\sm AREPO}. These results
confirm the visual impression obtained from
Fig.~\ref{fig:projections}, where we also saw that the hot atmosphere
is smaller in {\sm AREPO}, but the cold gas disk is larger. 

Next, we consider in Fig.~\ref{fig:gas_phases_volume} the
corresponding volume fractions of these gas phases.  The volume
fractions of the {\sm GADGET} runs are based on the specific volume of
individual SPH particles: $V_{\rm SPH} = m_{\rm SPH}/\rho_{\rm SPH}$.  However,
SPH is not volume-conserving in the sense that the sum of the specific
volumes of all SPH particles will not sum up to the total simulation
volume.  In finite-volume methods like the one used by {\sm AREPO},
this does not occur; all the cell volumes add up to the correct
simulation volume. For the comparisons of the volume fractions we
therefore do not calculate volume fractions with respect to the total
simulation box volume, but with respect to the sum of all individual
particle/cell volumes. This does not influence the volume fractions of
{\sm AREPO}, but changes the SPH volume fractions of {\sm GADGET},
which would otherwise not add up to unity.  The general trends we find
for the volume fractions in Fig.~\ref{fig:gas_phases_volume} are
similar to those for the mass fractions in
Fig.~\ref{fig:gas_phases_mass}. However, we can now clearly see that
the hot atmospheres are not only strongly reduced in mass in {\sm
  AREPO}, but also that their volume fraction is significantly smaller
compared to {\sm GADGET}. This is in agreement with the qualitative
results shown in Fig.~\ref{fig:projections} and
Fig.~\ref{fig:collage_gas_z2}. The same is true for the volume fraction
of the cold gas, which is however not as well-converged.

\begin{figure*} 
\centering
  \includegraphics[width=0.45\textwidth]{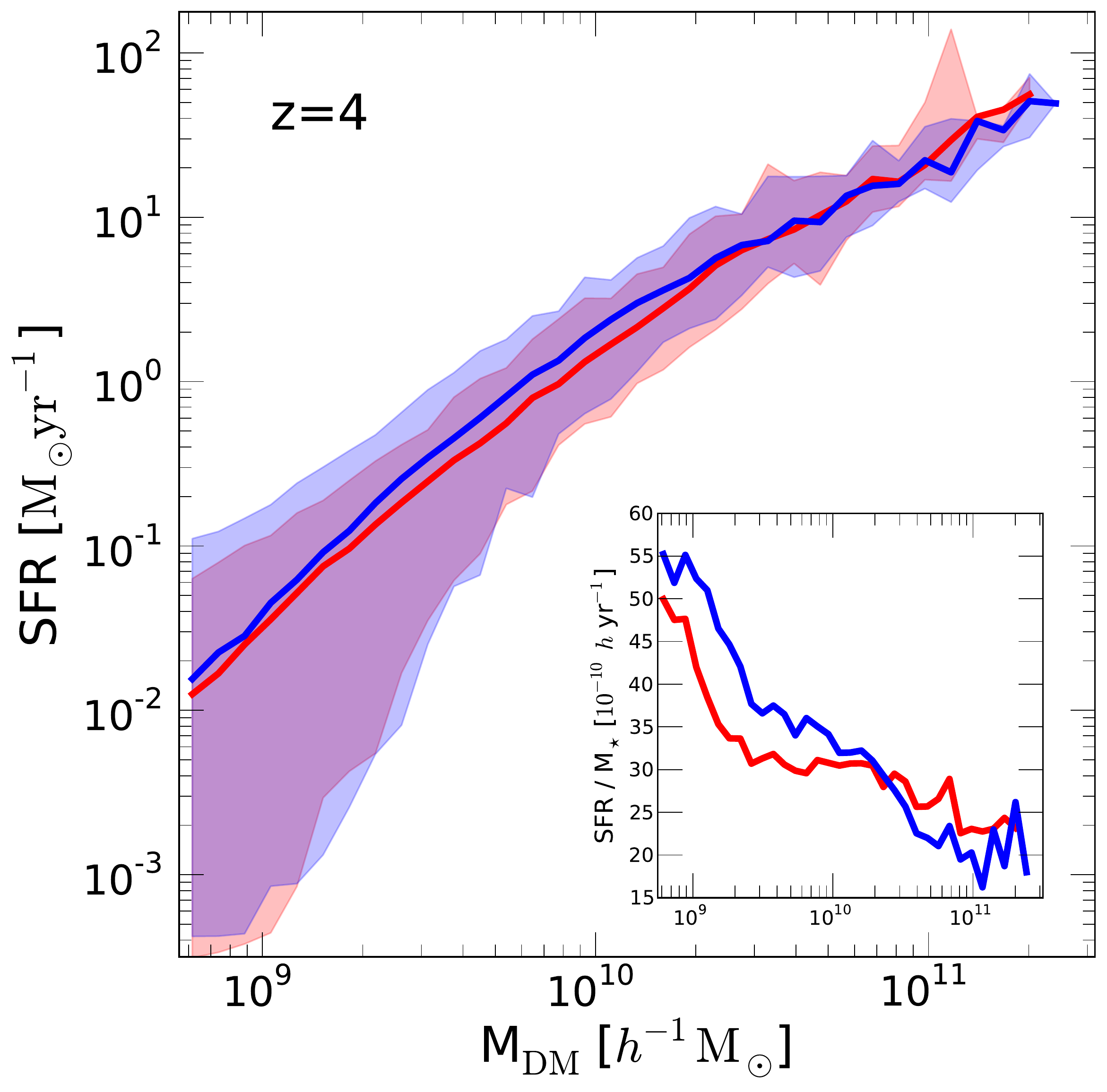} 
  \includegraphics[width=0.45\textwidth]{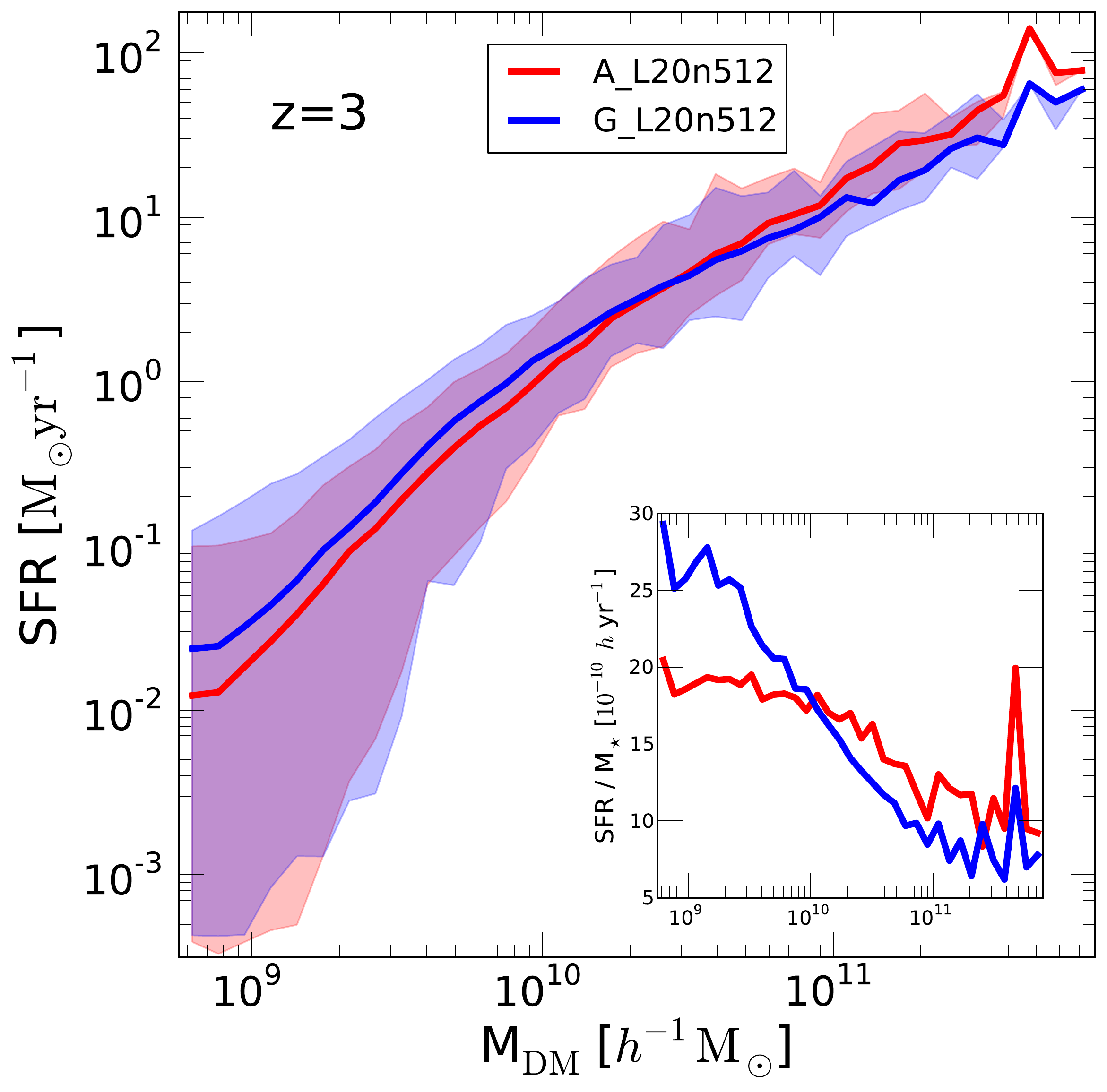} 
  \includegraphics[width=0.45\textwidth]{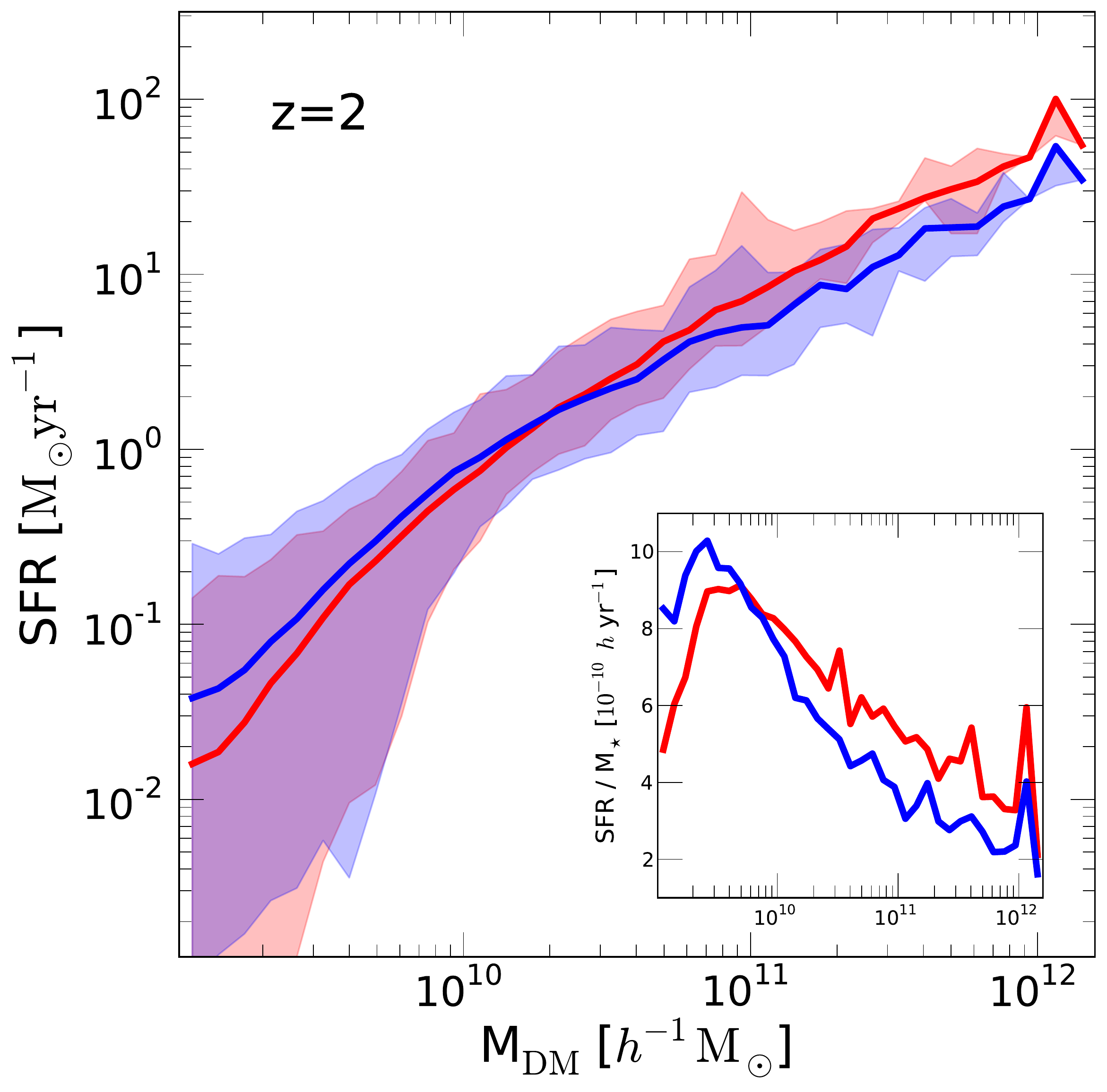} 
  \includegraphics[width=0.45\textwidth]{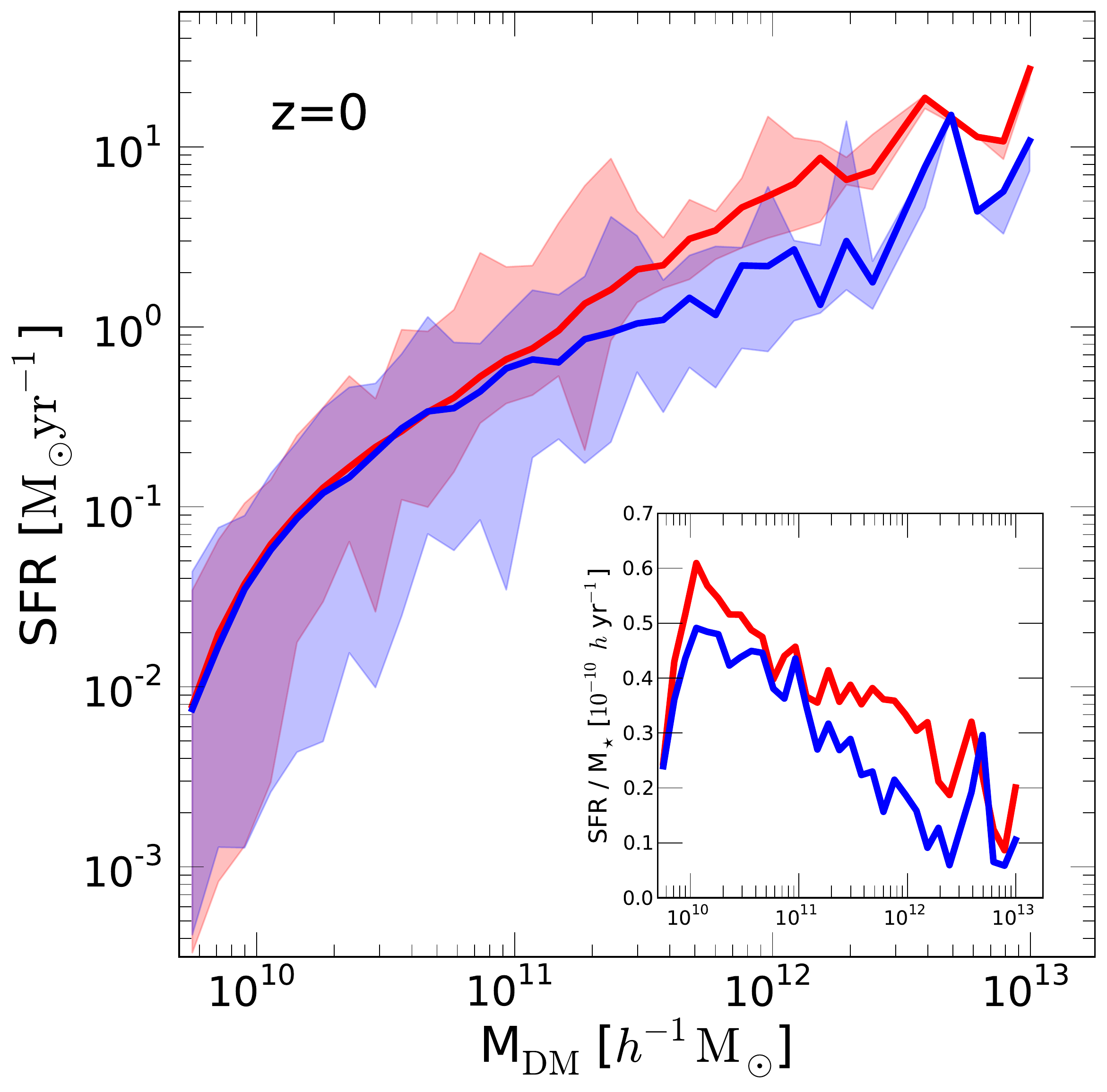} 
\caption{Star formation rates as a function of FoF group mass for the
  highest resolution simulations. Solid lines show the median values
  and the shaded regions represent the $1-99\%$ range of the
  distribution. For both cases and for all redshifts shown here the
  scatter is rather small. Towards lower redshifts a gradual trend
  shows up: both schemes start to deviate strongly towards the high
  mass end. At $z=0$ there is nearly perfect overlap of the star
  formation rates below $\sim 10^{11}\,h^{-1}\,\mathrm{M}_\odot$, but
  FoF groups above this mass form significantly more stars in the {\sm
    AREPO} simulation compared to the {\sm GADGET} run. Interestingly,
  this difference at the high mass end starts to show up first at
  around $z=3$. This coincides with the time that the
  global star formation rates deviate, as can be seen from the lower left
  panel of Fig.~\ref{fig:sfr_hist}. The massive end of the halo
  population is hence mainly responsible for the higher star formation
  rate in the moving mesh calculations at late times. The insets show
  specific star formations rates, where we divided the median curve
  of the main panel by the stellar mass $M_\star$ of the group.}
\label{fig:sfr_fof}
\end{figure*}

To focus more on the density structure, we marginalise
Fig.~\ref{fig:rho_T} over temperature and express the densities as
physical hydrogen number densities. We hence derive the hydrogen number
density probability distribution functions for the full simulation
volume, which are shown in Fig.~\ref{fig:density_PDF}.  The left
panel accounts for all gas particles/cells, whereas the right panel
shows only those with active star formation. In our star formation
model, the density threshold for the onset of star formation is set to
$n_{\rm H}=0.13~{\rm cm}^{-3}$, and therefore the right hand panel
occupies only this high density region. Interestingly, {\sm GADGET}
shows at all resolutions a significantly reduced gas fraction slightly
below the star formation density threshold.  We checked that the reduction
of the gas mass in this region is caused by the more extended
disks in {\sm AREPO}; i.e. the additional gas found in {\sm AREPO}
at these densities stems from disk material. Presumably, the `gap' phenomenon
found in SPH across strong contact discontinuities
\citep{2007MNRAS.380..963A}, such as the one we have here, also
contributes to the different breakdown in that density range.  The other
parts of the density distributions look rather similar for the two
simulation methods, however. We verified that the gas slightly below
the threshold is predominantly responsible for the more extended disks
in {\sm AREPO}. This gas effectively fills the gap which is present in
SPH.  The galaxies in {\sm GADGET} typically lead to a peak in the
density PDF above the star formation threshold instead of a more
plateau-like feature due to the extended gas disks in {\sm
  AREPO}. Therefore, the difference in the density PDF is a reflection
of significantly different galaxy radii for the different
hydro-schemes (see also Paper II).

\begin{figure} 
\centering
  \includegraphics[width=0.475\textwidth]{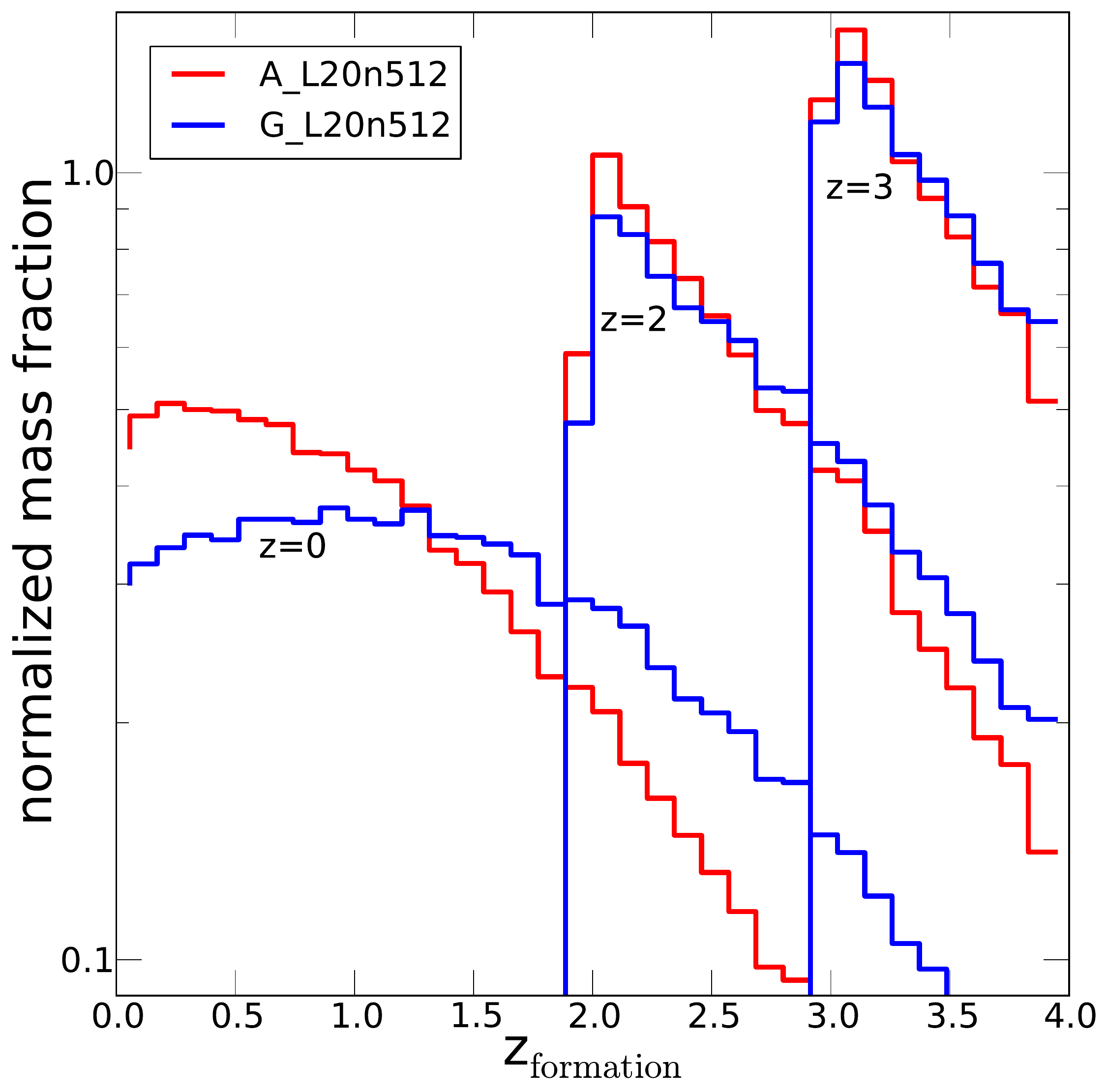} 
\caption{
Distribution of star formation times (redshifts) for the highest resolution {\sm AREPO}
and {\sm GADGET} simulations. The lines show the distribution at $z=0$, $z=2$
and $z=3$ as indicated. At higher redshift ($z=3$) the two curves agree reasonably
well in shape. But at $z=2$ there is already a trend visible, because the {\sm AREPO}
distribution is biased towards slightly younger stars. This effect becomes even more pronounced
towards lower redshift and is clearly visible in the distribution at $z=0$. At that
time the {\sm AREPO} result is significantly different from {\sm GADGET} and shows a
substantially larger fraction of young stars. The reason for this is the larger
overall star formation rate at late times, as shown in Fig.~\ref{fig:sfr_hist}.
The different stellar age distributions are also responsible for the bluer appearance of
the stellar populations of the {\sm AREPO} galaxies shown in Fig.~\ref{fig:collage_stars_z2}
and Fig.~\ref{fig:collage_stars_z0}.
\label{fig:stellar_age}}
\end{figure}

\subsection{Cosmic star formation history}

An important prediction of our simulations is the global star
formation history (SFH), which encodes key information about the
overall efficiency of the galaxy formation process. The SFHs of our
different simulation runs are presented in Fig.~\ref{fig:sfr_hist}.
The top left panel shows the SFH as a function of lookback time,
whereas the lower left panel plots it on a logarithmic scale as a
function of redshift. The corresponding integrated star formation
histories are shown in the right panels. The most obvious trend with
increasing resolution is that more high-$z$ star formation in small
haloes is resolved \citep{2003MNRAS.339..312S}.  This is simply due to
the better mass and spatial resolution for smaller haloes, which are
in part not even seen in the coarser simulations. This trend with
resolution also shifts the peak in the star formation rate density to
higher redshift with increasing resolution. Note however that there is
little cosmic time at these high redshifts, so that the mass of
stars formed there is small compared to the present stellar
density. Once the simulations resolve all haloes with virial
temperature of $10^4\,{\rm K}$ and above (necessitating a dark matter
particle resolution of about $10^6\,{\rm M}_\odot$), we expect however
that the trend with resolution will stop and all ordinary star
formation will be resolved (apart from additional Pop-III star formation,
which is not included in our models).

After the high-redshift peak of star formation, the star formation
rates decline and converge particularly well at lower redshifts for
the SPH-based {\sm GADGET} simulations. In this regime, most of the
star formation is contributed by higher mass haloes
\citep{2003MNRAS.339..312S}, for which our subresolution model
produces converged results already at moderate resolution. This
excellent convergence does not quite carry over to the moving mesh
calculations with {\sm AREPO}, where all resolutions slightly vary in
their late time star formation rates (for further explanation see 
Section 4.3 and Paper III). However, the mesh-based results
are still very close to one other, and more importantly, they lie
significantly higher than the SPH simulations, which is clearly
visible from the inset in the upper left panel of
Fig.~\ref{fig:sfr_hist} showing the ratio of the individual SFHs to
the mean.  We emphasise once more that this systematic difference is
a result only of the different hydrodynamical schemes, since both
codes use the same star formation model and calculate identical star
formation rates for a given gas density. There is a minor
technical difference in how stellar particles are created, but this
does not affect the outcomes, as we have explicitly checked in
numerous test calculations of isolated haloes and galaxies (see Paper III). The latter
do show very good agreement in the star formation rates between {\sm
  GADGET} and {\sm AREPO}, so the difference we find in
Fig.~\ref{fig:sfr_hist} must have a cosmological origin.

To understand better whether this increased star formation rate
affects all halo masses or whether it is dominated by a certain halo
population, we plot in Fig.~\ref{fig:sfr_fof} the star formation rates
as a function of FoF group mass for our highest resolution simulations.
Solid lines in Fig.~\ref{fig:sfr_fof} represent the median values and
shaded regions show the $1-99\%$ range of the distribution. The insets show
the specific star formation rates. In both
cases, and for all redshifts shown, the scatter around the median
relation is rather small.  At high redshift ($z=4$), the differences
between the SPH and moving-mesh calculations mainly occur at the
intermediate and low mass end, where {\sm GADGET} produces more stars
than {\sm AREPO}. This is in agreement with the general trend found
for the global SFH, where {\sm GADGET} shows a slightly higher star
formation rate at high redshift. We note that the higher specific star formation
rates in {\sm AREPO} at late times are due to a significantly more efficient
cooling.

We point out that one reason why {\sm GADGET} shows at high-$z$ a
larger star formation rate (SFR) at a given resolution than {\sm
  AREPO} is related to the fact that already at this early time the
disks in {\sm AREPO} are slightly more extended and hence have a lower
gas surface density (as demonstrated in Paper II).  Therefore, the
star formation rate is lower compared to the denser, more blob-like
galaxies formed in the SPH calculations.  We note that this
discrepancy between {\sm GADGET} and {\sm AREPO} decreases once the
resolution in SPH becomes high enough, but it still induces a
noticeable delay in {\sm AREPO}'s SFR-peak compared to {\sm GADGET},
although the peak amplitudes agree quite well.

\begin{figure}
\centering
  \includegraphics[width=0.48\textwidth]{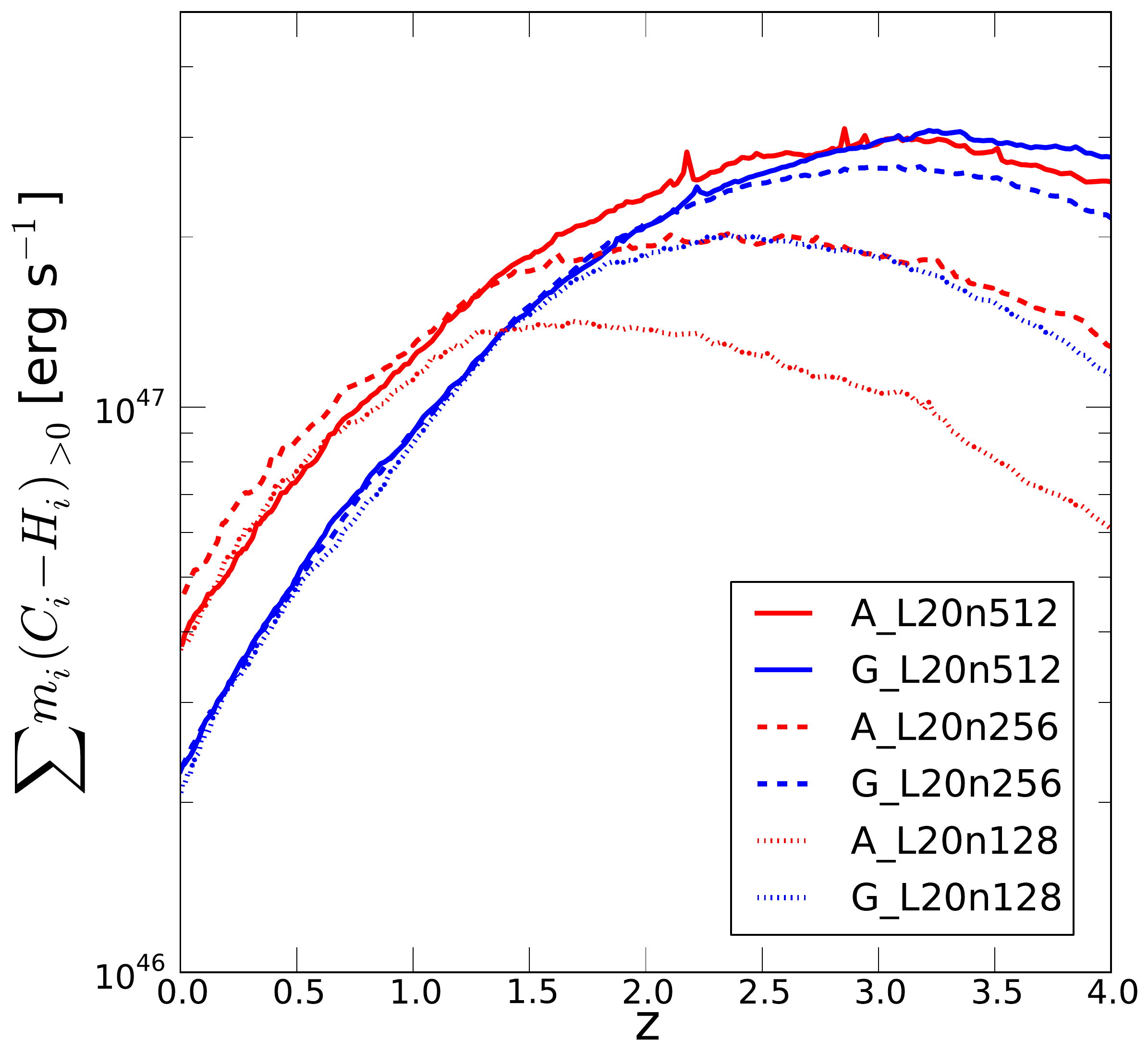}
\caption{Summed net cooling rate ($\sum m_i (C_i - H_i)$) of the gas
  for the different simulations, where $C_i$ and $H_i$ are the cooling
  and heating rates per unit mass.  We set the net cooling to zero for
  particles/cells that are subject to net heating.  The rates of all {\sm
    AREPO} runs are higher than those of the {\sm GADGET} simulations
  at late times.  This is directly related to the higher SFRs of all
  {\sm AREPO} runs compared to {\sm GADGET}, as seen at late times in
  Fig.~\ref{fig:sfr_hist}. At high redshift, the cooling rates of {\sm
    AREPO} are lower than those of {\sm GADGET} for the intermediate
  and the lowest resolution. This is a consequence of the more compact
  disks in {\sm GADGET} at these high redshifts.  This cooling rate
  difference decreases once we reach the highest resolution.  In that
  case the cooling rates of {\sm GADGET} and {\sm AREPO} agree better
  at high redshifts, but still show quite large differences at low
  redshifts, which also produces the different SFR history.
\label{fig:cooling_rates}}
\end{figure}

Towards lower redshifts, a gradual trend of a different nature appears:
while {\sm AREPO} and {\sm GADGET} start to agree better at low
masses, the schemes deviate more and more strongly towards the high
mass end. At $z=0$, there is a nearly perfect overlap of the star
formation rates below $\sim 10^{11}\,h^{-1}\,\mathrm{M}_\odot$, but FoF
groups above this mass form significantly more stars in the {\sm
  AREPO} simulations compared to {\sm GADGET} runs. Interestingly,
this difference in large haloes first starts to show up at around $z=3$
(upper right panel), which roughly coincides with the time when the
global star formation rates start to deviate, as can be seen from the
lower left panel of Fig.~\ref{fig:sfr_hist}.  This implies that the
massive end of the halo population is mainly responsible for the
higher star formation rate in the moving mesh calculations at late
times. These massive systems form at late times and systematically
create more stars in {\sm AREPO} than the corresponding {\sm GADGET}
calculations, driving the global SFH to a different behaviour in
the two codes.

\begin{figure*}
\centering
  \includegraphics[width=0.45\textwidth]{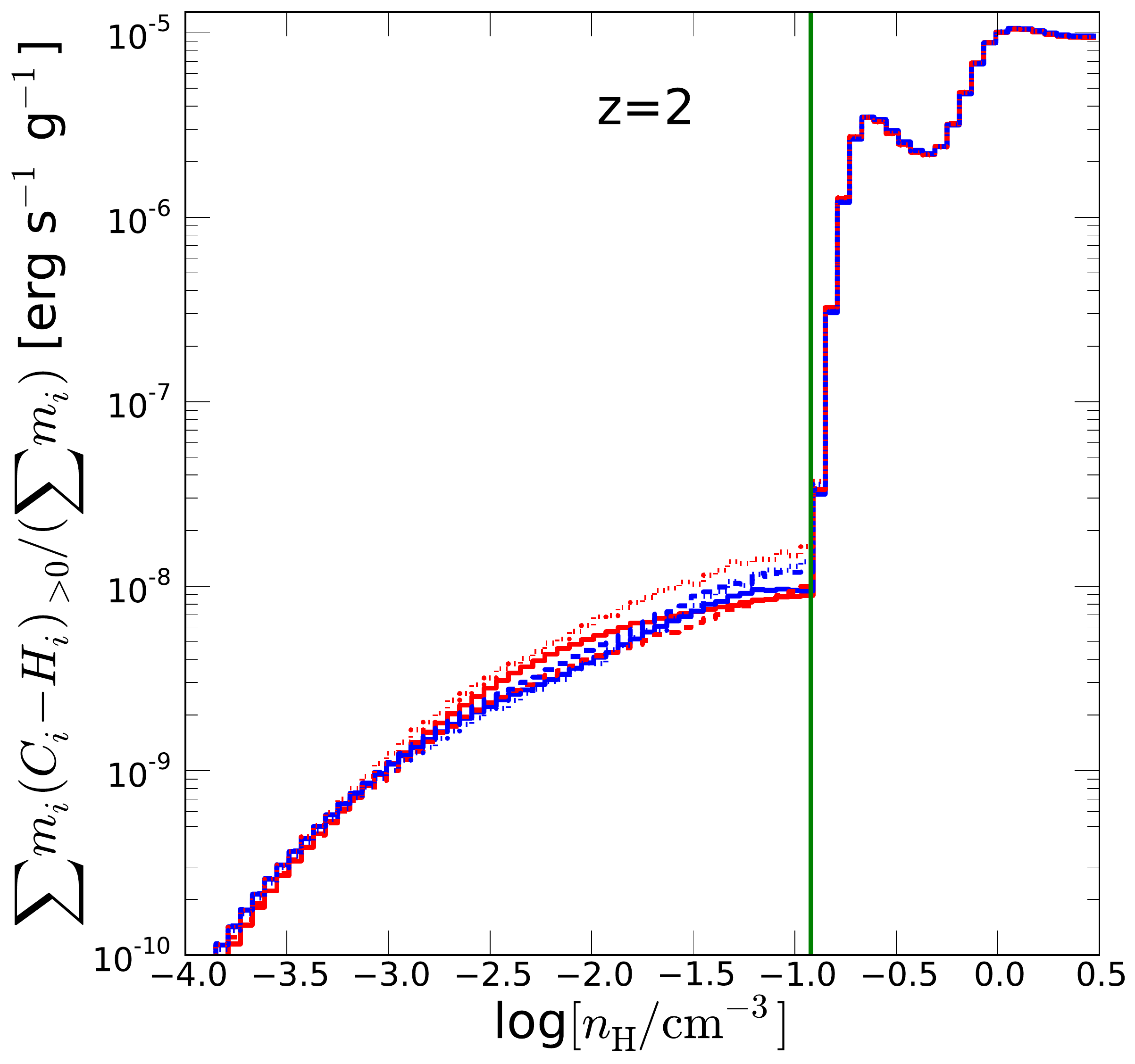}
  \includegraphics[width=0.45\textwidth]{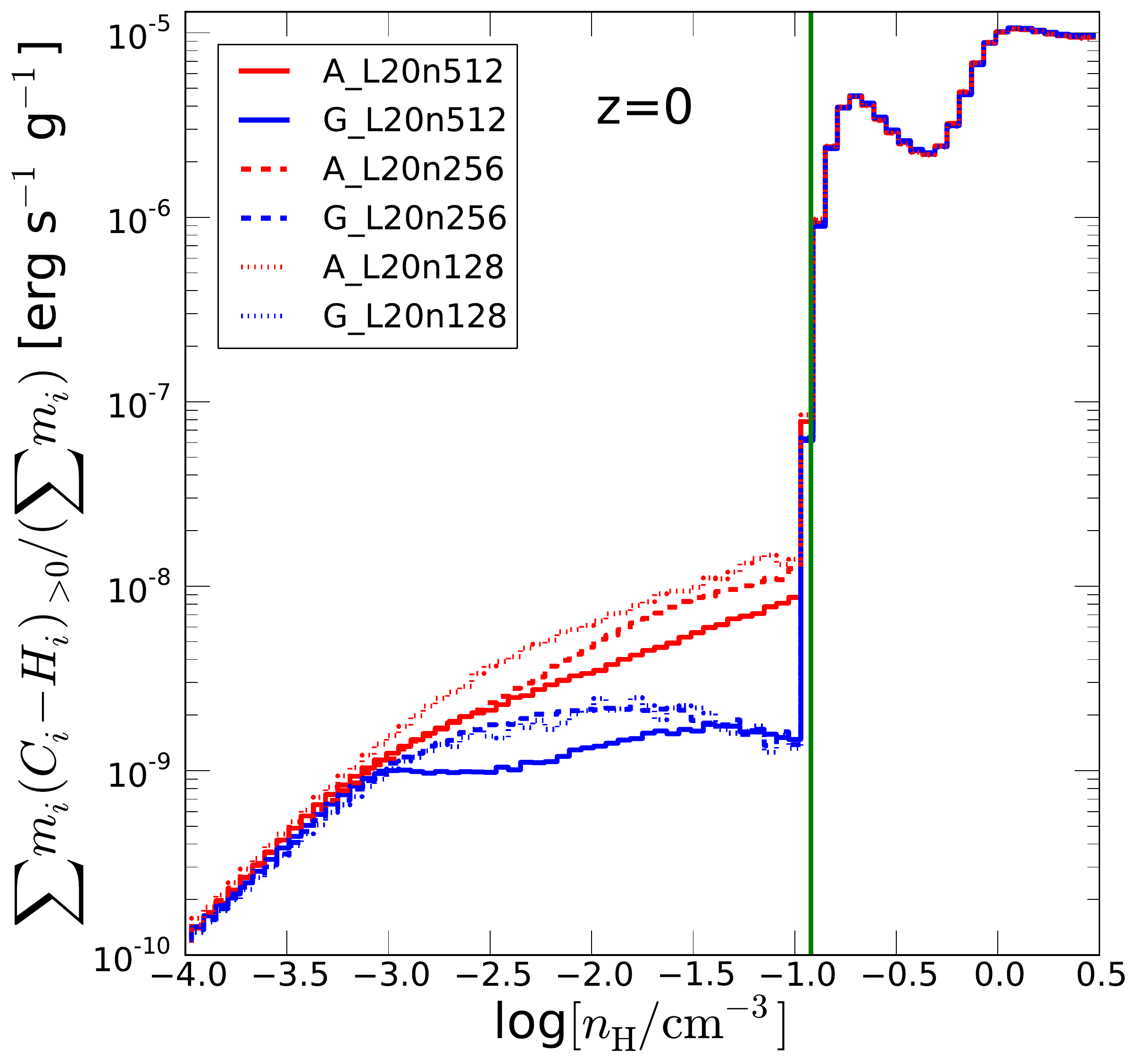}
\caption{The two panels show the mean mass-weighted net cooling rate
  ($(\sum m_i (C_i - H_i))/(\sum m_i)$), where $C_i$ and $H_i$ are the
  cooling and heating rates per unit mass, as a function of hydrogen number
  density at $z=2$ (left) and $z=0$ (right). We set the net cooling to
  zero for particles/cells that experience net heating. At
  high redshift the cooling rates agree well between different
  codes, but towards lower redshifts SPH shows systematic differences
  compared to the moving mesh approach. At $z=0$, {\sm AREPO} shows a
  significantly higher mean cooling rate directly below the star
  formation density threshold.  We note that the cooling rates agree
  well outside of the shown density range, i.e. the discrepancies in
  cooling occur primarily just below the
  threshold.
\label{fig:density_vs_coolingrate} }
\end{figure*}

The only way such a difference in the SFH can occur is a more
efficient cooling behaviour in {\sm AREPO} in haloes above $\sim
10^{11}\,h^{-1}\,\mathrm{M}_\odot$, allowing more gas to accumulate at the
bottom of halo potentials and above the density threshold for star
formation.  This interpretation also agrees with our findings for the
mass fractions and volume fractions in the different gas phases, where
{\sm GADGET} shows at all resolutions significantly more hot gas than
{\sm AREPO}. Furthermore it agrees with the gas density PDFs in
Fig.~\ref{fig:density_PDF}, which show that {\sm AREPO} has a larger
fraction of dense gas slightly below the star formation density
threshold.  As we will discuss in the next section, the culprit of the
reduced cooling in {\sm GADGET} lies in different hydrodynamical
dissipation rates in the outer parts of haloes compared with {\sm
  AREPO}.  In Paper III of this series, we furthermore show that {\sm
  AREPO}'s more accurate treatment of hydrodynamical fluid
instabilities strongly affects the stripping of gas out of gas clumps
that fall into haloes. This in turn leads to significant differences in
how the thermodynamic structure of haloes is affected by the
hierarchical merging process, contributing to the cooling differences.

We note that the different star formation histories also result in a
different distribution of stellar ages in individual galaxies. This
can already be seen in Fig.~\ref{fig:collage_stars_z2} and
Fig.~\ref{fig:collage_stars_z0}, where not only the morphology of the
stellar distribution is different, but also the average age as can be
inferred from the colour scale. The {\sm GADGET} galaxies look clearly
redder overall than the {\sm AREPO} galaxies, which are on average
bluer. To quantify this in more detail, we show in
Fig.~\ref{fig:stellar_age} distribution functions of the formation
redshifts of the stellar populations at $z=0$, $z=2$ and $z=3$, for
our two highest resolution simulations. This figure demonstrates that
at high redshift ($z=3$) the distributions agree reasonably well in
shape. But at $z=2$, a clear trend is already visible, because the
{\sm AREPO} distribution becomes biased towards slightly younger
stars. This has to do with the slight shift of the peak of the SFH and
the different late time behaviour of the SFH found in
Fig.~\ref{fig:sfr_hist}. The difference in the distributions becomes
even more pronounced towards lower redshift and is clearly evident at
$z=0$. At that time the {\sm AREPO} distribution is quite distinct
from {\sm GADGET} and shows a significantly larger fraction of young
stars. The reason for this lies in the larger star formation rate at
late times, as discussed above.

\section{Gas cooling}

\subsection{Cooling emission and temperature evolution}

The differences between {\sm AREPO} and {\sm GADGET} discussed so far
point towards a different cooling behaviour between the two codes. To
demonstrate this more clearly, Fig.~\ref{fig:cooling_rates} shows
the total net cooling rate ($\sum m_i (C_i - H_i)$), where $C_i$ and $H_i$ are the
cooling and heating rates per unit mass as a function of redshift, accounting for all gas in our
simulations.  In the case of net heating we set this rate to zero. 
The figure demonstrates that at late times, the cooling rates of all {\sm AREPO} runs
are higher than those of corresponding {\sm GADGET} runs. We note that
this is true for all three resolutions, yielding a clear separation
between the set of blue curves (representing the SPH simulations) and
the red curves (representing the moving mesh calculations).
Interestingly, the difference partially reverses at high redshift,
where the cooling rates of {\sm AREPO} at the intermediate and the low
resolution are lower than those of {\sm GADGET}.  This is a
consequence of the more compact galaxies in {\sm GADGET} at these high
redshifts, which result in larger densities and therefore more cooling
radiation.  This high-$z$ cooling rate difference decreases and nearly
disappears once we reach the highest mass resolution. In that case the
cooling rates of {\sm GADGET} and {\sm AREPO} agree reasonably well at
high redshifts, as shown in Fig.~\ref{fig:cooling_rates}. In contrast,
the quite large and systematic difference at low redshifts persists
independent of resolution.

Further insight into the origin of the cooling difference is provided
by Fig.~\ref{fig:density_vs_coolingrate}, which shows the mean
mass-weighted net cooling rates ($(\sum m_i (C_i - H_i))/(\sum m_i)$) at redshifts $z=2$ (left panel) and $z=0$
(right panel) as a function of hydrogen number density. If the temperature structure of the gas at fixed
density would agree between all simulations, then the mean mass-weighted
cooling rates should also agree for a given hydrogen number density.
But as Fig.~\ref{fig:density_vs_coolingrate} shows, this is not the
case and some systematic differences between {\sm GADGET} and {\sm
  AREPO} are present.  Interestingly, the curves in
Fig.~\ref{fig:density_vs_coolingrate} differ however only over a
limited density range corresponding to the diffuse gas in haloes,
beginning at the star formation density threshold (marked by the
vertical green line) and extending to lower densities. In that
density range, {\sm AREPO} shows a significantly larger mean cooling
rate at all resolutions, which must be due to a slightly lower gas
temperature in the mean at these densities. We note that the 
magnitude of
the difference is however redshift dependent and becomes smaller as we
go to higher redshifts, as seen by comparing the $z=0$ and $z=2$
panels in Fig.~\ref{fig:density_vs_coolingrate}.

At other hydrogen densities, including the range not shown in
Fig.~\ref{fig:density_vs_coolingrate}, the SPH and moving-mesh cooling
rates agree well for all our runs.  Above the threshold for star
formation this is however not surprising, because here the effective
equation of state produces  a tight correlation between density and
temperature (as seen in Fig.~\ref{fig:rho_T}). This leads to a nearly
one-to-one mapping of density to temperature in this regime, and
explains the two-peak structure of the curves in
Fig.~\ref{fig:density_vs_coolingrate} at high densities, which simply
reflects the primordial cooling curve with its characteristic hydrogen
and helium peaks.

\begin{figure}
\centering
  \includegraphics[width=0.475\textwidth]{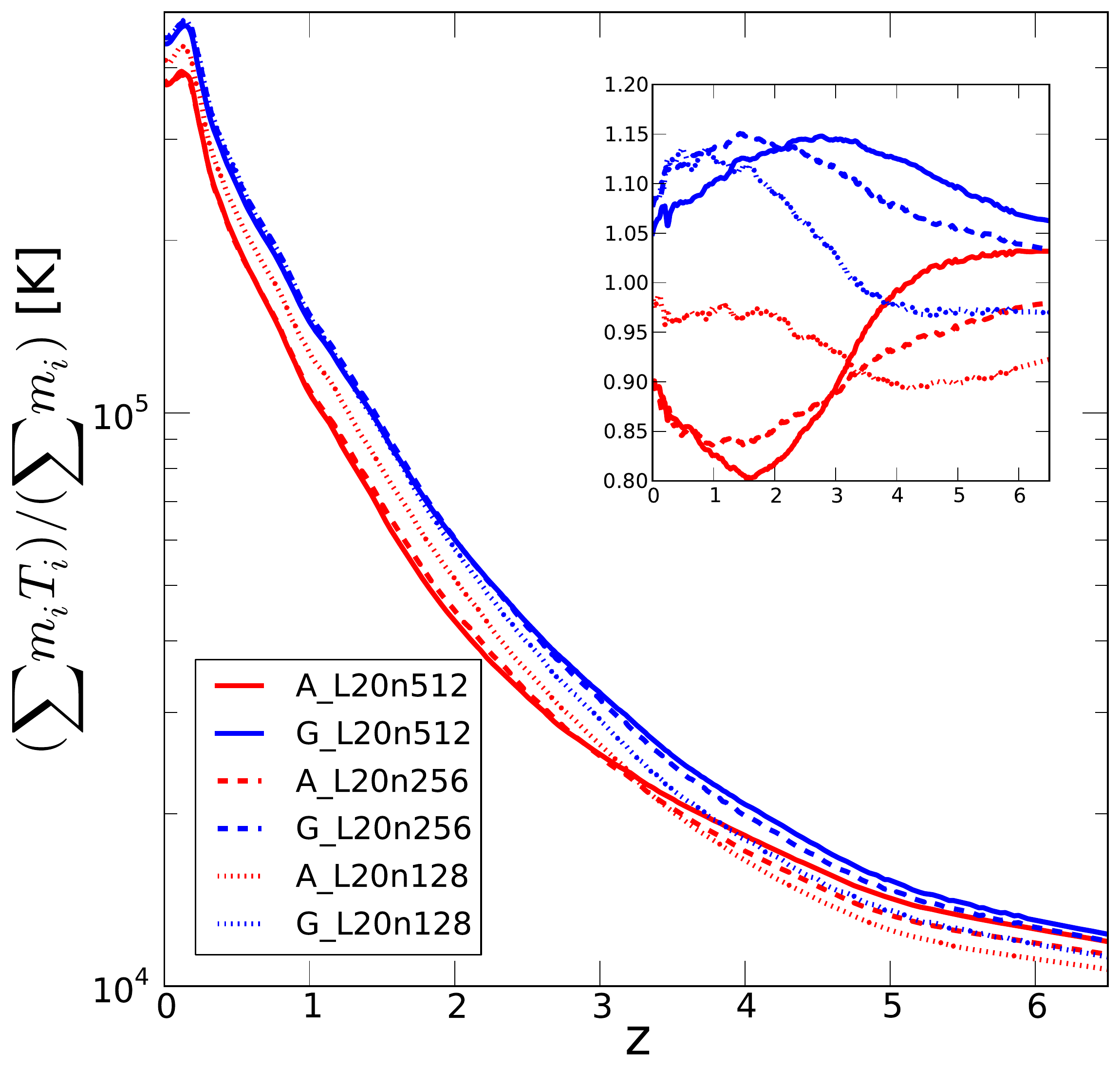}
\caption{Mean mass-weighted temperature of the gas in the different
  simulations. In the inset we divide each curve by the mean of all
  curves. In agreement with the findings for the SFR and the cooling
  rates, the mean temperature of all {\sm GADGET} simulations is
  higher than that of the {\sm AREPO} runs at nearly all redshifts.}
\label{fig:mean_temp}
\end{figure}

Further evidence for a different temperature structure in the two
simulation runs is provided by Fig.~\ref{fig:mean_temp}, which plots
the mean mass-weighted temperature in the whole simulation volume as a
function of time. In the inset we divide each curve by the mean of all
the curves to emphasise deviations. We find that the mean temperature
of {\sm GADGET} simulations is $\sim 15\%$ higher than that of the
{\sm AREPO} runs. The higher temperature can explain the reduction in
cooling emission of the SPH simulations relative to our moving-mesh
calculations, and it is consistent with our above findings for the SFR
evolution and the cooling emission.  It thus appears that the origin
of the low redshift discrepancy must lie in a different efficiency of
non-adiabatic heating processes in the gas, as this is required to
explain differences in the temperature distribution at a given
density.
 
\begin{figure*}
\centering
  \includegraphics[width=0.33\textwidth]{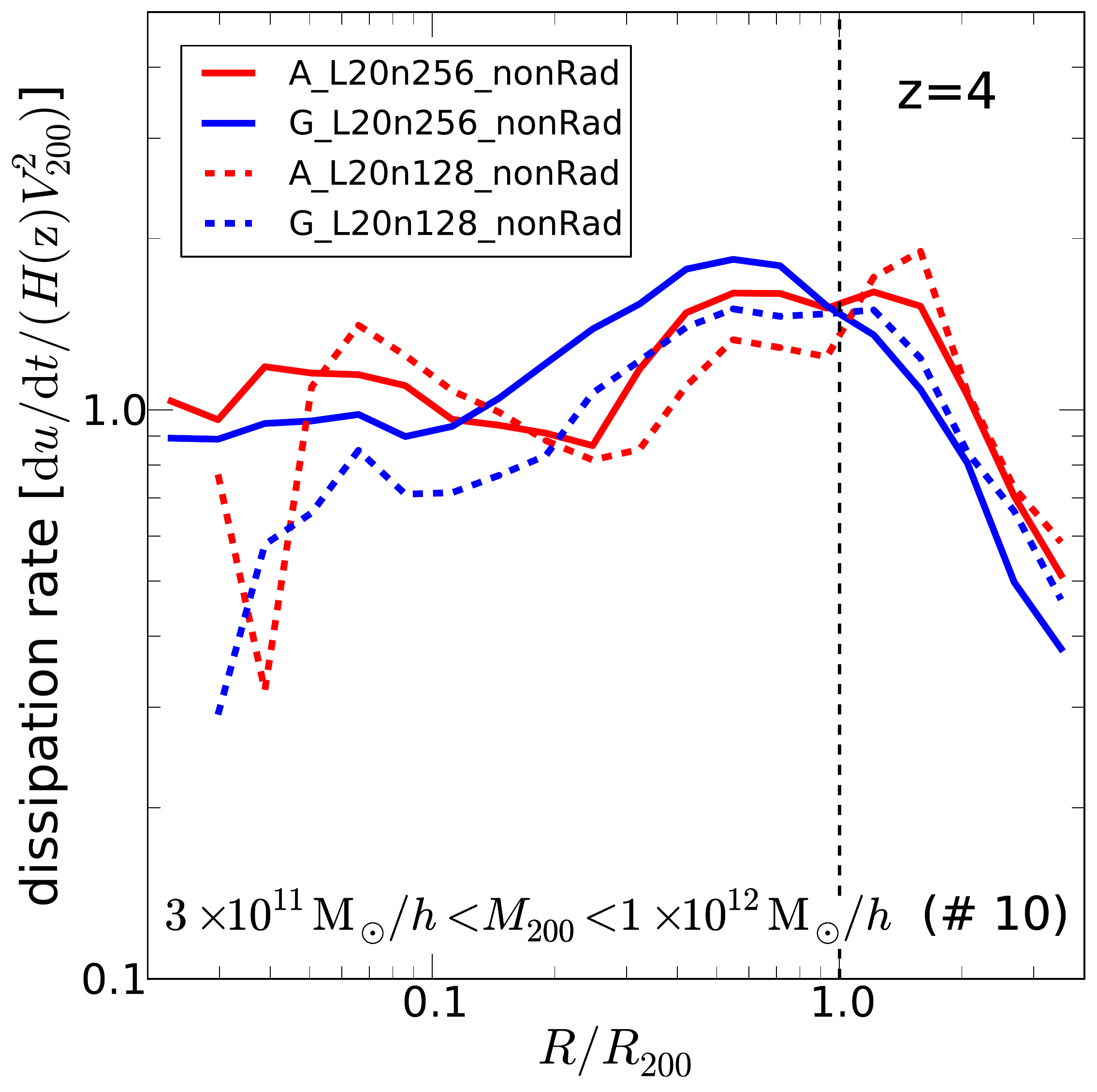}
  \includegraphics[width=0.33\textwidth]{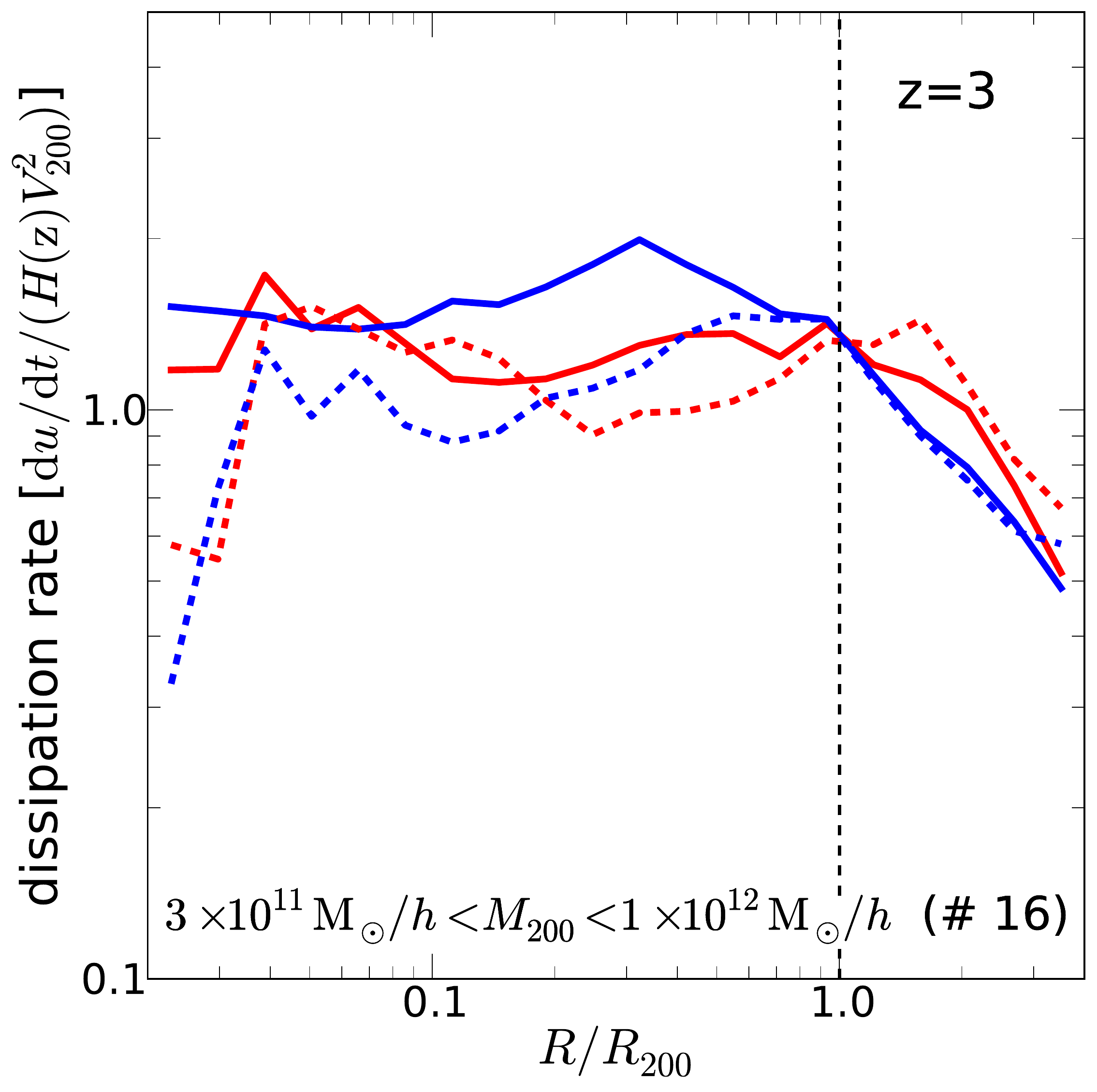}
  \includegraphics[width=0.33\textwidth]{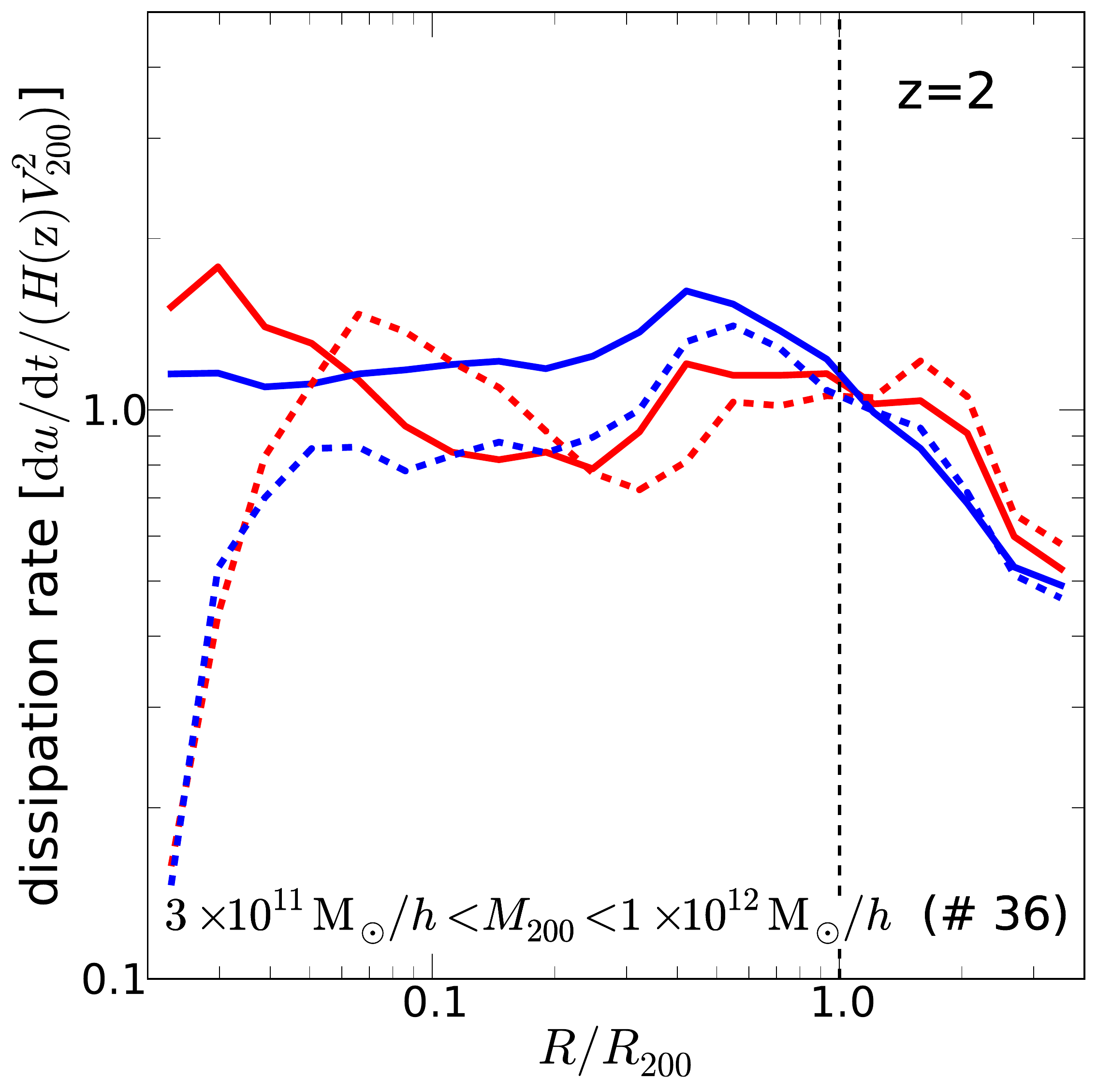}\\
  \includegraphics[width=0.33\textwidth]{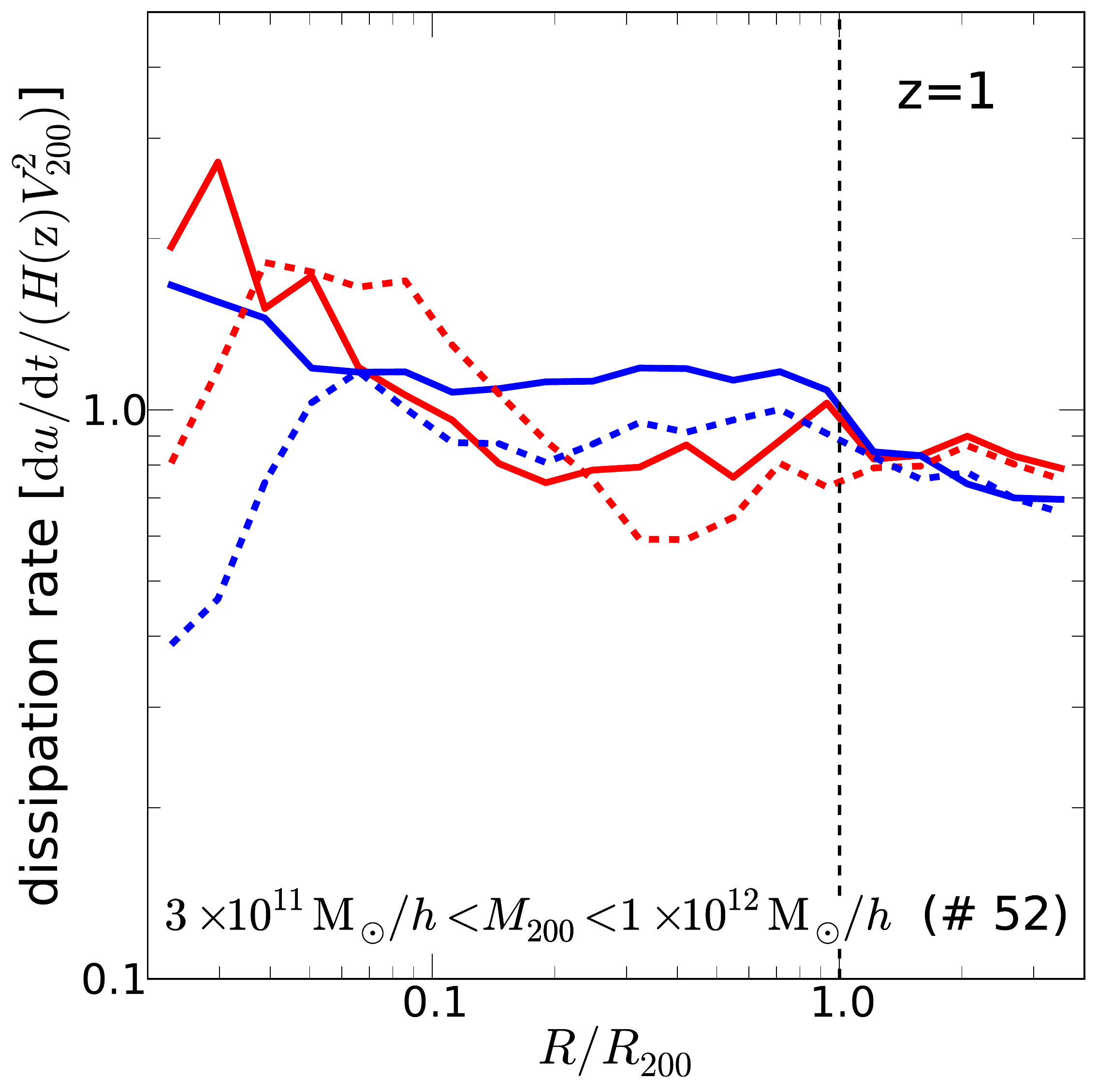}
  \includegraphics[width=0.33\textwidth]{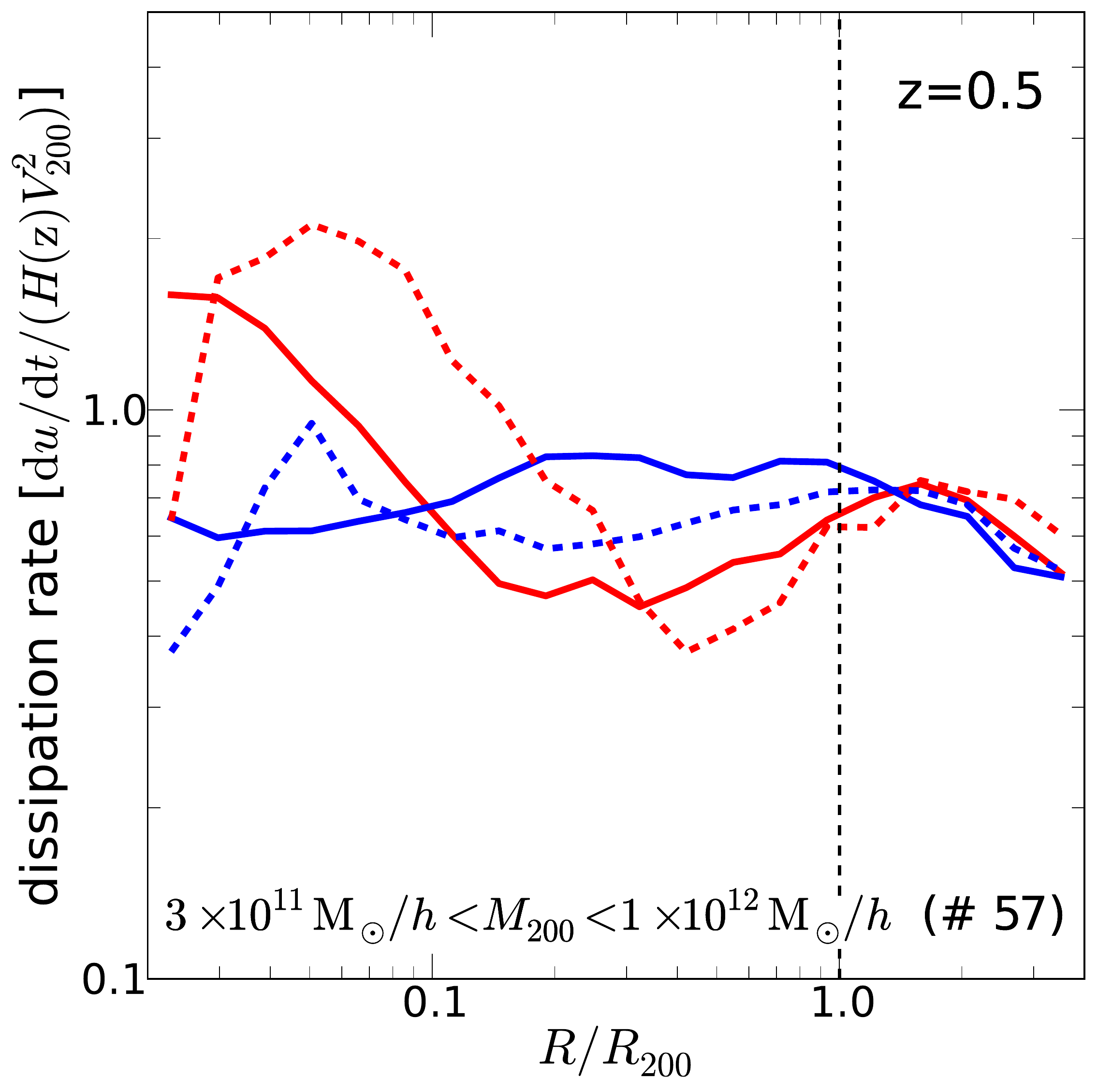}
  \includegraphics[width=0.33\textwidth]{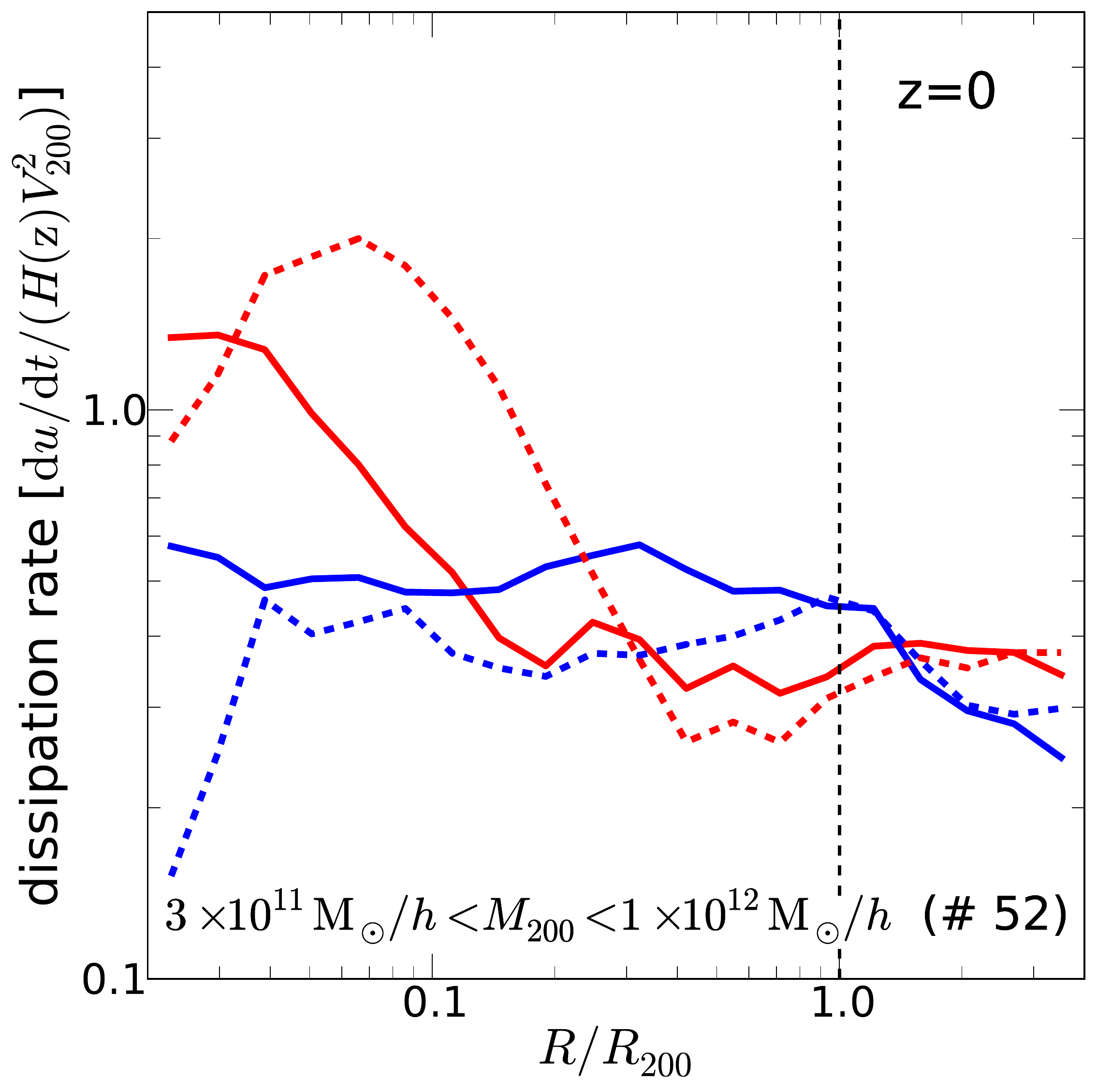}\\
\caption{Dissipation rate profiles for haloes formed in non-radiative
  simulations.  Each panel is for a given redshift between $z=4$ and
  $z=0$ (as labelled), and stacks the results for all haloes contained
  in the mass range $3.0\times 10^{11}\,h^{-1}\,{\rm M}_\odot \le {\rm M}_{\rm 200}
  \le 1.0\times 10^{12}\,h^{-1}\,{\rm M}_\odot$. The number of haloes that are
  averaged over is indicated in each panel. The red lines show
  results for the $2\times 256^3$ non-radiative simulation with {\sm
    AREPO}, the  blue lines the corresponding results for the
  {\sm GADGET} simulation. Dashed lines give the equivalent
  $2\times 128^3$ results.
\label{fig:dissipationrate1}}
\end{figure*}

\subsection{Dissipative heating in haloes}

It is widely appreciated that the intracluster medium of galaxy
clusters is partially supported by subsonic turbulence
\citep{Schuecker2004}, which is created by curved accretion shocks
around the haloes, and more importantly, by the hierarchical infall of
structures. Indeed, a number of numerical studies have analysed
turbulence in the intracluster medium
\citep{Dolag2005,VazzaBrunetti2009,Valdarnini2011, 2011MNRAS.414.2297I}.  A similar level
of turbulence can also be expected in smaller haloes that are large
enough to support quasi-hydrostatic atmospheres. It has further been
shown that shock heating is not only important in strong shocks at the
outer accretion radius, but is also significant in weaker flow shocks
that occur in large parts of the halo volume
\citep{Pfrommer2006}. Physically, the dissipation associated with
shocks and with the decay of turbulence occurs through microphysics on
very small scales. In our numerical approach, we neglect the
physical viscosity (which is assumed to be effectively zero on all
resolved scales; this is why we employ the Euler and not the
Navier-Stokes equations; see e.g. 
\cite{Munoz2012} for a Navier-Stokes implementation of {\sm AREPO})
and account for the necessary dissipation
through numerical viscosity. In SPH this is provided explicitly in
terms of an artificial viscosity, whereas in {\sm AREPO} it is
introduced implicitly through the solutions of the Riemann solver and
cell-averaging.

The good conservation properties of SPH allow it to capture
one-dimensional shock waves quite well, even though the post-shock
velocity field typically shows substantial noise
\citep{2010ARA&A..48..391S,2011MNRAS.413..271A}. This already hints
that SPH may not be particularly accurate for subsonic flow phenomena,
such as the turbulence we expect in the virialised gas of newly formed
haloes. Indeed, \cite{Bauer2011} have
systematically compared driven isothermal subsonic turbulence for {\sm
  GADGET} and {\sm AREPO}, using the same versions of the codes we
employ here. They find that SPH fails quite badly to account for a
turbulent cascade in the subsonic case, whereas a Kolmogorov
scaling is obtained for the moving-mesh code. This happens despite
identical driving fields in both cases, and is ultimately caused by
inaccurate gradient estimates in SPH and different dissipation as a
function of scale. The SPH simulations dissipate
most of the energy already close to the driving scale in the subsonic case, whereas in {\sm
  AREPO} efficient dissipation happens only on much smaller scales, so
that a self-similar turbulent cascade can develop over some inertial range. 
In addition, \cite{Bauer2011} find that SPH exhibits
a second maximum in the dissipation on very small scales of order the
mean interparticle separation. Here the subsonic noise of SPH is dissipated
by the artificial viscosity. Interestingly, the total amount of
dissipation on these scales is quite independent of the artificial
viscosity value itself. While a higher viscosity reduces the amplitude
of the small-scale SPH noise, the dissipation rate on these scales
still remains roughly the same, suggesting that the cause of the
small-scale noise, which originates in errors in the pressure gradient
estimates, cannot be reduced effectively with a different viscosity
setting.

There are good reasons to expect that these differences in the
dissipation properties of the hydrodynamical schemes may also induce
important effects for the thermodynamic structure of cosmological
haloes, which in turn can easily give rise to variations of the amount
of gas that cools out of these haloes. To examine this further, we have
carried out two auxiliary non-radiative  simulations at resolutions of
$2\times 128^3$ and $2\times 256^3$ particles/cells. These simulations
use the same box-size and parameters as our galaxy formation
runs, except that cooling and star formation were not
included. The use of non-radiative simulations allows us to cleanly
employ the method of \cite{Bauer2011} for measuring the
instantaneous dissipation rate of particles and cells,
respectively. In the case of SPH, this can simply be done by measuring
the work per unit time done by the artificial viscosity forces,
which represents the sole source of entropy generation in this method.
For the moving-mesh code, we instead also advect the thermodynamic
entropy between the cells, and then measure the rate of entropy
generation by comparing the state of a cell at the end of a timestep
computed adopting energy conservation with the state expected if the
entropy would have remained constant. The inferred rate of entropy
production can then also be converted into a fiducial heating rate.

\begin{figure}
\centering
  \includegraphics[width=0.39\textwidth]{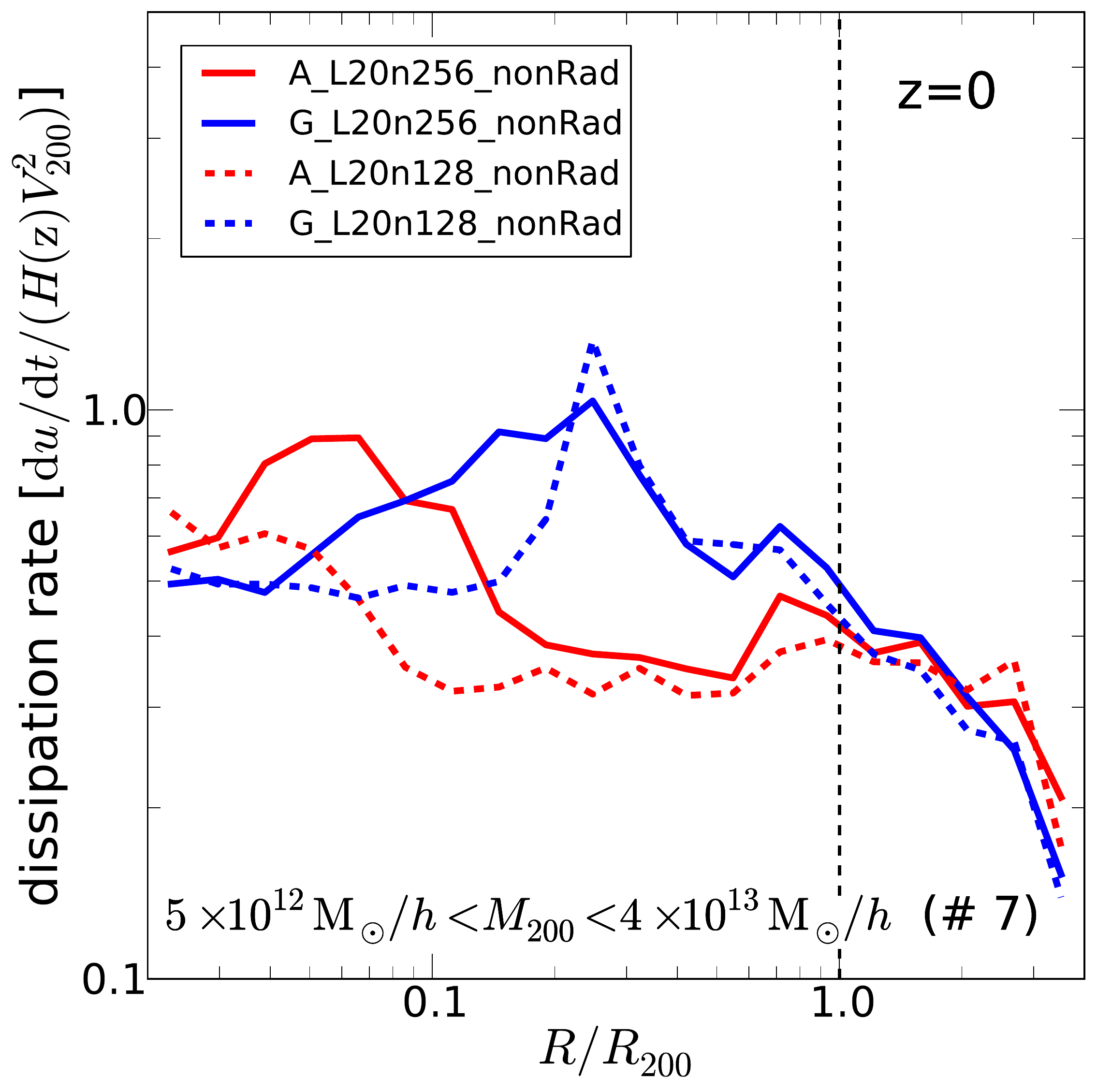}\\
  \includegraphics[width=0.39\textwidth]{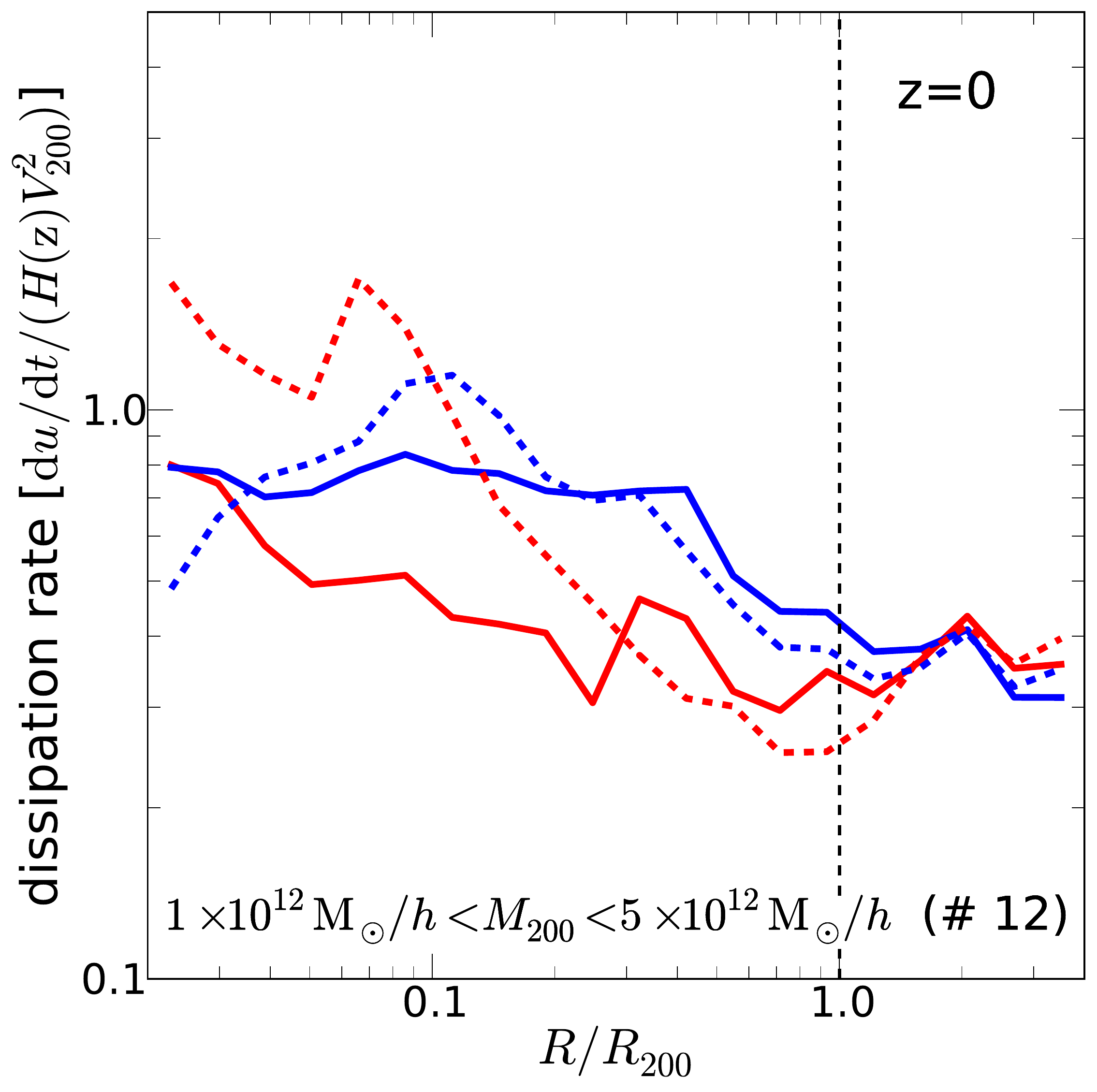}\\
  \includegraphics[width=0.39\textwidth]{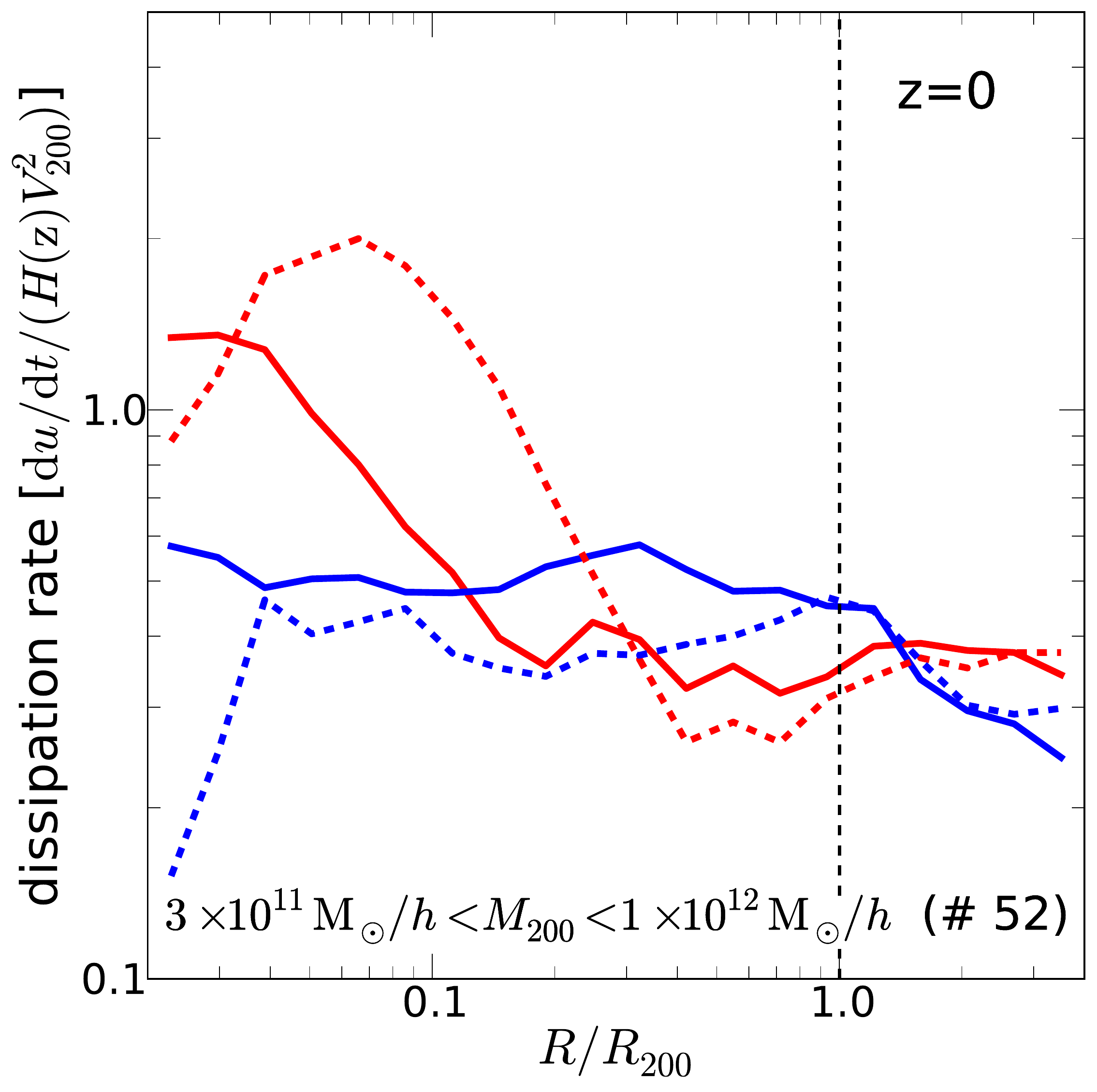}\\
\caption{Dissipation rate profiles for haloes formed in non-radiative
  simulations.  The different panels are for $z=0$ haloes contained in
  different mass ranges, from more massive (top) to less massive
  (bottom), as labelled. Note that the bottom panel is identical to
  the last panel in Fig.~\ref{fig:dissipationrate1}, which is here
  repeated for ease of comparison. The solid lines show results for
  the $2\times 256^3$ non-radiative simulations, while the dashed
  lines give the equivalent $2\times 128^3$ results.
\label{fig:dissipationrate2}}
\end{figure}

In Fig.~\ref{fig:dissipationrate1}, we compare stacked profiles of
the spherically averaged dissipative heating rate of haloes obtained in
this way. For the plots, we have selected all haloes in the virial mass
range $3\times 10^{11}\,h^{-1}\,{\rm M}_\odot < {M}_{\rm 200} < 1\times
10^{12}\,h^{-1}\,{\rm M}_\odot$, where ${M}_{\rm 200}$ refers to the mass inside a
radius ${R}_{200}$ that encloses 200 times the critical density. We have
then placed 20 logarithmic bins onto the radial range $0.01\times
R_{200}$ to $4.0 \times R_{200}$, and produced mass-weighted averages
of the dissipation rate per unit mass, ${\rm d}u/{\rm d}t$, that we
measured for each particle/cell. In order to allow a calculation of an
average profile by stacking the haloes, we express the dissipation rate
in dimensionless units by multiplying it with the Hubble time at the
given redshift, and by normalising it to ${V}_{200}^2$ of the particular
halo, where ${V}_{\rm 200} = \sqrt{G {M}_{\rm 200}/ {R}_{\rm 200}}$ is the circular
velocity at the virial radius. In the individual panels of
Fig.~\ref{fig:dissipationrate1}, we show results for different
redshifts, ranging from $z=4$ to $z=0$, and we compare {\sm AREPO}
with {\sm GADGET} at the two resolutions considered here.

We see that there is a clear systematic difference in the dissipative
heating rates as a function of halo-centric distance.  {\sm AREPO}
produces more heating in the infall region directly outside the virial
radius whereas {\sm GADGET} shows more heating in the outer parts of
virialised haloes, where most of the gas mass is located. This
systematic difference is present at all redshifts in large haloes. We
also note that the nature of the difference is robustly preserved at
different resolutions, even though there seem to be small residual
trends with spatial resolution. Interestingly, at fixed halo mass, the
relative level of dissipation in the innermost parts of haloes increases
in the mesh-code with time, which may be related to the build-up of a
turbulent entropy core in these haloes. In
Fig.~\ref{fig:dissipationrate2}, we consider dissipation profiles as
a function of halo mass at $z=0$, which however does not reveal any
clear trend with halo mass.

Our interpretation of the systematic difference revealed by
Figs.~\ref{fig:dissipationrate1} and \ref{fig:dissipationrate2} is
that: (1) shocks are captured more efficiently by {\sm AREPO} in the
accretion regime of haloes, and (2) {\sm AREPO} strips gas out of
infalling objects more efficiently there; both effects contribute to a
stronger dissipation in this region. In contrast, SPH stops infalling
gas less efficiently, leading to more dissipation at smaller radii. A
further contributor to the heating there lies in the dissipation of
the subsonic noise in SPH, which is constantly recreated in this
region by tapping free energy from the disturbances that impinge on
the haloes from the outside. As a net result of this dissipation of
subsonic noise in the outer parts of haloes, the SPH simulations end
up hotter overall (see also Fig.~\ref{fig:mean_temp}). Furthermore,
this difference in the heating occurs in a region where one expects
the cooling radius of many haloes in cosmological radiative
simulations, thereby directly affecting the amount of cold gas
produced in the quasi-hydrostatic cooling flows in large haloes.
We note that the measurement of the dissipation rate is difficult 
due to its high time variability. We use above stacked 
profiles of small sets of halos to mitigate this problem, but even 
then the inner logarithmic bins average over much smaller gas mass 
than the outer parts. This implies that the profiles in the inner 
$10\%$ of the virial radius are relatively noisy. But this inner
part is largely irrelevant as the impact on the cooling rates comes 
from larger radii as we discussed above.

\subsection{Mixing in haloes}

Another reason why differences in the global star formation rate occur
between {\rm AREPO} and {\sm GADGET} lies in the different accuracy
with which hydrodynamical fluid instabilities are treated. In
particular, we expect that for the moving-mesh code cold gas can be
stripped more efficiently from galaxies as they fall into larger
haloes, and this gas is then mixed with the halo's diffuse gas, which
also lowers its temperature.

\begin{figure}
\centering
  \includegraphics[width=0.475\textwidth]{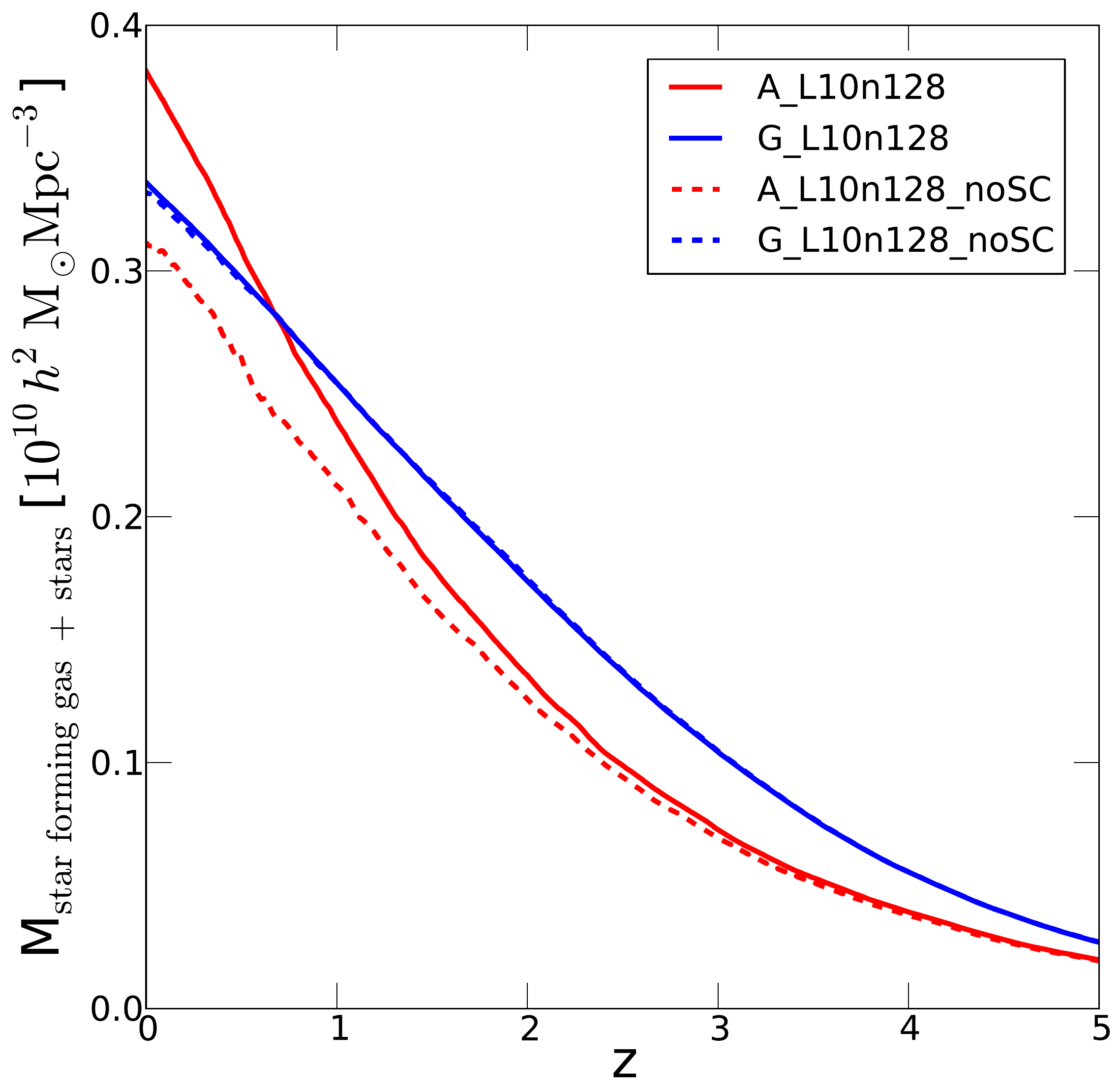}
\caption{Test simulation of a $10\,h^{-1}\,\mathrm{Mpc}$ box with
  $128^3$ SPH particles/cells. The $y$-axis shows the total amount of
  star-forming gas and stellar mass in the simulation volume. Solid
  lines show the result of a calculation with our standard star
  formation prescription. For the dashed lines we turned off the
  creation of stellar particles; i.e. there are no stars formed but
  the gas still cools and is supported by the multiphase equation of
  state.  The solid and dashed curves overlap well for {\sm GADGET},
  which shows that once gas overcomes the density threshold it either
  forms stars or stays above the threshold. There is no loss of
  star-forming gas mass.  For {\sm AREPO}, the amount of stellar mass
  and star-forming gas mass is larger than the amount of star-forming
  gas if stellar particle creation is turned off. This is because {\sm
    AREPO} resolves fluid instabilities better and therefore allows
  significant gas mass loss from infalling satellites. Here gas that
  once has been above the threshold for star formation may well be
  mixed with halo gas and reach much lower densities again.  On the
  other hand, once stellar particles are created from dense gas, the
  total amount of baryonic material that can be lost from subhaloes is
  reduced, which explains the difference seen in the two moving-mesh
  calculations.}
\label{fig:gas_balance}
\end{figure}

To demonstrate this difference more explicitly in our cosmological
runs, we performed special test simulations of a
$10\,h^{-1}\,\mathrm{Mpc}$ box with $128^3$ SPH particles/cells.  In
Fig.~\ref{fig:gas_balance}, we show the sum of the total amount of
star-forming gas mass and stellar mass in this box as a function of
redshift.  We note that an SPH particle/cell is star-forming if its
density is higher than the threshold of the subresolution star
formation model. The solid lines show the result of calculations with
our standard star formation prescription, i.e.~these simulations are
equivalent to the main runs presented in the paper. However, for the
dashed lines we turned off the creation of stellar particles, keeping
everything else the same.  In this case, the cold gas accumulates in
the galaxies and is supported by the effective equation of state
expected for gas at these densities, except that the gas is not
depleted and does not turn into stars at the normal rate. For these
runs, the mass in the figure therefore refers to the total amount of
star-forming gas in these simulations (no stars are
present). Interestingly, the solid and dashed curves overlap well for
{\sm GADGET} (blue) in Fig.~\ref{fig:gas_balance}, showing that once
gas has overcome the density threshold it will never return to lower
density. Stripping of gas out of galaxies does not appear to happen in
any significant way, otherwise we would expect that the run without
star formation should end up with a smaller amount of collapsed
baryons. Instead, the SPH result is consistent with no stripping at
all, i.e.~once a gas particle has cooled, it never returns to lower
density even though the galaxy may be subject to substantial shear
flows upon halo infall. This behaviour has also been found by He\ss~\&
Springel (2011, submitted) in a comparison of galaxy-wind interactions
in SPH and the VPH technique \citep{2010MNRAS.406.2289H}.

For {\sm AREPO} the situation is very different. As
Fig.~\ref{fig:gas_balance} shows, the sum of stellar mass and
star-forming gas mass is larger than the amount of star-forming gas in
the run without star particle creation.  Here the relevant fluid
instabilities for stripping (like the Kelvin-Helmholtz instability)
are significantly better resolved and facilitate a substantial mass
loss of infalling overdense blobs/subhaloes. Because of this, the sum
of the mass of stars and of star-forming gas is higher in {\sm AREPO}
for the run with star particle creation than for the fiducial run
where this is suppressed.  In the latter simulation, there is simply
more gas around that can be stripped again, while in the run with star
particle creation, baryons that have been converted to stars do of
course not suffer from fluid instabilities anymore. Apart from
affecting the cooling in haloes, this difference in the stripping and
mixing efficiency also modifies the dynamical friction timescale of
infalling gas clumps, as we examine in more detail in Paper III of
this series.

\section{Generic issues with SPH}
 
Various modifications have been proposed to the conventional
implementation of SPH as in {\sm GADGET} to improve its reliability in
certain cases.  For example, changes to the computation of smoothed
pressure gradients \citep[][]{2011MNRAS.413..271A} or the addition of
an artificial thermal conductivity to the equations of motion
\citep[e.g.][]{2008JCoPh.22710040P} enable SPH to better handle
shearing interfaces and the onset of Kelvin-Helmholtz instabilities.
Also, a number of studies focused on the development of improved
artificial viscosity parameterisations in SPH codes \citep{Morris1997,
  Cullen2010}, with the goal of reducing the numerical viscosity away
from shocks.

Given the results presented in this paper, it is tempting to argue
that one or several of these modifications of ``standard SPH'' may
lead to much better agreement with the moving-mesh results or even
resolve the discrepancies.  While we cannot exclude this possibility, we
note the above refinements are ad hoc in the sense that they are designed
to mitigate against particular problems in some circumstances, and may
have unwanted side-effects in other situations.  While these
side-effects may often be benign, the proposed modifications of SPH do
not address other, more generic problems with this technique. In
this section, we summarise these issues in order to put our findings
into context with the recent literature on SPH techniques.

One issue that has rarely been discussed in the cosmological community
concerns the convergence properties of SPH.  This is manifested in at
least two ways, the first having to do with local smoothed estimates
in SPH, and the second is related to the fact that SPH is not formally
Lagrangian.  Below, we first address the question of under what conditions
convergence is expected in SPH and how this influences the relation
between spatial resolution and the total particle number.  Then, we
discuss the pseudo-Lagrangian character of SPH and its implications for
defining convergence with this method.  Finally, we briefly suggest
ways in which these issues might be dealt with. Incidentally, they all would
have the effect of making SPH resemble a moving-mesh scheme like {\sm
  AREPO}.

\subsection{Convergence in SPH and nearest neighbour number}

For a real fluid, we define the density, ${\rho({\bf x})}$, to be a
continuous function of space, determined by averaging over many
molecules locally around ${\bf x}$.  In order to formulate a discrete
SPH counterpart to this (and this applies to all fluid variables, as
well as the equations of motion) a two-step procedure is applied.

First, we introduce a smoothed version of the original density field,
$\left<\rho\right>\!({\bf x})$, by convolving ${\rho({\bf x})}$ with a
smoothing kernel ${W({\bf x}, h)}$ according to
\begin{equation}
\left<\rho\right>\!({\bf x}) \, = \, \int \rho({\bf x^\prime}) W({\bf x - x^\prime}, h) \, \mathrm{d}V^\prime,
\end{equation}
where the integral is over all space and $h$ is
the smoothing length.

Second, the continuous smoothed field $\left<\rho\right>\!({\bf
x})$ is replaced by a discrete quantity, ${\left<\rho\right>_d\!({\bf
x})}$, so that it can be represented computationally.  This is done by
regarding ${\rho({\bf x^\prime}) \mathrm{d}V^\prime}$ as a mass
element ${\mathrm{d}m^\prime}$ and partitioning the fluid into a set
of $N$ discrete mass elements so that the above integral can be
approximated by a discrete sum:
\begin{equation}
\left<\rho\right>\!({\bf x}) \, \rightarrow \, \left<\rho\right>_d\!({\bf x}) \, = \, \sum_{j=1}^{N_{\rm ngb}} m_j W({\bf x - x}_j, h({\bf x})),
\end{equation}
where $m_j$ is the mass of fluid element $j$, ${\bf x}_j$ is its
spatial location, and the expression allows different fluid
locations/elements to have different smoothing lengths.  In principle,
the sum extends over all $N$ fluid elements.  However, in order that
the smoothed density ${\left<\rho\right>\!({\bf x})}$ approaches the
continuum limit represented by ${\rho({\bf x})}$, it is necessary that
${W({\bf x}, h)}$ is spatially localised.  So in practice a localised
kernel with compact support is adopted, implying that the discrete sum
extends not over all $N$ fluid elements but over a subset ${N_{\rm
    ngb}}$ of those neighbouring a certain point in space.

Now, we can ask about the conditions that must be satisfied for the
smoothed, discrete version of the density field to asymptote to the
actual continuum limit.  That is, under what circumstances is the
following true?
\begin{equation}
\left<\rho\right>_d\!({\bf x}) \, \rightarrow \, \left<\rho\right>\!({\bf x}) \, \rightarrow \rho({\bf x}) \, .
\end{equation}
Consider the second of these requirements first.  In order for
${\left<\rho\right>\!({\bf x}) \, \rightarrow \rho({\bf x})}$ we must have
 ${W({\bf x}, h) \, \rightarrow \, \delta({\bf x})}$ as ${h
  \rightarrow 0}$, as implied by the definition of
${\left<\rho\right>\!({\bf x})}$ in the above integral.  This also
requires that ${N\rightarrow \infty}$, otherwise there exist not enough
particles around a given spatial location to form smoothed
averages.

Turning to the first requirement above, ${{\left<\rho\right>_d\!({\bf x}) \,
    \rightarrow \, \left<\rho\right>\!({\bf x})}}$ will be satisfied
only if at the same time we enforce the condition ${N_{\rm
    ngb}\rightarrow \infty \,}$.  This has not been emphasised in the
cosmological literature, but from the defining relation above for
${\left<\rho\right>_d\!({\bf x})}$ it is clear that the discrete
approximation will not asymptote to the smoothed density field unless
this limit is taken.  Or, to put it another way, if the number of
neighbours is held fixed as $h\rightarrow 0$, there will be a constant
source of error present on the most well-resolved scales that will not
vanish as the resolution and number of particles is increased
indefinitely \citep[see also][]{Rasio2000, Read2010}.  (For a
discussion of this requirement in the context of elasticity, see
e.g. \cite{Belytschko1998}.)

To the best of our knowledge, the optimal way to enforce this
requirement has not been established.  However, there is little doubt
that this must be addressed in order that SPH provides properly
convergent solutions, as argued recently by
\cite{2011arXiv1101.2240R}: ``When performing a convergence study
using SPH, it is important to vary both the number of particles ... as
well as the ratio of smoothing length to particle spacing [to increase
the number of neighbours].  [Otherwise] the error due to the summation
interpolant [remains] constant [as the total particle number is
increased].''  We emphasise that this requirement is not fundamentally
an issue with SPH, but simply reflects the fact that a numerical
approximation to an integral in the form of a discrete sum will
not converge to the true answer unless the number of points where
the integrand is sampled is made larger.

An increase in the number of neighbours is also warranted to reduce
the gradient errors in SPH, which give in fact rise to a sizable
``zero-th order error'' \citep{Read2010}. These gradient errors
seriously degrade the accuracy of SPH in subsonic flow problems
\citep{2010ARA&A..48..391S} and are a primary cause for problems in
the representation of subsonic turbulence \citep{Bauer2011}.  However,
in practice, simply increasing the number of neighbours is met by a
serious obstacle, because it invokes the so-called clumping
instability in which particles located in the inner parts of the
kernel are pushed together by the pressure forces of the surrounding
particles, forming little clumps that frustrate attempts to reduce the
error of the kernel sums in this way. Partially circumventing this
problem and allowing a higher neighbour number requires different
kernel shapes than are commonly employed \citep{Read2010, Price2012},
but such kernels merely have different stability bands that still
impose severe restrictions on the number of neighbours that may be
used. In essence, due to the clumping instability, establishing
formally convergent SPH solutions is an unsolved problem because the
path of increasing the neighbour number is blocked in practice.

\subsection{Implications for SPH convergence rates and efficiency}

The requirement that the number of neighbours should increase to
reduce errors as the total number of particles is made larger has
also serious implications for the efficiency of SPH.  As an
example, consider a uniform fluid in a cube of side-length $L$
represented by $N$ SPH particles.  (For more general circumstances the
argument generalises if we think about small scales on which the fluid
properties are nearly uniform.)  The mean separation between the
particles, $r_0$, is
\begin{equation}
r_0 \, = \, L \, \left ( \frac{3}{4\pi}\right ) ^{1/3} \, N ^{-1/3} \, .
\end{equation}
If we take ${h \propto r_0}$, as has been done conventionally in SPH
codes used in cosmology, then the smoothing length will shrink in
proportion to $r_0$ and, so, the number of neighbours will be nearly
constant.  In a sense, this maximises the spatial resolution, but
leaves a fixed source of error in the discrete sums
\citep[][]{Read2010,2011arXiv1101.2240R}, so this path is in principle
not appropriate for achieving proper convergence with SPH.  

If, instead, we set ${h \propto r_0 ^\beta}$ then the fundamental
relation between $N$ and $h$ becomes ${N \propto h^{-3/\beta} \,}$.
What are the conditions on $\beta$ so that we can meaningfully
construct a procedure that converges numerically?  Clearly, ${\beta <
  1}$, otherwise $h$ will not decrease more slowly than $r_0$ as $N$
increases.  Also, we must have $\beta > 0$, as the choice ${\beta =
  0}$ would mean that the spatial resolution is constant, independent
of $N$.  The optimal value for $\beta$ is undetermined.  One 
physically appealing
choice would be ${\beta = 1/2}$, because then $h$ would be the
geometric mean between the mean interparticle separation, $r_0$, and
the size of the system, $L$.  In that case, the relation between $N$
and $h$ becomes ${N \propto h^{-6} \,}$, implying a steeply rising
computational cost with resolution. It is possible that a slightly
larger value of $\beta$ could be preferred in terms of CPU costs, but
it is clear, however, that in order to extend the dynamic range in
spatial scales resolved by SPH in a way that guarantees eventual
convergence, it is absolutely necessary to increase the number of
particles more aggressively than has been advocated previously. The
dilemma, of course, is that the clumping instability makes this highly
problematic.

Ignoring the clumping issue for the moment, the implications for the
convergence rate of SPH are in any case important. When the common
practice in the field is followed and the number of neighbours is held
constant, \citet{2010ARA&A..48..391S} reported a global L1 error for a
vortex flow (the Gresho test) that scales as $h^{0.7}$ with
resolution, as opposed to ${\rm L1}\propto h^{1.4}$ for {\small
  AREPO}. We can then ask how the ratio of the CPU-time cost of
simulations with the two schemes varies with the size of the
error. For simulations in 3D, the computational cost scales for both
codes as $h^{-4}$, where three powers of $h$ come from the spatial
dimensions, and one additional power enters due to the reduction of
timesteps for better resolution. This then implies that the cost ratio
scales as ${\rm CPU}_{\rm GADGET} / {\rm CPU}_{\rm AREPO} \propto {\rm
  L1}^{-2.86}$ with the error of the calculation.  Reducing the error
by a factor of 10 requires therefore about a $\sim 700$ times higher
effort in SPH than in {\small AREPO}. This conclusion is problem
dependent and can also be affected by specific details of the SPH
algorithm, like the kernel shape. However, the question of
computational efficiency of different numerical schemes is best
phrased in terms of such convergence rate comparisons. Here the Monte
Carlo character in the approximation of the SPH kernel sum invariably
introduces a serious disadvantage for SPH compared with the moving
mesh approach, as outlined above.

A related conclusion, but focusing on the efficiency impact of the
artificial viscosity parameterisation in SPH, was reached by
\citet{Bauer2011}. If the goal is a representation of subsonic
turbulence with a certain Reynolds number ${\cal R}_e$, they showed that
the computational cost of SPH scales at least as ${\rm CPU} \propto
{\cal R}_e^4$. In contrast, in the mesh-based treatment of {\sm
  AREPO}, the dissipation scale is tied to the mesh resolution,
implying a computational cost for reaching a certain Reynolds number
that scales as $\propto {\cal R}_e^3$. Given that our cosmological
simulations with {\sm AREPO} are only about $30\%$ slower than
{\sm GADGET} for simulations involving self-gravity, for the same
number of resolution elements, this means that at a given
computational cost, {\sm AREPO} resolves much larger Reynolds numbers
than SPH. In fact, to reach the same Reynolds numbers as in our
present moving-mesh simulations, it appears plausible that a factor ${\sim
  100 - 1000}$ more SPH particles than {\sm AREPO} cells would be
required.

\subsection{The pseudo-Lagrangian nature of SPH}

\begin{figure}
\centering
  \includegraphics[width=0.475\textwidth]{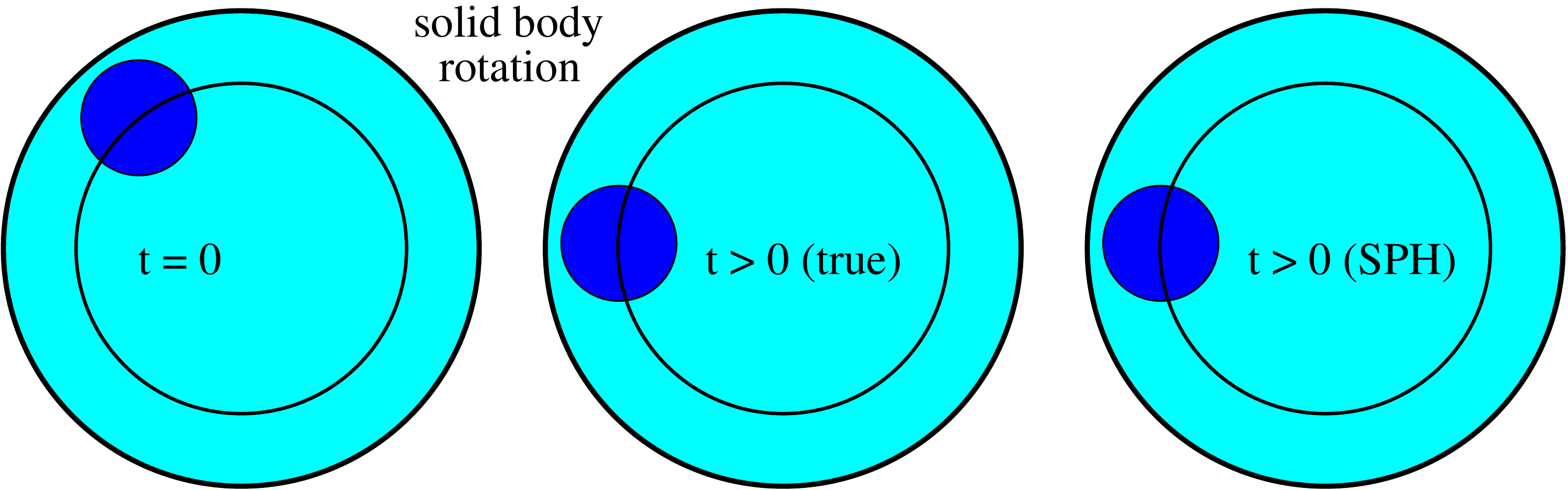}
  \vskip0.25cm
  \includegraphics[width=0.475\textwidth]{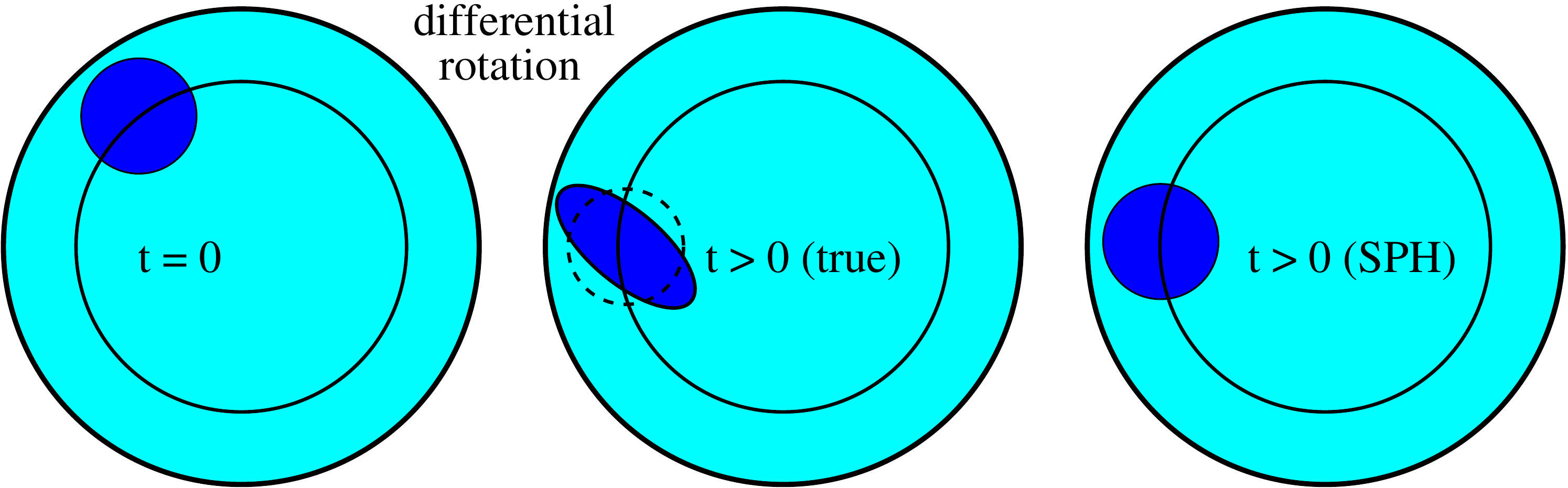}
  \caption{Schematic representation of a thin disk with gas evolving
    on circular orbits. The dark blue region shows the gas that is
    within the smoothing region of one SPH particle in this disk. The
    upper panels show the situation for solid body rotation, whereas
    bottom panels depict the situation for differential rotation. The
    leftmost panels give the initial setup at time ${t=0}$. The middle
    panels show the true gas distribution of the disk a short amount
    of time later. For solid body rotation, the initial blue region
    retains its shape. This is different for the differentially
    rotating disk, where the blue gas patch experiences the shear in
    the disk and gets distorted. The panels on the right-hand side
    show the corresponding SPH representations of the disk at this
    later time. Clearly, the distortion in the differentially rotating
    case is not captured correctly by SPH since the smoothing region
    does not change in shape. SPH is therefore only pseudo-Lagrangian,
    as opposed to {\sm AREPO}, which implements a quasi-Lagrangian
    scheme that allows for mass exchange between cells in a manner
    consistent with the hydrodynamic equations of motion.  The dashed 
    circle in the bottom middle panel illustrates a re-partitioning of
    the gas into SPH particles at the updated time and demonstrates
    why mixing is suppressed in SPH at the resolution scale (see text).}
\label{fig:SPH_scheme}
\end{figure}

SPH is often referred to as a Lagrangian algorithm, but this is not
formally correct and we suggest it is perhaps better to characterise
it as a ``pseudo-Lagrangian'' technique.  To appreciate the reasons
for this, consider the following example, which also highlights that
this distinction is intimately related to issues of (suppressed)
mixing in SPH.

Suppose gas is revolving in a thin disk on circular orbits, and
imagine that the gas is partitioned at some instant into regions that
are then represented by SPH particles, with the smoothing done over a
circular area.  Each SPH particle will comprise gas from an area
around its centre, extending outwards in radius of order the local
smoothing length, as shown for one such SPH particle in the left
panels of Fig.~\ref{fig:SPH_scheme} at time zero.  Consider now how
the system will look like a short time later, ${t>0}$, after the disk
has rotated by a small angle.

If the disk is in solid body rotation, as in the upper panels of
Fig.~\ref{fig:SPH_scheme}, the fluid initially contributing to the SPH
particle will occupy an area identical to the smoothing region of the
SPH particle at time ${t>0}$, as indicated by comparing the ``true''
situation in the upper middle panel with the SPH version in the upper
right panel.  In this case, the SPH representation is faithful to the
actual equations of motion because the fluid contributing to the SPH
particle at time zero is the same as that at $\Delta t$, and the region
bounded initially by the SPH smoothing volume has not changed its
shape.

However, suppose instead that the disk is rotating differentially,
turning around more rapidly in the inner parts than in the outer ones
(lower panels of Fig.~\ref{fig:SPH_scheme}).  The ``true'' situation
shown in the middle panel tells us that the gas contributing to the
area bounded by the SPH smoothing volume at time zero will be
stretched out owing to shear.  However, in the SPH formulation, shown
in the lower right panel, the material initially in the SPH smoothing
volume is forced to remain tied to this area and is not allowed to
shear, inconsistent with the equations of motion of the real fluid.

A true Lagrangian picture would allow fluid elements to be deformed in
the presence of shear, but this is inhibited by the SPH approach on
scales smaller than those set by the smoothing procedure.  Hence,
while SPH retains some characteristics of a Lagrangian method, it does
not evolve the fluid entirely in a manner consistent with the
equations of motion and should, in this sense, be termed
``pseudo-Lagrangian.''

Unfortunately, the error incurred by the approximations inherent to
the SPH formalism is highly problem dependent and cannot be estimated
based solely on the discretisation procedure.  For example, there is
no error made in the case of a disk in solid body rotation, or for a
uniform disk rotating differentially, or for gas motions that
are limited to expansion or contraction in three-dimensions.  However,
this is not true for shearing flows involving fluids with distinct
internal properties; here the implicit ``remapping'' involved in
enforcing a spherical shape for the SPH smoothing volume, which would
change the local composition of the internal properties, is ignored.
This limitation is ultimately the reason why SPH does not handle
mixing accurately, as we illustrate in the bottom, middle panel of
Fig.~\ref{fig:SPH_scheme}.  Suppose at the updated time we were to
re-partition the gas into new SPH particles, as indicated by the
dashed circle in this frame.  The gas associated with this particle
should include gas from the original particle, but also gas from the
surrounding flow.  This would allow mixing to occur between the
gas in the original particle and nearby parts of the flow, but
is not allowed to occur in SPH, by construction.

We note that {\sm AREPO} does not suffer from this defect.  In this
code, cells are not allowed to become arbitrarily distorted, in the
interests of efficiency and accuracy, and so {\sm AREPO} is also not
formally Lagrangian.  However, it still evolves fluids correctly when
deviations from strict Lagrangian behaviour occur by allowing for mass
exchange between cells, in a manner consistent with the equations of
motion.  Therefore, this method should be termed ``quasi-Lagrangian.''


A related issue arises in the context of N-body simulations of
collisionless systems.  There, a six-dimensional phase fluid is
partitioned into particles of fixed size that move through space in a
manner determined by the equations of motion.  Because the phase fluid
initially associated with a particular particle is always tied to that
particle, the small-scale dynamics of the system will not be
represented properly.  However, in this case, unlike for the example
discussed above, forces are strictly long-range and are less prone to
inaccuracies in the small-scale distribution of the material than
pressure gradients or viscosity.  In that sense, the local distortions
in the phase space fluid have fewer dynamical consequences than for a
hydrodynamical fluid in three dimensions.  Nevertheless, if an
accurate description of the small-scale structure of collisionless
systems is essential, these effects must be accounted for at some
level \citep[e.g.][]{Vogelsberger2008,Vogelsberger2011,Abel2012}.

\subsection{Discussion}

We should emphasise that whether the limitations inherent to SPH lead
to significant inaccuracies in the solution depend in detail on the
circumstances.  SPH can be expected to provide reliable results for
flows that are kinetically dominated, or if the motions are controlled
mainly by long-range forces.  In these situations, errors in the local
quantities are sub-dominant.  For example, \cite{Bauer2011} show that
while SPH does not accurately describe subsonic turbulence, in the
supersonic regime, when the flow energy is mainly kinetic, {\sm
  GADGET} and {\sm AREPO} give similar answers. Also, comparisons between
SPH and AMR codes have yielded compatible results for the structure of
the intergalactic medium, as reflected in the properties of the
Lyman-alpha forest \citep[e.g.][]{Regan2007}.  At these low densities,
gravity dominates over internal energy and, moreover, the fluid is not
subject to strong shear, so the above considerations are not critical.
Even in some cases where shear is present, the approximations
underlying SPH will not be significant, provided that gravitational
forces are more important than local ones and the relevant
evolution in flow properties occurs rapidly.  Hayward et al. (2012, in
preparation) have demonstrate this explicitly by simulating galaxy
mergers using both {\sm GADGET} and {\sm AREPO}.  They find that when
the gas is treated using an effective equation of state, star
formation rates during the course of a merger are nearly identical
between these two codes.  Thus, conclusions drawn from studies of
galaxy collisions with SPH, such as the stellar profiles of merger
remnants and their evolution \citep[e.g.][]{Hopkins2008,Hopkins2009a,
Hopkins2009b,Hopkins2009c} or the survivability
of disks during mergers \citep{2005ApJ...622L...9S}, are likely robust
to variations in the hydro solver.

Unfortunately, this good level of agreement does not extend to the
more complex flows associated with halos in cosmological environments,
as we highlight in this paper. There, various phases of gas will be
present in close proximity, often shearing relative to one another,
leading to errors in simulations done with SPH like those we have
identified here. Also, in the hydrostatic atmospheres of halos the
hydrodynamic forces are comparable to gravity forces, and the
subsonic turbulence present in these regions is affected by gradient
errors in SPH.  Without a formal convergence criterion, the
consequences of these errors are difficult to assess, because even
if a solution may plateau as the number of SPH particles is increased,
this by no means guarantees that the correct answer is being
obtained.  In particular, the solution may exhibit ``false
convergence,'' appearing to reach a converged solution, while
fixed sources of error from e.g. summation interpolant are
still present \citep{2011arXiv1101.2240R}.
A case in point is the long-standing discrepancy identified
for the central cluster entropy predicted by non-radiative SPH and
mesh-based simulations, first identified in the Santa Barbara
cluster comparison project \citep{1999ApJ...525..554F}.

The above considerations motivate thinking about ways to cure the
defects inherent to SPH so that it can provide solutions of comparable
accuracy and resolution to those obtained with a moving mesh approach
like {\sm AREPO}.  For example, the artificial viscosity typically
used in SPH codes to handle shocks could, in principle, be eliminated
by incorporating a Riemann solver into SPH.  Pioneering efforts along
these lines have been made by \cite{2002JCoPh.179..238I} and
\cite{2011MNRAS.417..136M}.  However, these implementations solve the
Riemann problem for each particle-neighbour pair, and this approach
will become prohibitively expensive as long as the smoothing procedure
requires averaging over an increasingly large number of neighbours to
achieve convergence as the total particle number is increased. We note
that alternatively a grid could be introduced into SPH just for the
purposes of solving the Riemann problem around each particle.  This
grid would necessarily have to be adaptive, and hence resemble the
Voronoi tessellation used in {\sm AREPO}.

In order to eliminate the pseudo-Lagrangian character of SPH and allow
for fluid elements to become distorted on scales small compared to a
few smoothing lengths, as should occur in e.g.~shearing flows, it is
necessary to allow for mass exchange between particles in a manner
consistent with the equations of motion.  This would have the added
benefit of eliminating the mixing problem in SPH. First suggestions in
this direction have recently been made \citep{2008MNRAS.387..427W,
  2008JCoPh.22710040P, Read2010, Read2011}, but the best way to
formulate such mixing terms is not clear.  One possibility to
unambiguously calculate the required mass flux between particles would
be to utilise some kind of grid.  Again, this grid would need to be
adaptive, like the unstructured mesh in {\sm AREPO}, reinforcing the
notion that our new moving-mesh approach quite naturally addresses
several of the generic issues with SPH.

\section{Conclusions}

In this study, we have presented the first cosmological hydrodynamical
simulations of structure formation carried out with the novel {\sm AREPO}
moving-mesh code. Compared to the widely employed SPH technique, this
method promises an important gain in accuracy in the numerical
treatment of gas dynamics for similar computational costs.
In order to understand the implications of
these differences for galaxy formation simulations, we have evolved
the same initial conditions both with {\sm AREPO} and the
well-established SPH code {\sm GADGET}. Both codes share the same
high-resolution gravity solver, and incorporate an identical radiative
cooling and star formation model. Differences in the outcome are
hence tied to
systematics of the hydrodynamical solvers that are
used. It is the primary goal of our paper to identify the main
differences and their magnitude, as well as providing evidence
for the primary sources of potential discrepancies.

\begin{figure}
\centering
  \includegraphics[width=0.475\textwidth]{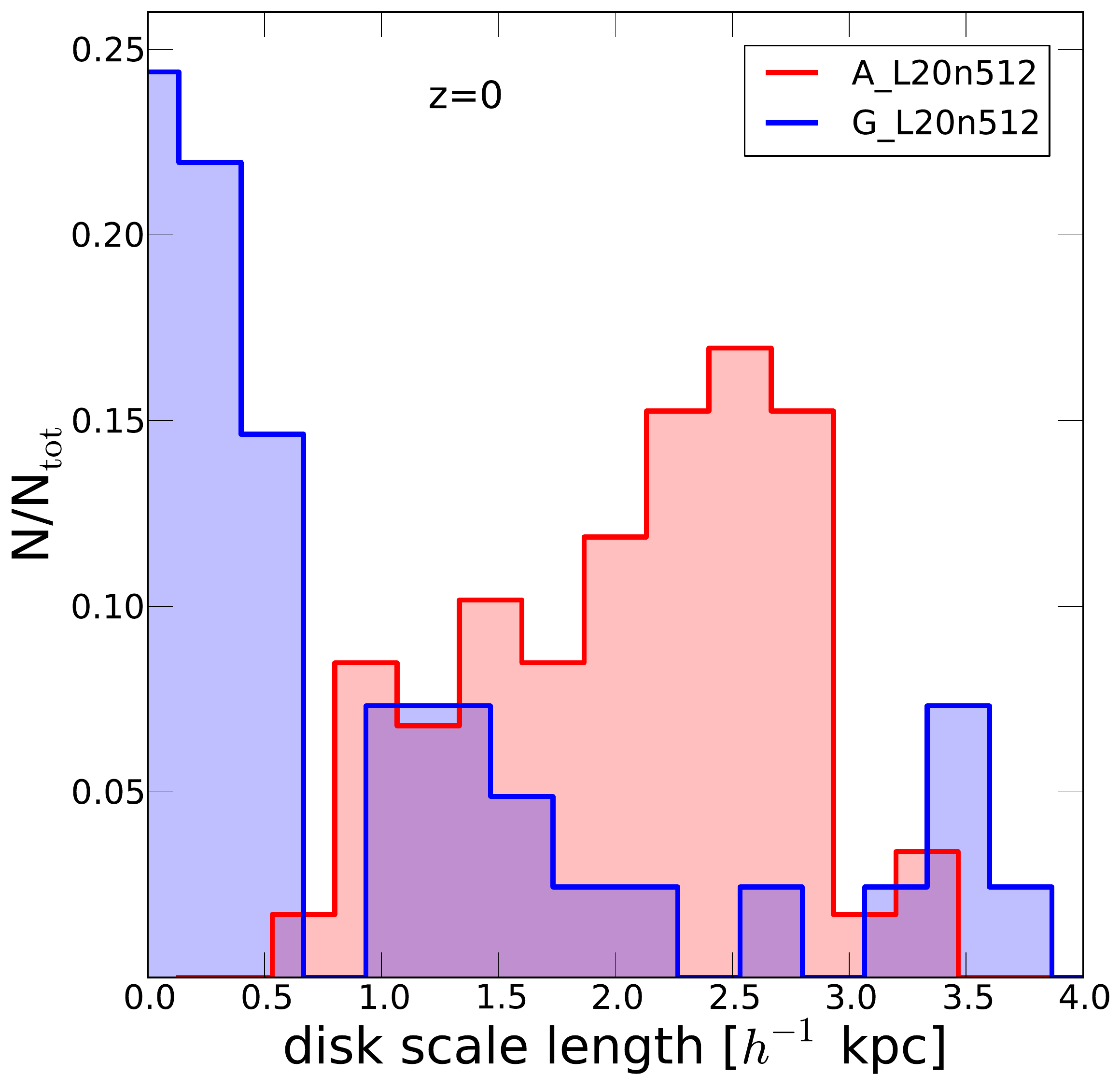}
\caption{Gas disk scale lengths
  obtained from exponential surface density fits of more than $50$ galaxies
  at $z=0$ in the halo mass range $\sim 4\times10^{11}\,h^{-1}\,{\rm M}_\odot$
  to $\sim 8\times10^{12}\,h^{-1}\,{\rm M}_\odot$. Clearly, 
  the gas disks forming in the {\sm AREPO} simulations
  are typically significantly larger than those in the corresponding SPH runs.
  }
\label{fig:scale_radii}
\end{figure}

To achieve this goal, we have considered a primary set of simulations
in boxes of $20\,h^{-1}\,{\rm Mpc}$ that include a basic treatment of
galaxy formation physics, at various resolutions ranging from
$2\times 128^3$ to $2\times 512^3$ particles/cells. In addition, we
have carried out a few auxiliary simulations either in smaller boxes,
or of non-radiative type, in order to more clearly measure particular
numerical effects.  For reasons discussed in Section~2.3, we have
chosen to perform the comparisons between the codes at the same
initial mass resolution, because then the computational costs
are similar.

Our principle findings may be summarised as follows:

\begin{itemize}

\item The overall distribution of gas in density and temperature is
  broadly in agreement between SPH and {\sm AREPO}, but there are some
  subtle and important differences. In particular, the mean
  mass-weighted temperature in {\sm AREPO} is slightly smaller than in
  {\sm GADGET}. Also, the hot gas extends to somewhat lower density in
  {\sm GADGET} and hence occupies a larger volume fraction.

\item The cosmic star formation rate density peaks at similar values,
  but at slightly lower redshifts in {\sm AREPO}. At high redshift and
  at high resolution, the SFRs between the two codes are in good
  agreement, but towards lower redshift there are significantly
  stronger cooling flows in {\sm AREPO}, causing a correspondingly
  larger star formation rate. This difference occurs primarily in
  haloes with mass larger than~$\sim 10^{11}\,h^{-1}\,{\rm M}_\odot$.

\item We find that the two codes differ significantly in the
  dissipative heating rates within haloes, with {\sm AREPO} producing
  more dissipation in the halo infall regions, whereas SPH produces
  higher dissipative heating throughout most of the outer regions of
  virialised haloes. This difference is mainly responsible for the higher
  temperatures found in the SPH simulations, and the correspondingly
  weaker cooling. We argue that this dissipation in SPH is likely to
  be of spurious nature, and is a combination of viscous damping of
  SPH's inherent noise and the unphysical damping of subsonic
  turbulence injected into haloes in the infall regions.

\item Visual comparison of simulated galaxies shows considerably
  larger stellar disks and more extended and disky stellar
  distributions in {\sm AREPO} compared with {\sm GADGET}. Also, the
  halo gas surrounding the galaxies looks smoother and less clumpy in
  {\sm AREPO}. Fig.~\ref{fig:scale_radii} shows gas surface density
  disk scale length radii obtained from exponential surface density
  fits of more than $50$ galaxies at $z=0$ in the halo mass range
  $\sim 4\times10^{11}\,h^{-1}\,{\rm M}_\odot$ to $\sim
  8\times10^{12}\,h^{-1}\,{\rm M}_\odot$. This clearly demonstrates
  that the gas disks forming in the {\sm AREPO} simulations are
  typically significantly larger than those in the corresponding SPH
  runs.  We further note that the gas surface density of disk galaxies
  in the {\sm AREPO} simulations follow very closely exponential
  profiles, which is typically not the case for the SPH runs. A more
  detailed analysis is presented in \cite{Torrey2011}. We note that
  the large disks in the {\sm AREPO} simulations result even without
  strong feedback in the form of galactic winds or modifications of
  the star formation prescription. This is also demonstrated in
  simplified test simulations in Paper III. The larger extent of
    disk galaxies is also demonstrated in Paper II, where we do not
    assume exponential profiles, but rather use a cut in surface
    density to determine a characteristic size. We also independently
    calculated half-mass radii, which show the same trends. All these
    results confirm the findings reported in
    Fig.~\ref{fig:scale_radii}, i.e.~that {\sm AREPO} produces in
    general larger disks than {\sm GAGDET}.
\end{itemize}

The above findings clearly suggest that there are significant
quantitative differences caused in galaxy formation simulations by
the choice of hydrodynamical technique. It appears that the limited
accuracy of SPH for subsonic flow phenomena such as Kelvin-Helmholtz
instabilities or subsonic turbulence induces important differences in
the predicted properties of galaxies at low redshift, affecting both
their morphology and stellar mass. Because of the accuracy problems of
SPH for certain subsonic test problems \citep[][Bauer \& Springel
  2011]{2007MNRAS.380..963A,2010ARA&A..48..391S} it appears that
SPH cannot be expected to reach highly accurate results in all regimes
relevant for cosmic structure formation. At the same time, {\sm AREPO}
performs significantly better on these same problems, and yields
generally a smaller error norm and higher convergence rate when
scrutinised against problems with known analytic solutions
\citep{2010MNRAS.401..791S,2010ARA&A..48..391S}. We are hence
confident that the {\sm AREPO} results presented in this paper entail
a more faithful treatment of cosmological hydrodynamics,
especially in view of the discussion in Section~5.3.

It is also reassuring that the {\sm AREPO} runs improve the predicted
morphologies of simulated galaxies, as our preliminary analysis
suggests. In Paper II of this series we will back up this finding with
a detailed analysis of the galaxy properties. Finally, in Paper III we
provide a more detailed analysis of the relevant numerical effects
responsible for the differences in mixing and cooling. The body of
these results makes it clear that {\sm AREPO} simulations provide a
significant opportunity for the development of next generation models
of galaxy formation that promise to achieve a so-far unknown
combination of accuracy, dynamic range, and faithfulness to the
relevant physics.

\section*{Acknowledgements}

The simulations in this paper were run on the Odyssey cluster supported by the 
FAS Science Division Research Computing Group at Harvard University.
Scaling tests were done on the Ranger
cluster at the Texas Advanced Computing Center (TACC). We thank Paul
Torrey for allowing us to show Fig.~\ref{fig:scale_radii} and 
Elena D'Onghia, Patrik Jonsson, Andrew MacFadyen, Diego J. Munoz 
for useful discussions.  We also thank the anonymous referee for helpful comments.
DS acknowledges NASA Hubble Fellowship through grant HST-HF-51282.01-A.
DK acknowledges support from NASA through Hubble Fellowship grant HSTHF-51276.01-A.

\appendix 

\section{Voronoi mesh statistics}

\begin{figure}
\centering
  \includegraphics[width=0.48\textwidth]{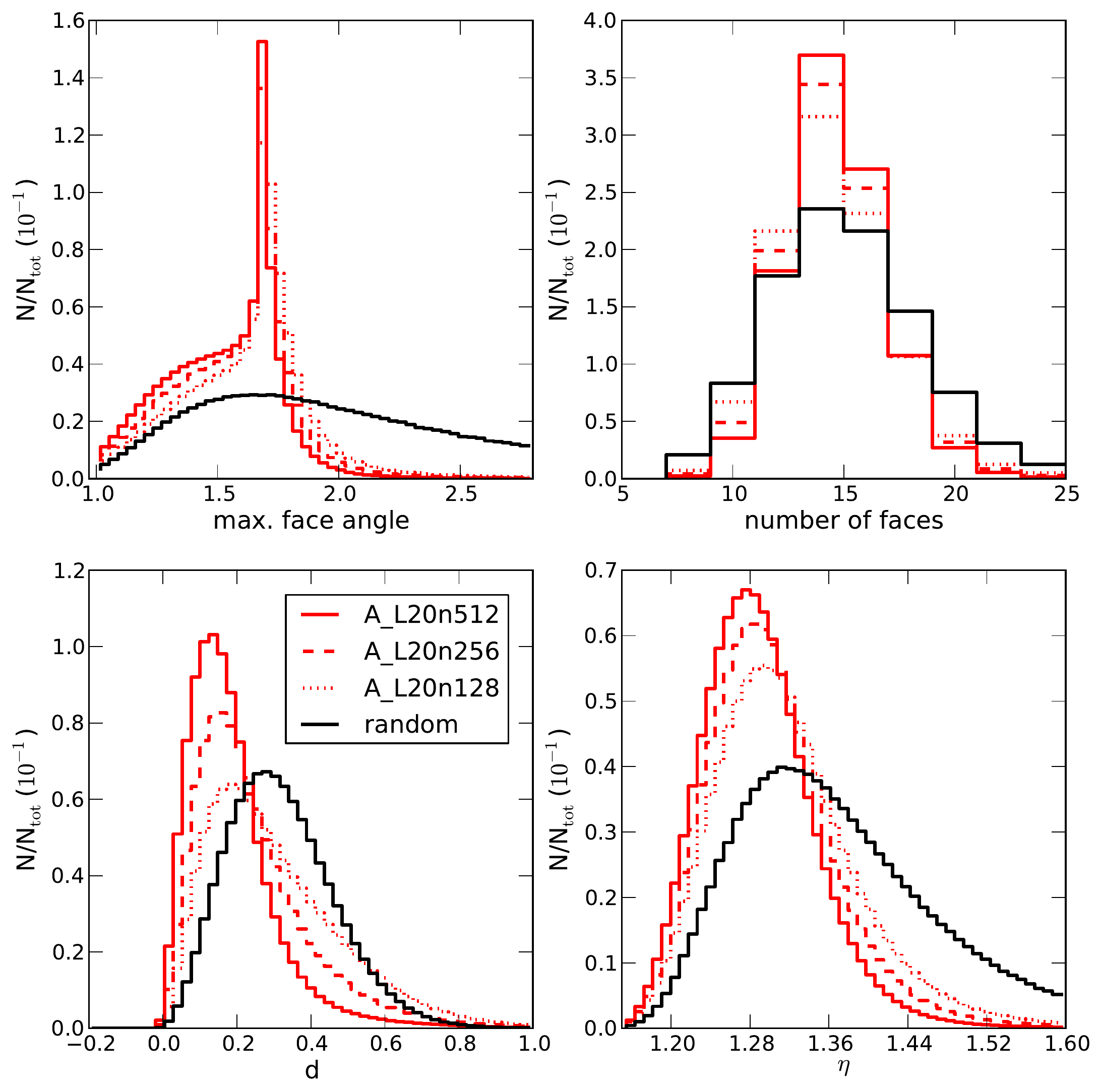}
\caption{Properties of the Voronoi mesh at $z=0$ for the different
  {\sm AREPO} runs. The black line shows the characteristics for a
  mesh where $128^3$ vertex points are randomly distributed. Our
  implemented regularisation scheme tries to keep the maximum face
  angle small to avoid very distorted Voronoi cells. This can clearly
  be seen in the upper left panel, where the random distribution
  deviates significantly from the regularised simulation
  distributions. The regularisation also affects the offset between
  geometrical centre of the cell and the vertex location shown in the
  bottom left panel. It is important to avoid too large offsets here
  since they introduce inaccuracies in the linear reconstruction step
  of the MUSCL-Hancock scheme.  The ``roundness'' of the cells is
  measured by $\eta$, which is closer to $1$ (the value of a sphere)
  for the regularised simulations. Again, this is a measure of the
  regularity of the Voronoi mesh.  For each face of a Voronoi cell, {\sm
    AREPO} needs to solve a Riemann problem. Since each face is only
  solved once, the average number of Riemann problems per cell is
  given by half the average number of its faces. The distribution of
  Voronoi faces per cell is shown in the top right panel.}
\label{fig:cell_shape_statistics}
\end{figure}

\begin{figure}
\centering
  \includegraphics[width=0.45\textwidth]{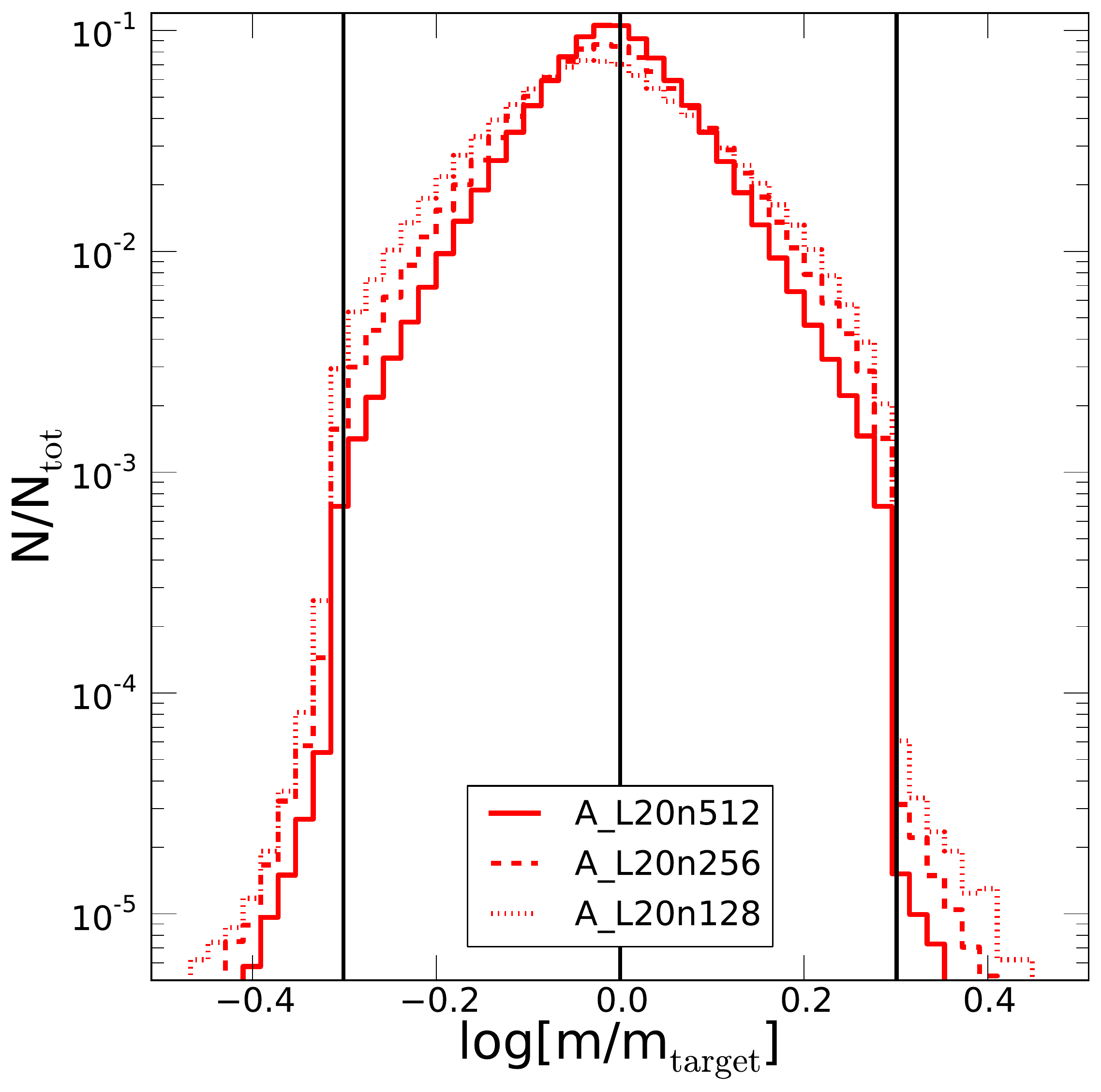}
  \includegraphics[width=0.45\textwidth]{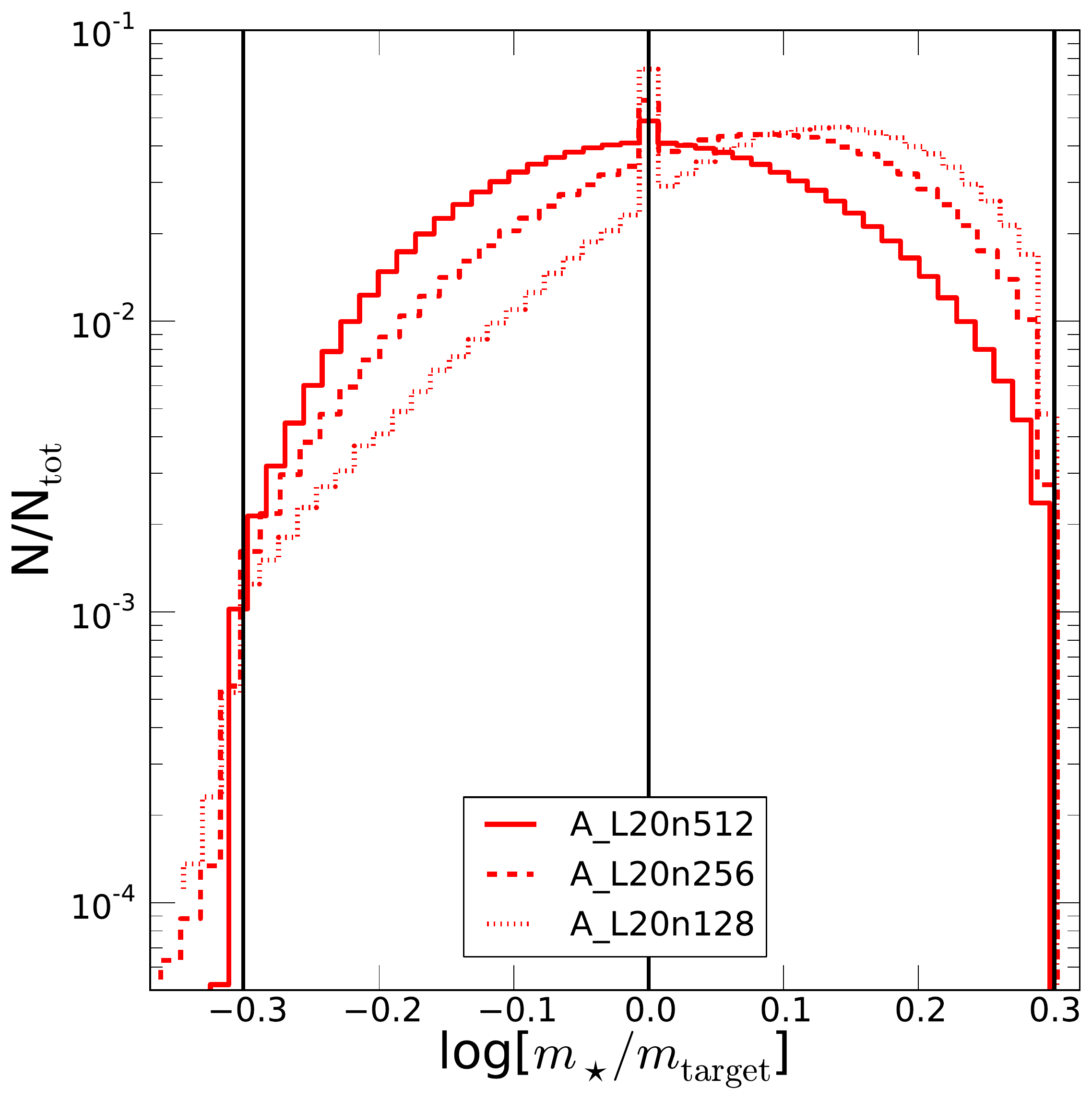}
\caption{Top panel: Distribution of cell masses divided by the target
  gas mass for each {\sm AREPO} simulation at $z=0$. Our
  re-/derefinement scheme guarantees that the mass of a Voronoi cell
  never deviates by more than approximately 
  a factor of 2 from the target gas mass,
  which is in our case chosen to be the mean baryon mass of a cell for
  a uniform distribution assuming all cells have equal volume. Bottom 
  panel: Distribution of masses of the stellar particles divided by
  the imposed target gas mass for the Voronoi cells.  Our constraints
  for the mass of cells and our scheme for creating star particles
  result in a quite narrow mass distribution of stellar particles,
  where each stellar particle  deviates at most a factor of $\sim
  2$ from the desired target gas mass, except for a tiny number of
  lighter star particles, whose mass is however always larger than a
  quarter of the target gas mass.  }
\label{fig:cell_mass_statistics}
\end{figure}

Here we present some basic geometric properties of the Voronoi mesh in 
our cosmological simulations.  As discussed in Section~2, it is
beneficial for the accuracy of {\sm AREPO} for the Voronoi mesh to remain
as regular as possible throughout the simulation. To this end, the
motion of the mesh vertices is slightly modified for highly distorted
cells in order to obtain a mesh that has more ``roundish'' cells and is
closer to a centroidal Voronoi tessellation where geometrical the centres
of the cells coincide with the vertex points of the individual Voronoi
cells.

We can quantify the quality of a mesh in a similar way as was done
in S10. There, mesh quality indicators were calculated for a
non-radiative simulation of a massive cluster \citep[the ``Santa
  Barbara Cluster'',][]{1999ApJ...525..554F}, and it was demonstrated
that the mesh indeed preserved its desired properties of reasonably
round cells during the simulation.  However, in this paper we have
used a modified regularisation scheme based on the maximum face-angle,
and we include cooling, star formation, and feedback, which
drastically increases the dynamic range the simulations need to
address. It is therefore important to verify whether our
regularisation scheme performs sufficiently well in these more
demanding simulations.

In Fig.~\ref{fig:cell_shape_statistics}, we first show in the top
left panel the maximum face angle distribution at $z=0$ for all {\sm
  AREPO} simulations. Our new regularisation scheme should guarantee
that these angles do not get too large. We have set up the
regularisation such that the mesh is steered towards face angles
smaller than $1.68$.  As the upper left panel of
Fig.~\ref{fig:cell_shape_statistics} demonstrates, the mesh indeed
avoids very large face angles with a maximum in the distribution
around the desired value, i.e.~the regularisation works as expected.
The remaining three panels of Fig.~\ref{fig:cell_shape_statistics}
show distribution functions for the number of faces of the Voronoi
cells, for the distance $d$ of mesh-generating points to the geometric
centres of their cell in units of the fiducial cell radius $R=(3 V/ 4
\pi)^{1/3}$ of each cell, and finally, for the $\eta=S^{3/2}/(6
\sqrt{\pi} V)$ parameter which measures how ``roundish'' each cell is
(here $S$ is the total surface area of a cell). The corresponding
distributions for a random Poisson sample of $128^3$ vertex points
are shown as black lines in all panels, for comparison. Note that we
do not explicitly control the offset $d$ in our regularisation scheme,
but the face angle criterion is of course correlated with this
quantity. This allows us to obtain also quite small values for the
distance of the geometric centre of the cell from the location of the
mesh generating points. This is important for the numerical accuracy
of the linear reconstruction.  Overall, our Voronoi mesh is
significantly ``rounder'' and much closer to a centroidal Voronoi
tessellation than the mesh of a random point distribution.

The goal of our re- and derefinement strategy is to keep all cell
masses close to a predefined target cell mass, which we have chosen to
be the total baryonic mass in the box divided by the total number of
cells at the initial time. In this way, a narrow mass spectrum for the
formed stellar particles is obtained, and a straightforward and 
even-handed
comparison to the SPH calculations done with {\sm GADGET} becomes
possible.  As we have described in Section~2, we guarantee the
approximately constant mass resolution by splitting cells that reach a
mass larger than $2 \times m_{\rm target}$ into two cells, while cells
dropping in mass below $0.5 \times m_{\rm target}$ are dissolved. We
note that thanks to the Lagrangian mesh motion in {\sm AREPO}, these
re-/derefinement operations are invoked only rarely.  In the
top panel of Fig.~\ref{fig:cell_mass_statistics} we demonstrate
that the re- and derefinement scheme successfully keeps the cell
masses in the desired mass range, yielding approximately a
$\log$-normal distribution around the target mass within the desired
bounds.

The mass distribution of stellar particles at $z=0$ is shown in the
bottom panel of Fig.~\ref{fig:cell_mass_statistics}. Since our star
formation implementation does not allow star particles to be formed
with a mass larger than $2\times m_{\rm target}$, the stellar particle
mass distribution is cut off at that value. On the other hand, the
derefinement operations are constrained by the requirement that
neighbouring cells cannot both be derefined in the same time step. As
a result, it is possible that some cells can temporarily have masses
below $1/2 \times m_{\rm target}$ for a few time steps, and hence also
some star particles with masses smaller than this value may form in
principle.  To protect against the formation of unreasonably low mass
star particles, which may be prone to two-body effects, we however do
not allow cells with mass less than $1/4 \times m_{\rm target}$
to form any star particles, imposing a lower limit in the star
particle mass distribution. As the bottom panel of
Fig.~\ref{fig:cell_mass_statistics} shows, star particles with masses
between $1/4 \times m_{\rm target}$ and $1/2 \times m_{\rm target}$
make up only a tiny fraction, as desired, such that the suppression of
the formation of extremely small star particles does not lead to any
appreciable error.  Note that star-forming cells with a mass above $2
\times m_{\rm target}$ will form stars with exactly $m_{\rm target}$,
keeping the rest of the gas mass in the cell. This explains the small
spike in the distribution at that mass value. Our analysis thus
confirms that star particles with a narrow range of masses are formed,
which helps to limit two-body relaxation effects and is well matched
to the fixed gravitational softening we use for collisionless
particles.

\section{Gravitational softening and regularisation scheme}

\begin{figure}
\centering
  \includegraphics[width=0.45\textwidth]{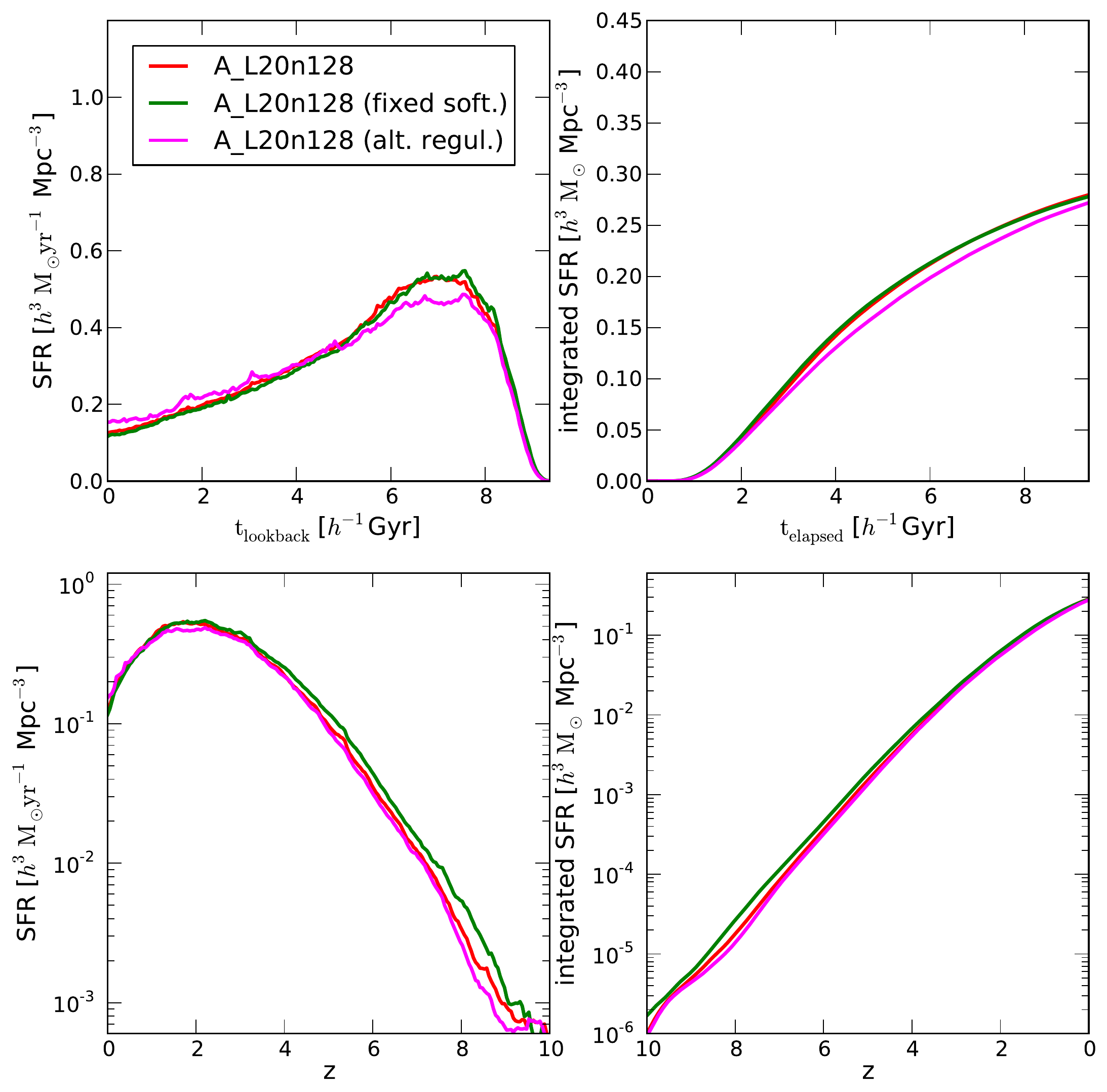}
\caption{Star formation rates per unit volume as a function of
  lookback time (top left) and redshift (bottom left). The panels on
  the right show the corresponding integrated star formation rates as
  a function of lookback time (top right) and redshift (bottom
  right). We show the result of the A\_L20n128 simulation along with
  a the same run with a fixed (comoving) gravitational softening length
  and a run with a different mesh regularisation scheme.
  All runs show very similar star formation rates.}
\label{fig:sfr_hist_appendix}
\end{figure}

\begin{figure}
\centering
  \includegraphics[width=0.22\textwidth]{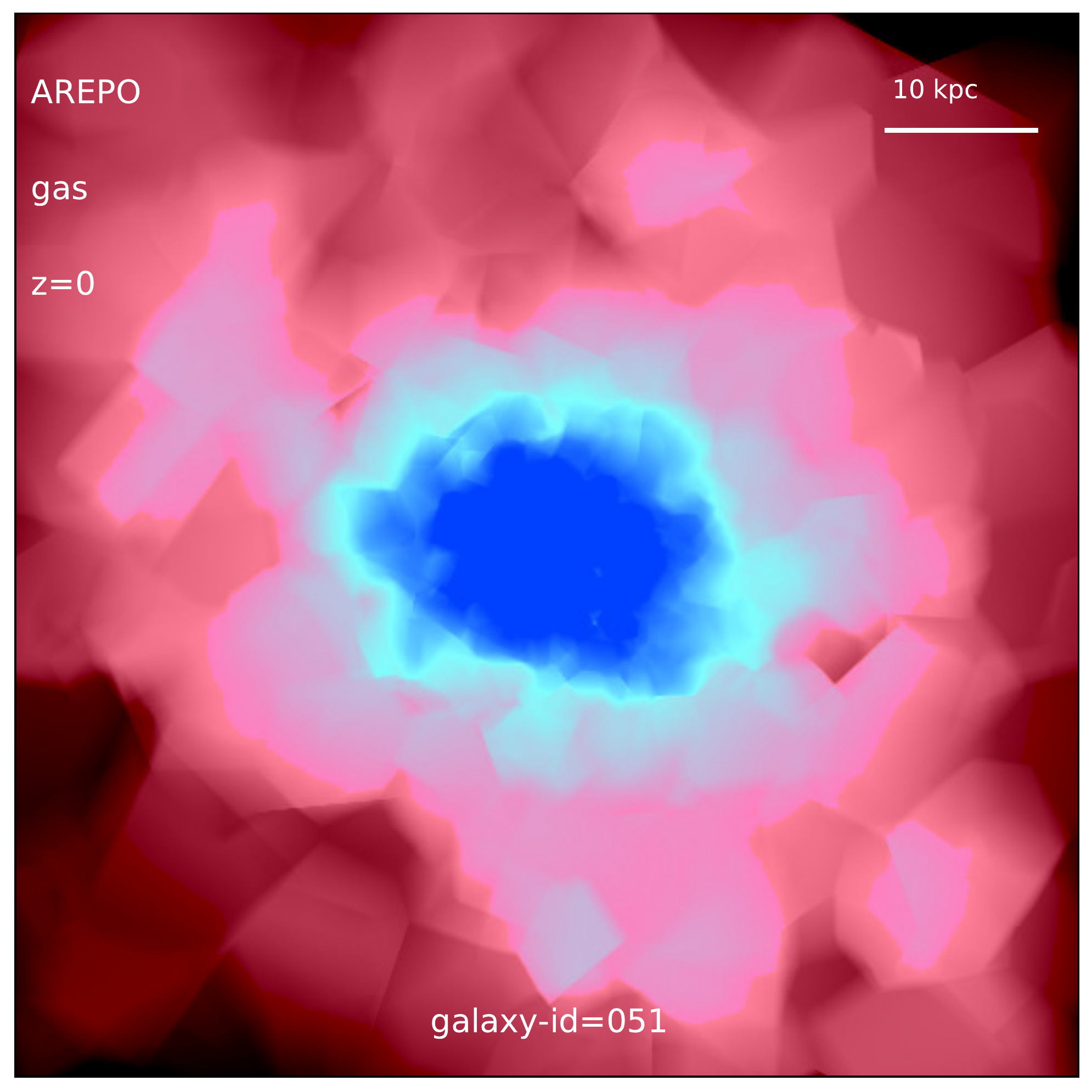}
  \includegraphics[width=0.22\textwidth]{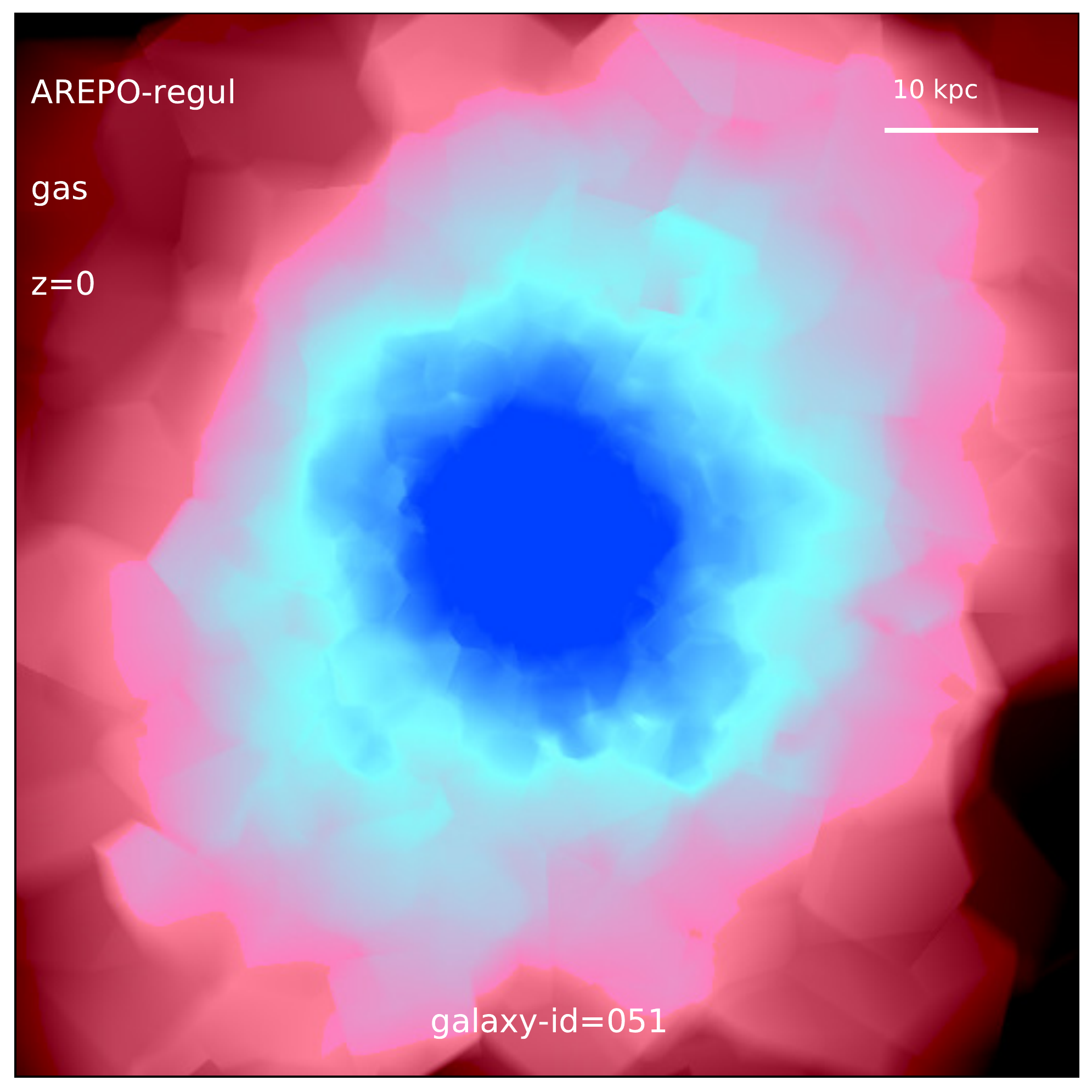}
\caption{Gas density maps of a matched galaxy of the simulation A\_L20n128
  and a similar simulation with a different mesh regularisation scheme (based on the displacement
  of the geometric centre of the cell). As expected,
  the detailed mesh geometry and the shape of cells change if we employ a different
  regularisation scheme. But the resulting density field looks similar, and especially
  the overall extent of the disk is not significantly affected by the way the mesh
  is regularised. We note that L20n128 is our lowest resolution and therefore most
  sensitive to details of how  exactly the mesh is treated. These effects are significantly
  smaller for our intermediate resolution and nearly vanish for the highest resolution
  simulation.
  }
\label{fig:gas_map_regul}
\end{figure}

The {\sm AREPO} simulations presented in this paper use an adaptive
gravitational softening length with a lower floor for the gas, whereas
the SPH simulations done with {\sm GADGET} employ a fixed comoving
softening length equal to the lower floor. We checked that this does
not bias our results in any way. To demonstrate this point we show in
Fig.~\ref{fig:sfr_hist_appendix} the star formation history of the
standard A\_L20n128 simulation together with a run where we held the
gravitational softening of the cells fixed at the same value as in the
SPH calculations. Clearly this leads only to very minor differences in
the star formation history. We also checked that disk half mass radii
are not affected by this. As stated above this is due to the fact that
star forming gas is pressurised by an effective EOS of the ISM and
therefore gravitational softening effects do not play an important
role in that regime.

The mesh regularisation in {\sm AREPO} can be done in different ways
and we have used a scheme that is based on the maximum face angle as
described above. We have also done simulations with the original
regularisation scheme presented in \cite{2010MNRAS.401..791S}, which
is based on the displacement of the geometric centre of the cell. The
resulting star formation history for the A\_L20n128 is also shown in
Fig.~\ref{fig:sfr_hist_appendix}. Again this leads only to minor
changes. We note that this difference becomes smaller with resolution
and essentially vanishes. Regularisation scheme differences are only
relevant for very low resolution simulations. Disk sizes are also
  not significantly affected by the details of the regularisation
  scheme.  In Fig.~\ref{fig:gas_map_regul}, we show gas density maps
  of a matched object in A\_L20n128 and the A\_L20n128 run, which
  features an alternative regularisation scheme. We stress again that
  L20n128 is our lowest resolution setup, and therefore most sensitive
  to details of how exactly the mesh is treated. But even in this
  regime the differences in the gas density maps are very small, and
  the size and overall extent of the gas disk do not change. Note that
  due to the low resolution of L20n128, individual Voronoi cells are
  clearly visible in the map, highlighting the changes in mesh
  geometry and cell shapes induced by different regularisation
  schemes.

\section{Code performance}

\begin{figure}
\centering
  \includegraphics[width=0.45\textwidth]{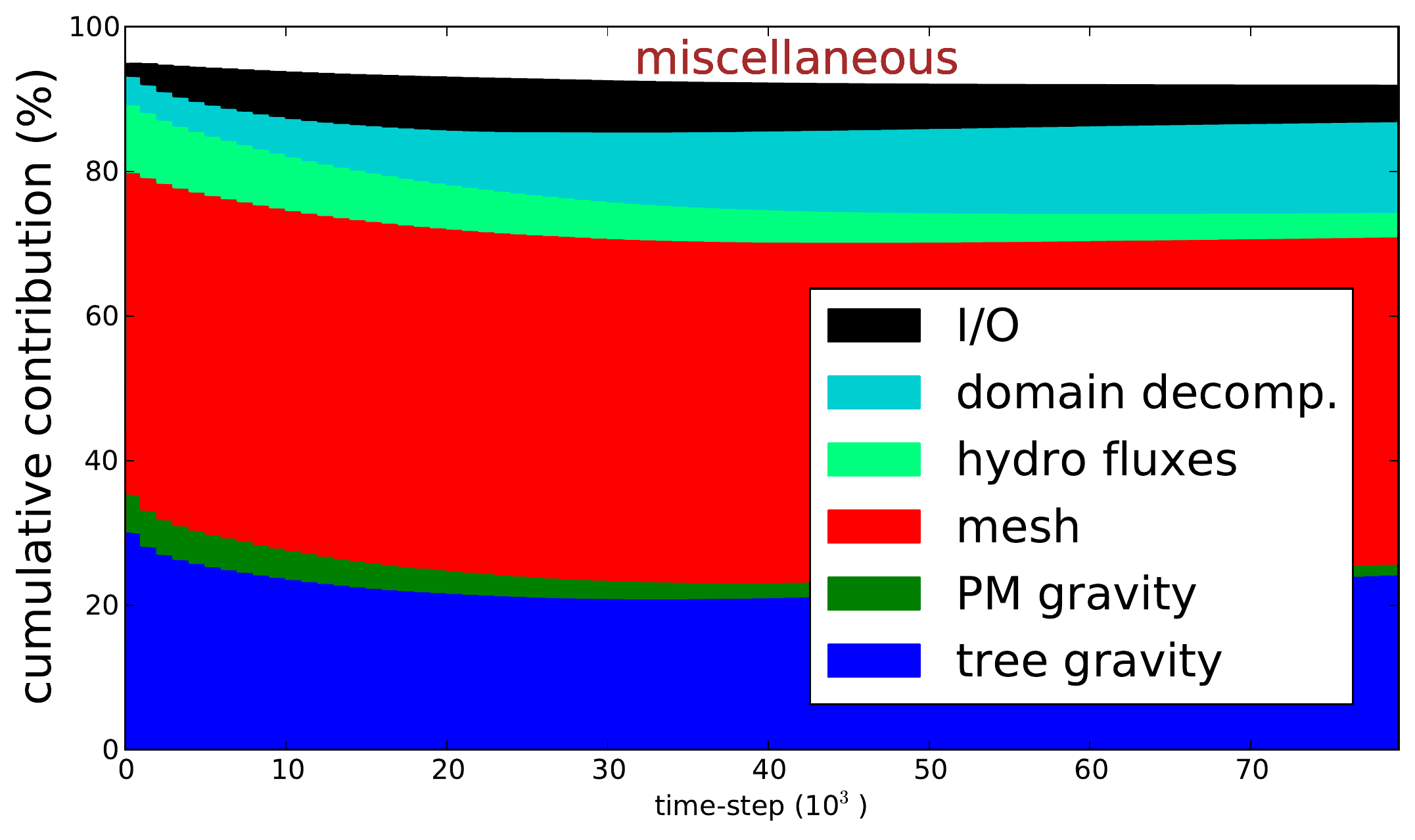}
  \includegraphics[width=0.45\textwidth]{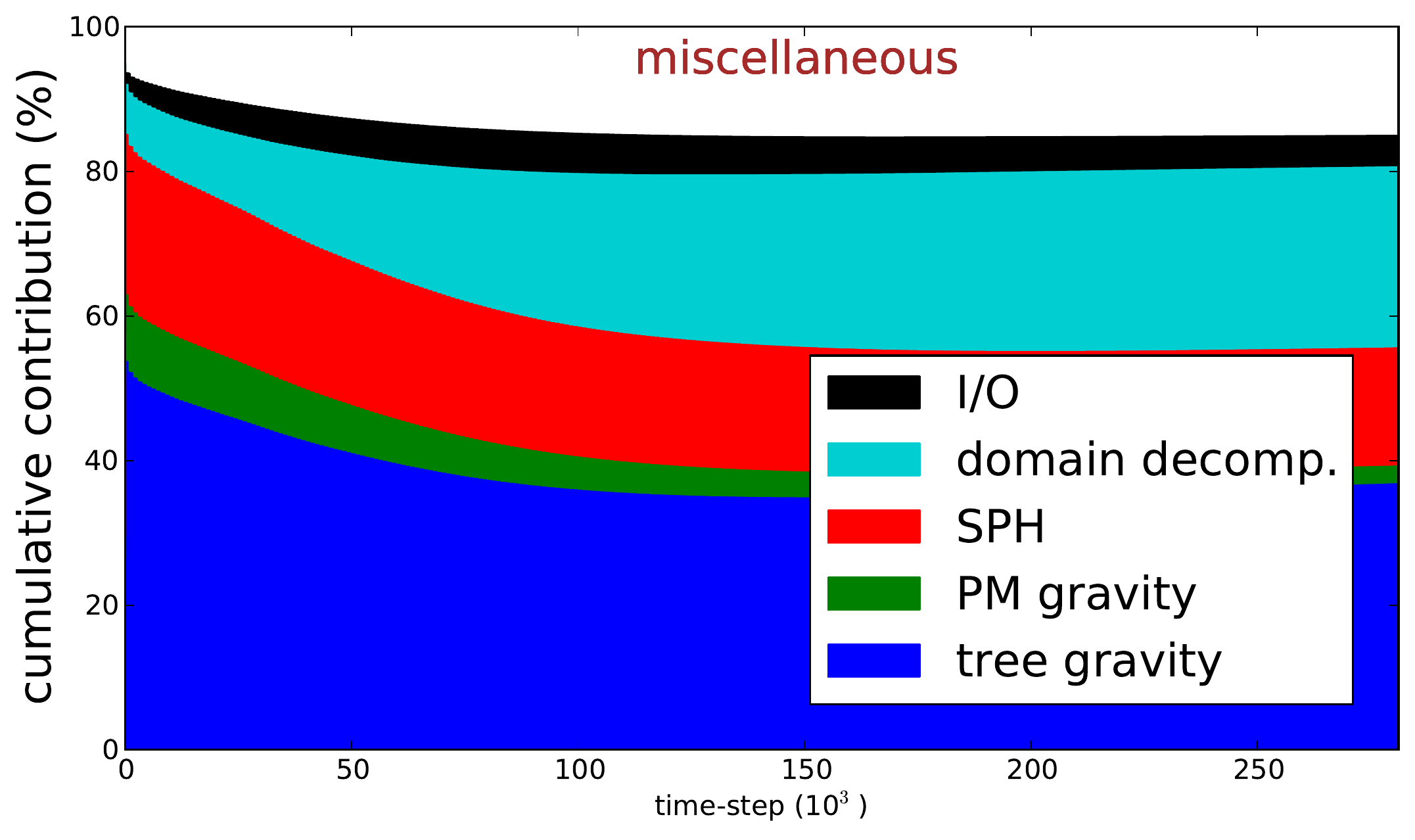}
\caption{Cumulative time spent in different parts of the code as a
  function of time-step for {\sm AREPO} (top panel) and {\sm GADGET}
  (bottom panel).  The Voronoi mesh contribution of {\sm AREPO} includes
  mesh constructions and mesh updates. The total number of time-steps
  in {\sm AREPO} is somewhat smaller because a larger CFL factor can
  be used, which makes the smallest time-step of {\sm AREPO} larger
  than that of {\sm GADGET}. The remaining fraction of computing time
  is spent in the time integration, star formation, cooling, and other
  minor contributions.}
\label{fig:runtime_stats}
\end{figure}

The total runtime of our {\sm AREPO} simulations is in the worst case
only about $30\%$ longer than the corresponding {\sm GADGET} run at
the same nominal resolution.  Fig.~\ref{fig:runtime_stats} shows in
which code parts most of this CPU time is spent. Obviously, the rather
complex mesh construction and mesh update in {\sm AREPO} takes up a
significant amount of time. However, in both codes the Tree-PM based
gravity calculation consumes a large amount of time, too, alleviating
speed differences in the hydrodynamical solvers. In any case, we argue
that {\sm AREPO} is actually surprisingly fast given the extremely
complex mesh management operations that are required. Given the small
difference in raw speed, one can rightly describe the overall
performance of both codes as similar. However, accounting for the fact that we here find
that SPH leads to a systematic and significant offset in the cooling
rates of haloes at late times, as well as in galaxy sizes, the
discussion of CPU time requirements at the same nominal resolution is
a bit moot. Ideally one would like to select the fastest method for a
given desired accuracy, and as it appears, this can be reached
with {\sm AREPO} more efficiently, whereas standard SPH's systematic bias may 
be difficult or even impossible to overcome. The white
parts in Fig.~\ref{fig:runtime_stats} account for the time spent in
the time integration, star formation, cooling, and other minor contributions.
We note that the I/O contribution is a bit larger than typical, because
we wrote out snapshots at a very high frequency.

\section{Strong and weak scaling}  

\begin{figure}
\centering
  \includegraphics[width=0.48\textwidth]{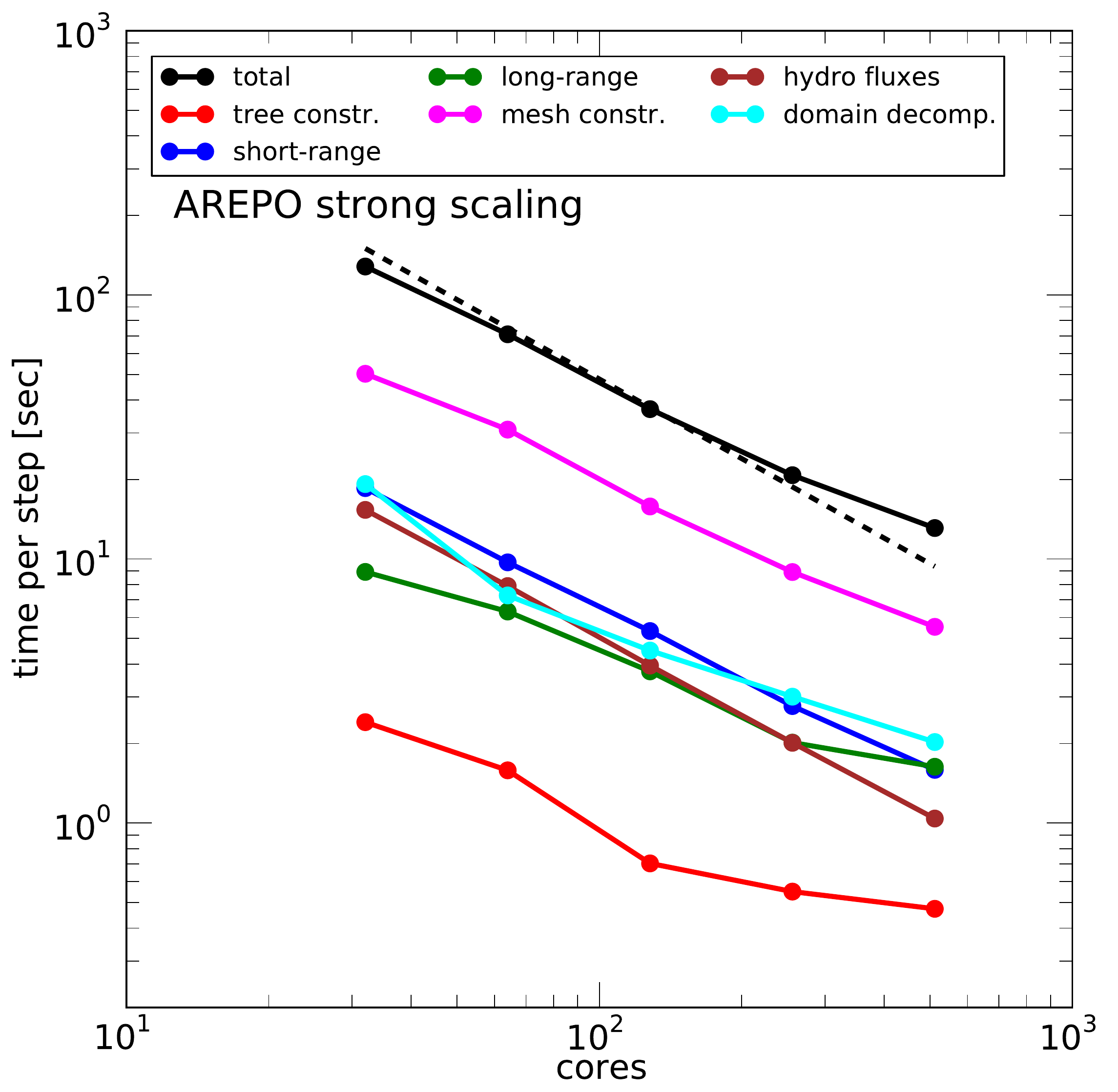}
\caption{
  Strong scaling test for the {\sm AREPO} code on
  Ranger, using between $32$ and $512$ cores. In all runs an identical
  simulation size of $2\times 256^3$ in a $12.5\,h^{-1}\,{\rm Mpc}$ box
  was used, with a $512^3$ FFT for the long-range gravity
  calculation. The black solid line shows the total wall-clock time for a full
  step as a function of the number of cores used, averaged over three
  steps at high redshift. The code was run with all physics enabled,
  and all code overhead was included in these averages. The other 
  measurements included in the figure give the times for the most
  important individual parts of the code, which are the Voronoi mesh
  construction, the short-range and long-range gravity calculations,
  the hydrodynamical flux calculation, the domain decomposition and
  the tree construction. The black dashed line indicates the ideal strong
  scaling.
}
\label{fig:strongscaling}
\end{figure}

\begin{figure}
\centering
  \includegraphics[width=0.48\textwidth]{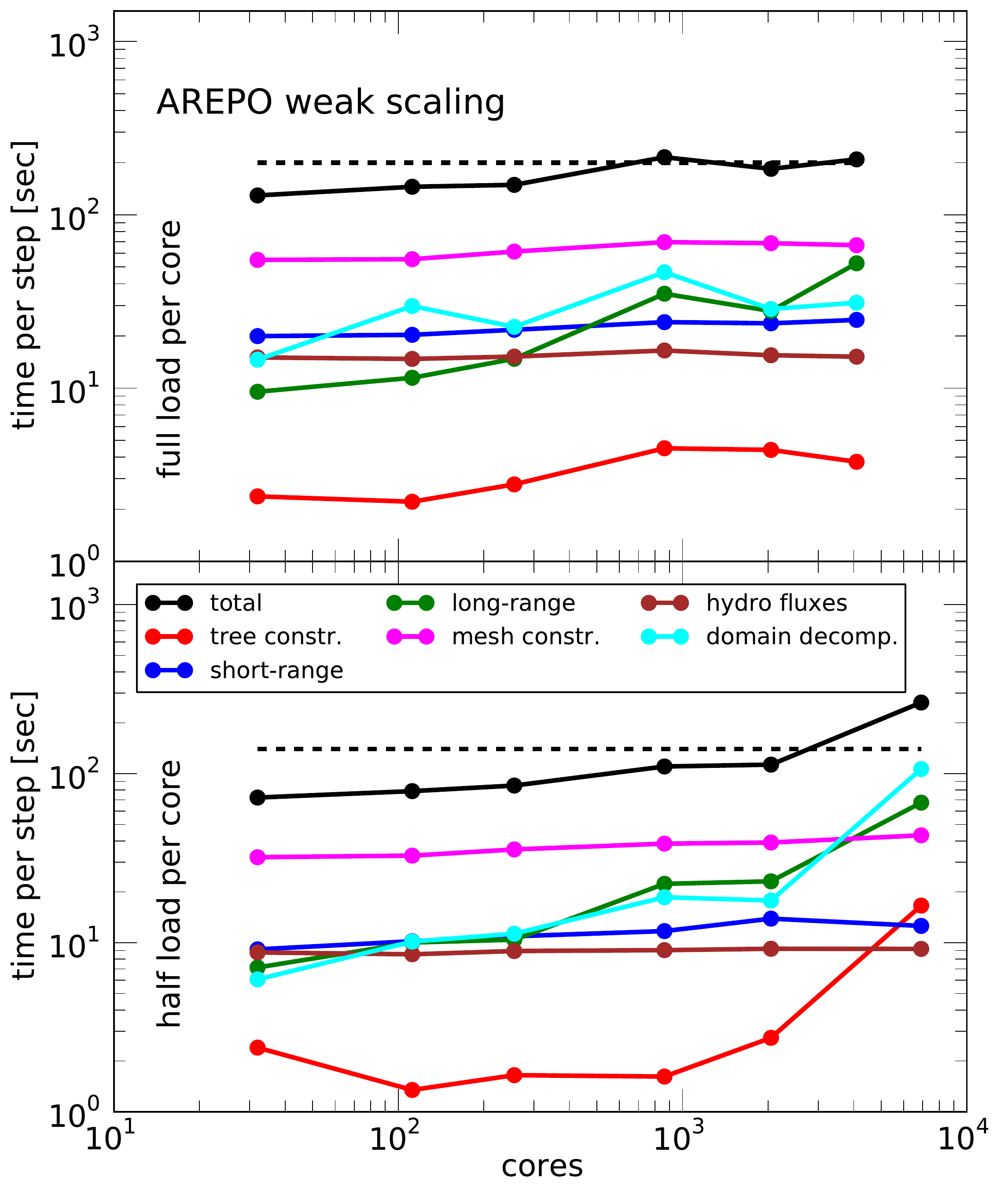}
\caption{
  The two panels show weak scaling plots for the {\sm AREPO} code on
  Ranger, using from 32 up to 6912 cores. In all runs of each of the
  two series, the load per core was held constant, and the simulation
  box-size and FFT size used was scaled in proportion to the core
  number. For ideal weak scaling, the code speed in all of its parts
  should then be constant, despite the fact that the simulation size
  grows by a factor of more than $200$ in these tests. We show in black the total times
  averaged for three full steps, and with different colours the most
  important code parts, as labelled. The top panel refers to a series
  with ``fully loaded'' nodes, where we use a load that is close to the
  maximum we can fit into the memory available per Ranger core. In the
  bottom panel, we have reduced the load by a factor of 2, allowing us to
  extend the tests to a core count of 6912 (for such large MPI
  partition sizes, less application memory remains available on the specific
  machine). Here, however, the communication costs in three parts of our
  code become quite substantial and lead to a noticeable negative impact of
  the scalability. The particle load of the top panel
  goes from $2\times256^3$ to $2\times1280^3$, whereas for the
  bottom panel it goes from $2\times214^3$ to $2\times1280^3$.
  The black dashed line indicates the ideal weak scaling.
}
\label{fig:weakscaling}
\end{figure}

In Fig.~\ref{fig:strongscaling}, we show a strong scaling test of {\sm
  AREPO}, from $32$ to $512$ cores, carried out on Ranger at the Texas
Advanced Computing Center (TACC).  Here, the simulation size has been
kept constant at $2\times 256^3$, and only the number of compute cores
has been increased. The reported times have been averaged for three
full time-steps at high redshift. In this regime, gravity accounts for
$\sim 33\%$ of the computational time, mesh-construction and mesh
bookkeeping consumes $\sim 40\%$ of the time, while the calculation of
the hydrodynamical fluxes amounts to $\sim 10\%$ of the time.  The
remainder is needed for miscellaneous items such as domain
decomposition, radiative cooling and star formation. In order to
clearly show the scalability of the most important different parts of
our code, we have included separate measurements in
Fig.~\ref{fig:strongscaling} for long-range gravity (by means of
FFTs), short-range gravity (done through a tree-walk), the necessary
tree construction, the mesh construction and management, the
hydrodynamic flux calculations, and the domain decomposition.  We see
that the code shows quite good strong scaling, despite the tightly
coupled nature of the system. Some losses are in particular apparent
for the mesh construction. These arise because the more cores are
used, the larger the number of spatial domains in which the simulation
volume is cut. This enlarges the surface area of domain boundaries,
and as a result the cumulative ``ghost'' region volume in which the
mesh has to be constructed twice on two neighbouring processes to
ensure seamless consistency across the domain boundary.

For large-scale cosmological applications it is more important that the code
parallelisation shows good weak scaling behaviour. {\sm GADGET} has been
used for many large-scale simulations and we will demonstrate here
that {\sm AREPO} performs also well in weak scaling tests.  For these tests the
computational load per core is kept constant, and the simulation volume
and particle/cell number is scaled accordingly. In
Fig.~\ref{fig:weakscaling}, we show two different weak scaling
series carried out on Ranger at the TACC. They differ in their
particle and cell load per core. In the results shown in the top panel,
this load was `maximal', corresponding to $\sim 1$ million cells and
particles per core, where the current version of {\sm AREPO} requires up to
1600 MB memory consumption per core in the peak. Because some memory is needed
for the operating system and the MPI communication library, not all of
the physical memory is available for the application code on the
Ranger platform. Unfortunately, the amount of memory consumed by the
MPI subsystem increases with increasing size of the MPI
partition. This in fact prevents us from running the ``full load'' 
configuration for partitions larger than 4096 cores. This also
prompted us to create a second weak scaling series shown in the bottom
panel of Fig.~\ref{fig:weakscaling}, where we reduced the load by almost a
factor of 2. This allows us to scale {\sm AREPO} up to $6912$ cores with
$6912$ MPI tasks.
 
It is apparent that the short-range gravitational tree calculation,
the Voronoi mesh construction, and the hydrodynamical flux calculation
show excellent weak scaling (which runs horizontally in the plots of
Fig.~\ref{fig:weakscaling}). However, there are deviations from
perfect scalability for the FFT-based long-range gravitational force
calculation, the domain decomposition, and the tree construction. This
is primarily because all three of these parts involve substantial
all-to-all communication. For larger processor counts, especially for
$6912$ cores, these communication costs start to affect the weak
scalability of our code and lead to losses in efficiency. One reason
for the sub-optimal scaling in this regime lies in the FFT part of the
code. The slab-based parallelisation of the FFTs we use for the
long-range gravity calculation does not scale to core numbers larger
than the size of the FFT, a regime we enter once more than 4096 cores
are used (because the FFT-size reaches $4096^3$ at this point and
cannot be grown further due to memory constraints). We note that this
can easily be optimised by either using a block-structured FFT
decomposition or by using a mixed approach for the parallelisation
with either threads or OpenMP. Overall we find that {\sm AREPO} shows
good weak scaling behaviour and can readily be applied to large-scale
cosmological hydrodynamics simulations.

\label{lastpage}

\end{document}